\documentclass[journal]{IEEEtran}

\usepackage{booktabs}

\usepackage{hyperref}
\hypersetup{hidelinks}

\usepackage{xspace}

\usepackage{prettyref}

\usepackage{comment}

\usepackage{amsfonts}

\usepackage{amssymb}
\usepackage{amsmath}

\usepackage{dsfont}
\usepackage{pifont}

\usepackage[normalem]{ulem}

\usepackage[hypcap]{caption}

\usepackage{enumitem}

\usepackage{balance}

\usepackage{float}

\usepackage[ruled,lined,noend,linesnumbered,procnumbered]{algorithm2e}

\usepackage{subcaption}

\usepackage{multirow}

\usepackage{color}

\usepackage{graphicx}

\usepackage{orcidlink}

\usepackage[noadjust]{cite}

\newrefformat{eq}{(\ref{#1})}
\newrefformat{sec}{Section~\ref{#1}}
\newrefformat{alg}{Algorithm~\ref{#1}}
\newrefformat{fig}{Fig.~\ref{#1}}
\newrefformat{tab}{Table~\ref{#1}}
\newrefformat{rmk}{Remark~\ref{#1}}
\newrefformat{clm}{Claim~\ref{#1}}
\newrefformat{def}{Definition~\ref{#1}}
\newrefformat{cor}{Corollary~\ref{#1}}
\newrefformat{lem}{Lemma~\ref{#1}}
\newrefformat{prop}{Proposition~\ref{#1}}
\newrefformat{app}{Appendix~\ref{#1}}
\newrefformat{ass}{Assumption~\ref{#1}}
\newrefformat{hyp}{Hypothesis~\ref{#1}}
\newrefformat{thm}{Theorem~\ref{#1}}
\newrefformat{assump}{Assumption~\ref{#1}}
\newrefformat{conj}{Conjecture~\ref{#1}}
\newrefformat{prob}{Problem~\ref{#1}}

\newtheorem{problem}{Problem}
\newtheorem{definition}{Definition}
\newtheorem{theorem}{Theorem}
\newtheorem{corollary}{Corollary}
\newtheorem{lemma}{Lemma}

\SetKw{Or}{or}
\SetKw{Break}{break}
\SetKw{And}{and}

\newcommand{\calA}{{\mathcal{A}}}

\newcommand{\calD}{{\mathcal{D}}}

\newcommand{\calG}{{\mathcal{G}}}
\newcommand{\calH}{{\mathcal{H}}}
\newcommand{\calI}{{\mathcal{I}}}

\newcommand{\calP}{{\mathcal{P}}}
\newcommand{\calQ}{{\mathcal{Q}}}

\newcommand{\calS}{{\mathcal{S}}}

\newcommand{\diff}{{\rm d}}
\newcommand{\Expect}{\mathbb{E}}
\newcommand{\var}{\mathsf{var}}
\newcommand{\bm}{\boldsymbol}

\newcommand{\ie}{i.e.\xspace}

\newcommand{\vnge}{{\mathcal{H}_{\rm{vn}}}}
\newcommand{\ose}{{\mathcal{H}_1}}
\newcommand{\vol}{{\rm{vol}}}
\newcommand{\tr}{{\rm tr}}
\newcommand{\bx}{{\bf{x}}}
\newcommand{\by}{{\bf{y}}}

\definecolor{mygreen}{RGB}{0, 100, 0}
\newcommand{\xmark}{\textcolor{red}{\ding{55}}}
\newcommand{\cmark}{\textcolor{mygreen}{\ding{51}}}

\newcommand{\overbar}[1]{\mkern 1.5mu\overline{\mkern-1.5mu#1\mkern-1.5mu}\mkern 1.5mu}

\begin{document}
    \title{On the Similarity between von Neumann Graph Entropy and Structural Information: Interpretation, Computation, and Applications}
    \author{Xuecheng~Liu\textsuperscript{\orcidlink{0000-0001-5937-5307}},
            Luoyi~Fu\textsuperscript{\orcidlink{0000-0001-7796-9168}},
            Xinbing~Wang\textsuperscript{\orcidlink{0000-0002-0357-8356}},
            and~Chenghu~Zhou
            \thanks{Manuscript received February 18, 2021; revised November 9, 2021; accepted January 2, 2022. Date of publication TBD; date of current version TBD.
            This work was supported by National Key R\&D Program of China (No. 2018YFB2100302), 
            NSF China (No. 42050105, 62020106005, 62061146002, 61960206002, 61822206, 61832013, 61829201), 
            2021 Tencent AI Lab RhinoBird Focused Research Program (No. JR202132), 
            and the Program of Shanghai Academic/Technology Research Leader under Grant No. 18XD1401800.
            \emph{(Corresponding author: Xinbing Wang.)}}
            \thanks{Xuecheng Liu and Xinbing Wang are with the Department of Electronic Engineering, Shanghai Jiao Tong University, Shanghai, 200240 China (email:liuxuecheng@sjtu.edu.cn, xwang8@sjtu.edu.cn).}
            \thanks{Luoyi Fu is with the Department of Computer Science and Engineering, Shanghai Jiao Tong University, Shanghai, 200240 China (email:yiluofu@sjtu.edu.cn).}
            \thanks{Chenghu Zhou is with the Institute of Geographical Science and Natural Resources Research, Chinese Academy of Sciences, Beijing, 100101 China (email:zhouch@lreis.ac.cn).}
            \thanks{Communicated by R. Talmon, Associate Editor for Signal Processing.}
            \thanks{Copyright (c) 2017 IEEE. Personal use of this material is permitted.  However, permission to use this material for any other purposes must be obtained from the IEEE by sending a request to pubs-permissions@ieee.org.}
    }
    \markboth{IEEE TRANSACTIONS ON INFORMATION THEORY}{}
    \maketitle

\begin{NoHyper}
    \begin{abstract}
        The von Neumann graph entropy is a measure of graph complexity based on the Laplacian spectrum.
        It has recently found applications in various learning tasks driven by the networked data.
        However, it is computationally demanding and hard to interpret using simple structural patterns.
        Due to the close relation between the Laplacian spectrum and the degree sequence, we conjecture 
        that the structural information, defined as the Shannon entropy of the normalized degree sequence,
        might be a good approximation of the von Neumann graph entropy that is both scalable and interpretable.

        In this work, we thereby study the difference between the structural information and the von 
        Neumann graph entropy named as \emph{entropy gap}. Based on the knowledge that the degree sequence 
        is majorized by the Laplacian spectrum, we for the first time prove that the entropy gap is between 
        $0$ and $\log_2 e$ in any undirected unweighted graphs. Consequently we certify that the structural 
        information is a good approximation of the von Neumann graph entropy that achieves provable accuracy, 
        scalability, and interpretability simultaneously. This approximation is further applied to two 
        entropy-related tasks: network design and graph similarity measure, where a novel graph similarity 
        measure and the corresponding fast algorithms are proposed. Meanwhile, we show empirically and theoretically
        that maximizing the von Neumann graph entropy can effectively hide the community structure, and then propose 
        an alternative metric called \emph{spectral polarization} to guide the community obfuscation.

        Our experimental results on graphs of various scales and types show that the very small entropy gap readily
        applies to a wide range of simple/weighted graphs. As an approximation of the von Neumann graph entropy, the 
        structural information is the only one that achieves both high efficiency and high accuracy among the 
        prominent methods. It is at least two orders of magnitude faster than SLaQ \cite{Anton2020WWW} with comparable accuracy.
        Our structural information based methods also exhibit superior performance in downstream tasks such as 
        entropy-driven network design, graph comparison, and community obfuscation.
    \end{abstract}


    \section{Introduction}

\IEEEPARstart{E}{vidence} has rapidly grown in the past few years that graphs are ubiquitous in our daily life;
online social networks, metabolic networks, transportation networks, and collaboration networks are 
just a few examples that could be represented precisely by graphs. One important issue in graph analysis is to 
measure the complexity of these graphs \cite{Bonchev2005,Li2016TIT} which refers to the level of organization of the structural features such as the 
scaling behavior of degree distribution, community structure, core-periphery structure, etc.
In order to capture the inherent structural complexity of graphs, many 
entropy based graph measures \cite{Li2016TIT,Rashevsky1955,Raychaudhury1984,Konstantinova1990,Dehmer2008,Samuel2006} are proposed, each of which is a specific form 
of the Shannon entropy for different types of distributions extracted from the graphs.

As one of the aforementioned entropy based graph complexity measures,
the von Neumann graph entropy, defined as the Shannon entropy of the spectrum of the trace rescaled Laplacian matrix of a graph (see \prettyref{def:VNGE}), is of special interests to scholars and practitioners \cite{Chen2019ICML,Anton2020WWW,Choi2020,Kontopoulou2020TIT,Domenico2016PRX,Domenico2015Nature,Lu2011,Joshua2016}.
This spectrum based entropy measure distinguishes between different graph structures.
For instance, it is maximal for complete graphs, minimal for graphs with only a single edge, and takes on intermediate values for ring graphs.
Historically, the entropy measure originates from quantum information theory and is used to 
describe the mixedness of a quantum system which is represented as a density matrix. It is Braunstein et al. that first use the von Neumann entropy to measure the complexity of graphs by viewing the scaled Laplacian matrix as a density matrix \cite{Samuel2006}.

Built upon the Laplacian spectrum, 
the von Neumann graph entropy is a natural choice to capture the graph complexity
since the Laplacian spectrum is well-known to contain rich information about the 
multi-scale structure of graphs \cite{Jamakovic2007,Ghosh2006CDC}.
As a result, it has recently found applications in downstream tasks of complex network analysis and pattern recognition.
For example, the von Neumann graph entropy facilitates the measure of graph similarity via Jensen-Shannon divergence, which could be used to 
compress multilayer networks \cite{Domenico2015Nature}
and detect anomalies in graph streams \cite{Chen2019ICML}.
As another example, the von Neumann graph entropy could be used to design edge centrality measure \cite{Joshua2016}, vertex centrality measure \cite{Rossi2017}, and entropy-driven networks \cite{Giorgia2018}.

\subsection{Motivations}

However, despite the popularity received in applications, the main obstacle encountered in practice is the computational inefficiency of the exact von Neumann graph entropy.
Indeed, as the spectrum based entropy measure, the von Neumann graph entropy suffers from computational inefficiency since the 
computational complexity of the graph spectrum is cubic in the number of nodes.
Meanwhile, the existing approximation approaches \cite{Chen2019ICML,Choi2020,Anton2020WWW} such as the quadratic approximation,
fail to capture the presence of non-trivial structural patterns that seem to be encapsulated in the spectrum based entropy measure.
Therefore, {\em there is a strong desire to find a good approximation that achieves accuracy, scalability, and interpretability simultaneously}.

Instead of starting from scratch, we are inspired by the well-known knowledge that there is a close relationship between the combinatorial
characteristics of a graph and the algebraic properties of its associated matrices \cite{Chung1997}.
To illustrate this, we plot the Laplacian spectrum and the degree sequence together in a same figure for four representative real-world graphs and four synthetic graphs.
As shown in \prettyref{fig:spectra-and-degree}, the sorted spectrum sequence and the sorted degree sequence almost coincide with each other.
The similar phenomenon can also be observed in larger scale free graphs, which indicates that 
it is possible to reduce the approximation
of the von Neumann graph entropy to the time-efficient computation of simple node degree statistics.
Therefore, we ask without hesitation the first research question,

  \vspace{1mm}
  {\bf RQ1:} {\em Does there exist some \uline{non-polynomial} function $\phi$ such that $\sum_{i=1}^n \phi\left(d_i/\sum_{j=1}^n d_j\right)$ is close to the
  von Neumann graph entropy? 
  Here $d_i$ is the degree of the node $i$ in a graph of order $n$.
  } 
  \vspace{1mm}


We emphasize the 
non-polynomial property of the function $\phi$ since most of previous works that are based on the polynomial approximations fail to fulfill the interpretability.
The challenges from both the scalability and the interpretability are translated directly into two requirements on the function $\phi$ to be determined.
First, the explicit expression of $\phi$ must exist and be kept simple to ensure the 
interpretability of the sum over degree statistics. Second, 
the function $\phi$ should be graph-agnostic to meet the scalability requirement, that is, 
$\phi$ should be independent from the graph to be analyzed. 
One natural choice yielded by the entropic nature of the graph complexity measure for the 
non-polynomial function $\phi$ is $\phi(x)=-x\log_2 x$.
The sum $-\sum_{i=1}^n \left(d_i/\sum_{j=1}^n d_j\right)\log_2\left(d_i/\sum_{j=1}^n d_j\right)$ has been named as one-dimensional {\em structural information} 
by Li et al. \cite{Li2016TIT} in a connected graph since it has an entropic form and captures the information of a classic random walker in a graph.
We extend this notion to arbitrary undirected graphs.
Following the question {\bf RQ1}, we raise the second research question,





  \vspace{1mm}
  {\bf RQ2:} {\em Is the structural information an accurate proxy of the von Neumann graph entropy?}
  \vspace{1mm}

To address the second question, we conduct to our knowledge a first study of 
the difference between the structural information and the von Neumann graph entropy, which we name as 
{\em entropy gap}.

\begin{figure*}[t]
  \begin{subfigure}{0.24\linewidth}
    \includegraphics[width=\textwidth]{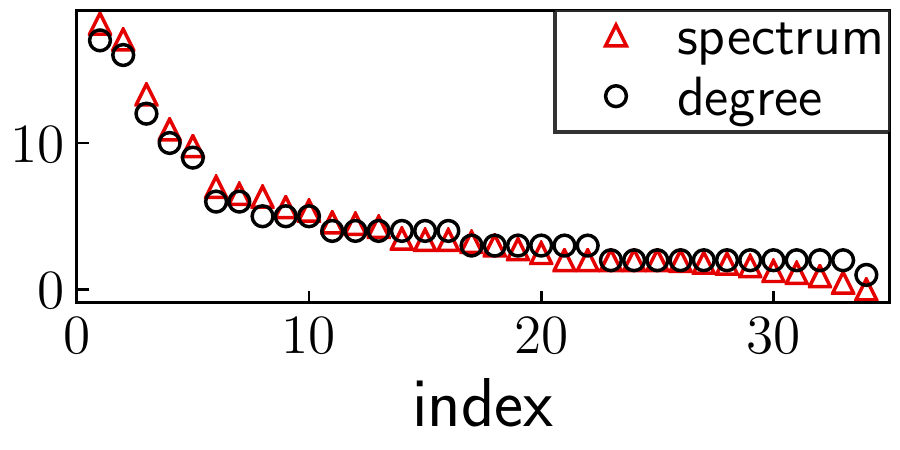}
    \caption{Zachary's karate club}
  \end{subfigure}
  \hspace{1mm}
  \begin{subfigure}{0.24\linewidth}
    \includegraphics[width=\textwidth]{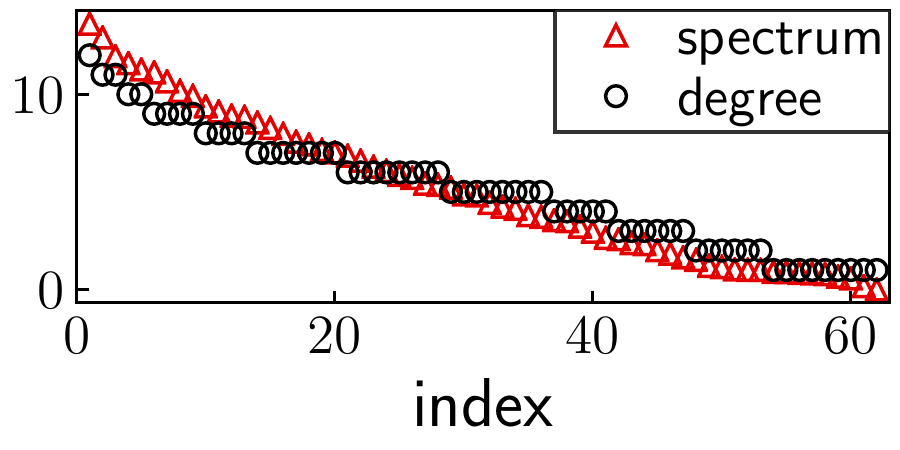}
    \caption{Dolphins}
  \end{subfigure}
  \begin{subfigure}{0.24\linewidth}
    \includegraphics[width=\textwidth]{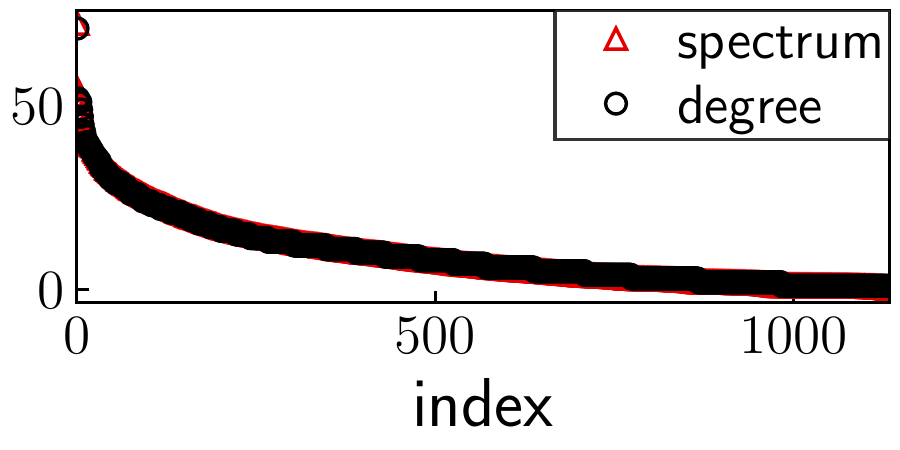}
    \caption{Email}
  \end{subfigure}
  \begin{subfigure}{0.24\linewidth}
    \includegraphics[width=\textwidth]{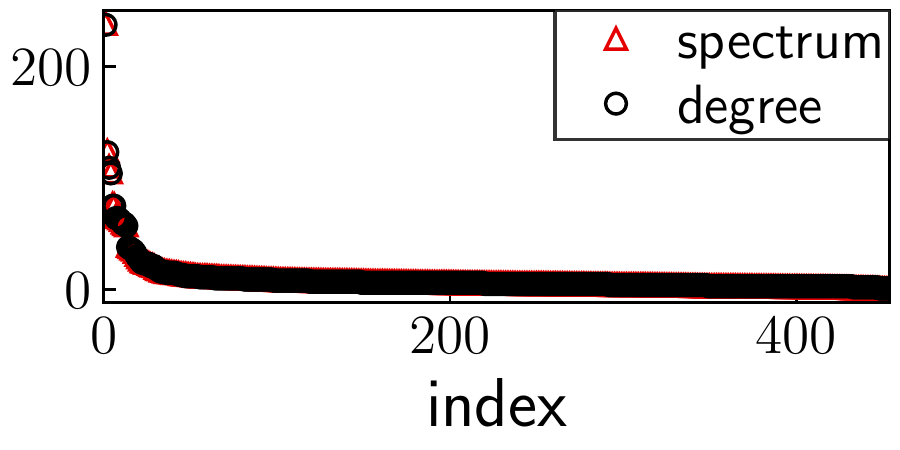}
    \caption{Celegans}
  \end{subfigure}
  \begin{subfigure}{0.24\linewidth}
    \includegraphics[width=\textwidth]{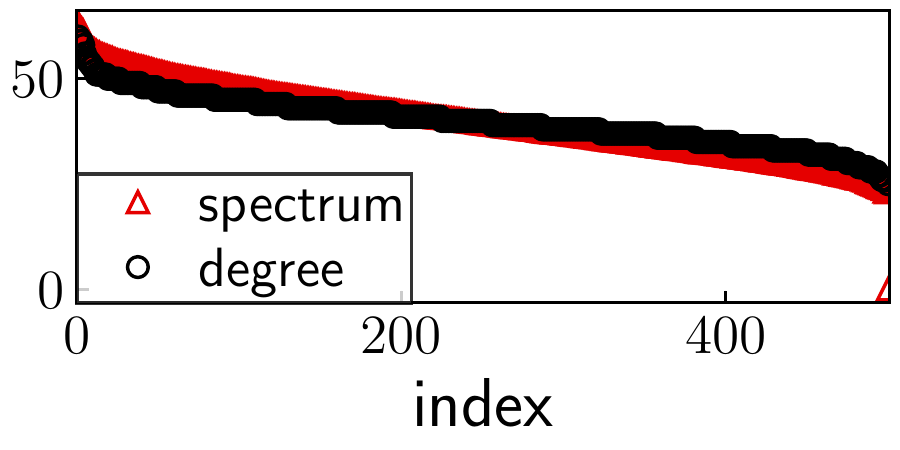}
    \caption{ER graph of order $500$}
  \end{subfigure}
  \begin{subfigure}{0.24\linewidth}
    \includegraphics[width=\textwidth]{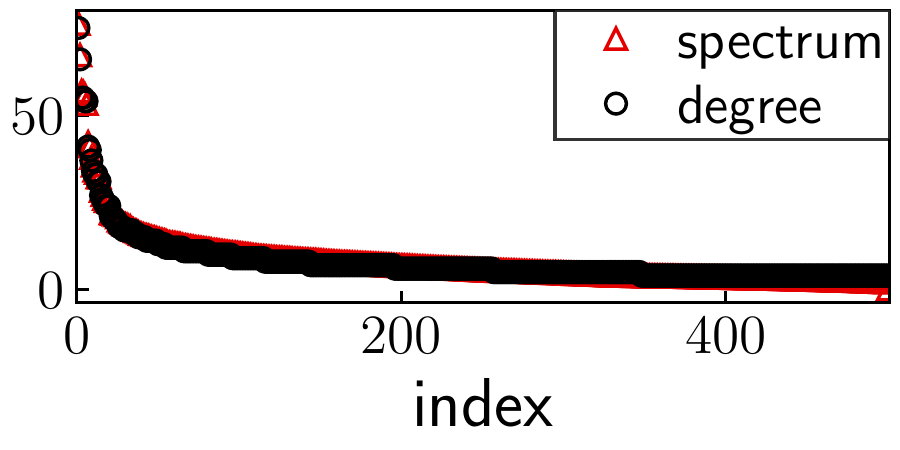}
    \caption{BA graph of order $500$}
  \end{subfigure}
  \begin{subfigure}{0.24\linewidth}
    \includegraphics[width=\textwidth]{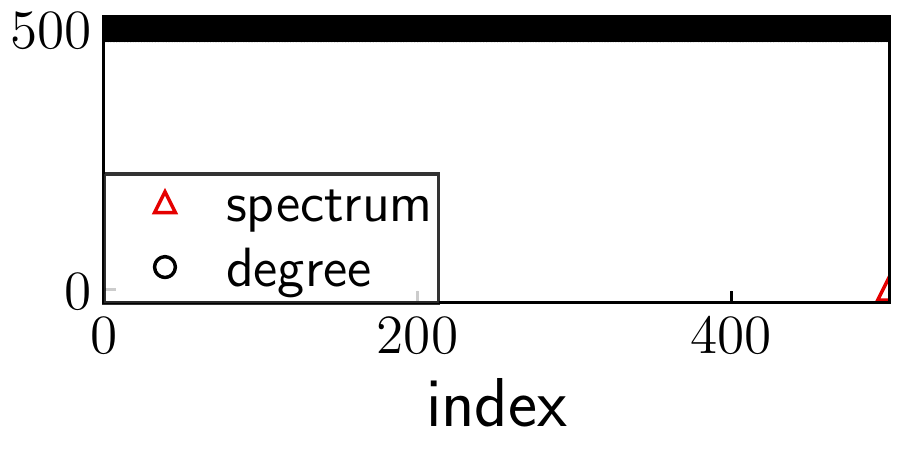}
    \caption{Complete graph of order $500$}
  \end{subfigure}
  \begin{subfigure}{0.24\linewidth}
    \includegraphics[width=\textwidth]{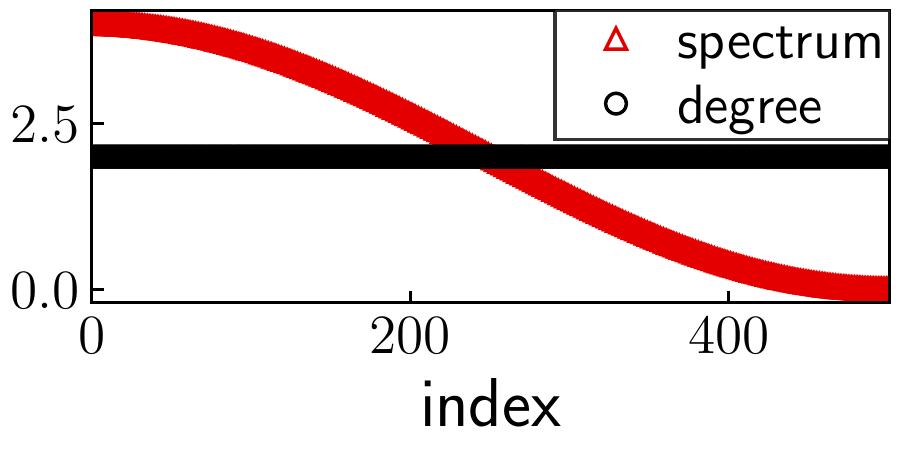}
    \caption{Ring graph of order $500$}
  \end{subfigure}
  \caption{The close relation between Laplacian spectra and degree sequence in four representative real-world graphs (a-d) and four common synthetic graphs (e-h). Both the Laplacian spectra and degree sequence are sorted in non-increasing order.
  The x-axis represents the index of the sorted sequences, and the y-axis represents the value of Laplacian spectrum and degree.}
  \label{fig:spectra-and-degree}
\end{figure*}

\begin{table*}[t]
  \centering
  \caption{\label{tab:accuracy} Structural information and von Neumann graph entropy of the graphs in \prettyref{fig:spectra-and-degree}.}
  \begin{tabular}{l||c|c|c|c|c|c|c|c}
    \toprule
    Measurements & Zachary & Dolphins & Email & Celegans & ER & BA & Complete & Ring \\
    \midrule
    structural information $\ose$ & $4.7044$ & $5.7005$ & $9.5665$ & $7.9257$ & $8.9497$ & $8.5739$ & $8.9659$ & $8.9658$ \\
    von Neumann graph entropy $\vnge$ & $4.5504$ & $5.5489$ & $9.5029$ & $7.8631$ & $8.9302$ & $8.4935$ & $8.9629$ & $8.5231$ \\
    entropy gap $\Delta\calH$ & $0.1540$ & $0.1516$ & $0.0636$ & $0.0626$ & $0.0195$ & $0.0804$ & $0.0030$ & $0.4427$ \\
    relative error $\frac{\Delta\calH}{\vnge}$ & $3.38\%$ & $2.73\%$ & $0.67\%$ & $0.80\%$ & $0.22\%$ & $0.95\%$ & $0.03\%$ & $5.19\%$ \\
    \bottomrule 
  \end{tabular}
\end{table*}

\subsection{Contributions}

To study the entropy gap,
we are based on a fundamental relationship between the Laplacian spectrum $\bm{\lambda}$ and the degree sequence $\mathbf{d}$ in undirected graphs: 
$\mathbf{d}$ is majorized by $\bm{\lambda}$. In other words, 
there is a doubly stochastic matrix $P$ such that 
$P{\bm\lambda}=\mathbf{d}$.
Leveraging the majorization and the classic Jensen's inequality, we prove that the entropy gap is strictly larger than $0$ in arbitrary undirected graphs.
By exploiting the Jensen's gap \cite{Liao2019} which is an inverse version of the classic Jensen's inequality, we further prove that the entropy gap is no more than
$\frac{\log_2 e\cdot\tr(A^2)}{\delta\cdot\vol(G)}$ for any undirected graph $G$, where $A$ is the weight matrix, $\delta$ is the minimum degree, and $\vol(G)$ is the volume of the graph.
The upper bound on the entropy gap turns out to be $\log_2 e$ for any unweighted graph.
And both the constant lower and upper bounds on the entropy gap can be further sharpened using more advanced knowledge about the Lapalcian spectrum and the degree sequence, such as the Grone-Merris majorization \cite{Bai2011TAMS}.

In a nutshell, our paper makes the following contributions:
\begin{itemize}[leftmargin=*]
  \item {\bf Theory and interpretability:} 
  Inspired by the close relation between the Laplacian spectrum and the degree sequence, we for the first time 
  bridge the gap between the von Neumann graph entropy and the structural information
  by proving that the entropy gap is between $0$ and $\log_2 e$ in any unweighted graph.
  To the best of our knowledge, the constant bounds on the approximation error in unweighted graphs are sharper than that of any existing approaches with provable accuracy, such as FINGER \cite{Chen2019ICML}.
  Therefore, the answers to both {\bf RQ1} and {\bf RQ2} are YES!
  As shown in \prettyref{tab:accuracy}, the relative approximation error is around $1\%$ for small graphs, which is practically good.
  Besides, the structural information provides a simple geometric interpretation of the von Neumann graph entropy
  as a measure of degree heterogeneity.
  Thus, 
  the structural information is a good approximation of the von Neumann graph entropy that achieves provable accuracy, scalability, and interpretability simultaneously.
  \item {\bf Applications and efficient algorithms:}
  Using the structural information as a proxy of the von Neumann graph entropy with bounded error (the entropy gap), we develop fast algorithms for two entropy based applications: network design and graph similarity measure.
  Since the network design problem aims to maximize the von Neumann graph entropy, we combine the greedy method and a pruning strategy to accelerate the searching process.
  For the graph similarity measure, we propose a new distance measure based on the structural information and the Jensen-Shannon divergence. 
  We further show that the proposed distance measure is a pseudometric and devise a fast incremental algorithm to compute the similarity between adjacent graphs in a graph stream. 
  \item {\bf Connection with community structure:} 
  While the two-dimensional structural information \cite{Li2016TIT} encodes the community structure, we find empirically that both the von Neumann graph entropy and the one-dimensional structural information are uninformative of the community structure.
  However, they are effective in adversarial attacks on community detection, since maximizing the von Neumann graph entropy will make the Laplacian spectrum uninformative of the community structure.
  Using the similar idea, we propose an alternative metric called {\em spectral polarization} which is both effective and efficient in hiding the community structure.
  \item {\bf Extensive experiments and evaluations:}
  We use 3 random graph models, 9 real-world static graphs, and 2 real-world temporal graphs to evaluate the properties
  of the entropy gap and the proposed algorithms. 
  The results show that the entropy gap is small in a wide range of simple/weighted graphs.
  And it is insensitive to the change of graph size. 
  Compared with the prominent methods for approximating the von Neumann graph entropy, the structural information is superior in both accuracy and computational speed.
  It is at least 2 orders of magnitude faster than the accurate SLaQ \cite{Anton2020WWW} algorithm with comparable accuracy.
  Our proposed algorithms based on structural information also exhibit superb performance
  in downstream tasks such as entropy-driven network design, graph comparison, and community obfuscation.
\end{itemize}

An earlier version of this work appeared in our WWW 2021 conference paper \cite{Liu2021WWW}.
In addition to revising the conference version, this TIT submission includes the following new materials:
\begin{itemize}
  \item More experimental results to illustrate the relationship between the Laplacian spectrum and the degree sequence (\prettyref{fig:spectra-and-degree});
  \item More straightforward results to illustrate the tightness of the approximation (\prettyref{tab:accuracy});
  \item A fine-grained analysis on the lower bound of the entropy gap to show that the entropy gap is actually strictly larger than $0$ (\prettyref{thm:bounds});
  \item A theoretical analysis on the entropy gap of several classes of graphs, including complete graph, complete bipartite graph, path graph, and ring graph (\prettyref{tab:various-graph-types} and \prettyref{app:proof-specific-graphs});
  \item An analysis and discussion over the connection between graph entropy and community structure (\prettyref{sec:comm} and \prettyref{app:appendix-comm}).
\end{itemize}

\noindent{\bf Roadmap:}
The remainder of this paper is organized as follows.
We review three related issues in \prettyref{sec:related-work}.
In \prettyref{sec:notation-and-definition} we introduce the definitions of the von Neumann graph entropy, structural information, and the notion of entropy gap.
\prettyref{sec:bound-on-entropy-gap} shows the close relationship between von Neumann graph entropy and structural information by bounding the entropy gap.
\prettyref{sec:applications-and-algorithms} presents efficient algorithms for two graph entropy based applications.
In \prettyref{sec:comm} we discuss the connection between von Neumann graph entropy and community structure.
\prettyref{sec:experiments} provides experimental results.
\prettyref{sec:conclusion-and-future-work} offers some conclusions and directions for future research.

    \section{Related Work}
\label{sec:related-work}

\begin{table}[t]
	\centering
	\small
  \caption{\label{tab:compare} Comparison of methods for approximating the von Neumann graph entropy in terms of fulfilled (\cmark) and missing (\xmark) properties.}
	\scalebox{0.9}{
	\begin{tabular}{l||ccc|c}
		\toprule
		& \cite{Chen2019ICML} & \cite{Anton2020WWW} & \cite{Choi2020} & Structural Information (Ours) \\
		\midrule
		Provable accuracy           &  \cmark          & \xmark    & \xmark &    \cmark               \\
		Scalability            & \cmark          & \cmark    & \xmark &      \cmark             \\
		Interpretability          & \xmark          & \xmark    & \xmark        &      \cmark             \\
		\bottomrule
	\end{tabular}
	}
\end{table}	

We review three issues related to the von Neumann graph entropy: computation, interpretation, and connection with the community structure.
The first two issues arise from the broad applications \cite{Joshua2016,Giorgia2018,Domenico2015Nature,Majtey2005PRA,Lamberti2008,Jop2009PRA,Lu2011,George2020} of the von Neumann graph entropy,
whereas the last issue comes from 
spectral clustering \cite{Ulrike2007} and two-dimensional structural information based clustering \cite{Li2016TIT}.

\subsection{Approximate Computation of the von Neumann Graph Entropy}
In an effort to overcome the computational inefficiency of the von Neumann graph entropy, 
past works have resorted to various numerical approximations.
Chen et al. \cite{Chen2019ICML} first compute a quadratic approximation of the entropy via Taylor expansion, then derive two finer approximations with accuracy guarantee by spectrum-based and degree-based rescaling, respectively.
Before Chen's work, the Taylor expansion is widely adopted to give computationally efficient approximations \cite{Cheng2014PRE}, but there is no theoretical guarantee on the approximation accuracy.
Following Chen's work, Choi et al. \cite{Choi2020} propose several more complicated quadratic approximations based on advanced polynomial approximation methods the superiority of which is verified through experiments.

Besides, there is a trend to approximate spectral sums using stochastic trace estimation based approximations \cite{Han2017SIAM}, the merit of which is the 
provable error-bounded estimation of the spectral sums.
For example, Kontopoulou et al. \cite{Kontopoulou2020TIT}
propose three randomized algorithms based on Taylor series, Chebyshev polynomials, and random projection matrices to approximate the von Neumann entropy of density matrices.
As another example, based on the stochastic Lanczos quadrature technique \cite{Ubaru2017}, Tsitsulin et al. \cite{Anton2020WWW} propose an efficient and effective approximation technique called SLaQ to estimate the von Neumann entropy and other spectral descriptors for web-scale graphs.
However, the approximation error bound of SLaQ for the von Neumann graph entropy is not provided.
The disadvantages of such stochastic approximations are also obvious; their computational efficiency depends on the number of random vectors used in stochastic trace estimation, and 
they are not suitable for applications like anomaly detection in graph streams and entropy-driven network design.

The comparison of methods for approximating the von Neumann graph entropy is presented in \prettyref{tab:compare}.
One of the common drawbacks of the aforementioned methods is the lack of interpretability, that is,
none of these methods provide enough evidence to interpret this spectrum based entropy measure in terms of structural patterns.
By contrast, as a good proxy of the von Neumann graph entropy, the structural information offers us the intuition that the spectrum based entropy measure is closely related to the degree heterogeneity of graphs.

\subsection{Spectral Descriptor of Graphs and Its Structural Counterpart}
Researchers in spectral graph theory have always been interested in establishing 
a connection between the combinatorial characteristics of a graph and the algebraic properties of its associated matrices.
For example, the algebraic connectivity (also known as Fiedler eigenvalue), defined as the second smallest eigenvalue
of a graph Laplacian matrix, has been used to measure the robustness \cite{Jamakovic2007} and synchronizability \cite{Yu2011TAC} of graphs.
The magnitude of the algebraic connectivity has also been found to reflect how well connected the overall graph is \cite{Ghosh2006CDC}.
As another example, the Fiedler vector, defined as the eigenvector corresponding to the Fiedler eigenvalue of a graph Laplacian matrix,
has been found to be a good sign of the bi-partition structure of a graph \cite{Ding2001ICDM}.
However, there are some other spectral descriptors that have found applications in graph analytics, but require more structural interpretations, such as 
the heat kernel trace \cite{Xiao2009,Anton2018KDD} and von Neumann graph entropy.

Simmons et al. \cite{Simmons2018} suggest to interpret the von Neumann graph entropy as the centralization of graphs, which is very similar to our interpretation using the structural information. 
They derive both upper and lower bounds on the von Neumann graph entropy in terms of graph centralization under 
some hard assumptions on the range of the von Neumann graph entropy.
Therefore, their results cannot be directly converted to accuracy guaranteed approximations of the von Neumann graph entropy for arbitrary simple graphs.
By constrast, our work shows that the structural information is an accurate, scalable, and interpretable proxy of the von Neumann graph entropy for arbitrary simple graphs.
Besides, the techniques used in our proof are also quite different from \cite{Simmons2018}.

\subsection{Spectrum, Detectability, and Significance of Community Structure}
Community structure is one of the most recognized characteristics of large scale real-world graphs in which similar nodes tend to cluster together.
Thus it has found applications in classification, recommendation, and link prediction, etc.
Started from the Fiedler vector, spectral algorithms have been widely studied for detecting the community structure in a graph \cite{Newman2006PRE,Krzakala2013PNAS} 
because they are simple and theoretically warranted.
Cheeger's inequality $\frac{\mu_2}{2}\leq h_G\leq\sqrt{2\mu_2}$ bounds the conductance $h_G$ of a graph $G$ using the second smallest eigenvalue of the normalized Laplacian matrix.
This is later generalized to multiway spectral partitioning \cite{Lee2014JACM} yielding the higher-order Cheeger inequalities
$\frac{\mu_k}{2}\leq\rho_G(k)\leq O(k^2)\sqrt{\mu_k}$ for each $k$, where $\mu_k$ is the $k$-th smallest eigenvalue of the normalized Lapalcian matrix
and $\rho_G(k)$ is the $k$-way expansion constant.
Since both $h_G$ and $\rho_G(k)$ measure the significance of community structure,
the graph spectrum is closely related to the community structure.

The coupling between graph spectrum and community structure has been empirically validated in \cite{Newman2006PRE} where Newman 
found that if the second smallest eigenvalue $\lambda_2$ of the Laplacian matrix is well separated from the eigenvalues above it,
the spectral clustering based on Lapalcian matrix often does very well.
However, community detection by spectral algorithms in sparse graphs often fails, because the spectrum contains no clear evidence of community structure.
This is exemplified under the sparse stochastic block model with two clusters of equal size \cite{Nadakuditi2012PRL,Krzakala2013PNAS},
where the second smallest eigenvalue of the adjacency matrix gets lost in the bulk of uninformative eigenvalues.
Our experiments complement the correlation between graph spectrum and community structure by showing that 
the spikes in a sequence of spectral gaps are good indicators of the community structure.

The significance and detectability of community structure has found its application in an emerging area called {\em community obfuscation} \cite{Nagaraja2010,Chen2017CCS,Fionda2018TKDE,Waniek2018Nature,Liu2019NIPS,Chen2019,Jia2020WWW},
where the graph structure is minimally perturbed to protect its community structure from being detected.
None of these practical algorithms exploit the correlation between graph spectrum and community structure except for the structural entropy proposed by Liu et al. \cite{Liu2019NIPS}.
Our work bridges the one-dimensional structural entropy in \cite{Liu2019NIPS} with the spectral entropy, elaborates both empirically and theoretically that
maximizing the spectral entropy is effective in community obfuscation, and thus provides a theoretical foundation for the 
success of the structural entropy \cite{Liu2019NIPS}.
    \section{Preliminaries}
\label{sec:notation-and-definition}

In this paper, we study the undirected graph $G=(V,E,A)$ with positive edge weights,
where $V=[n]\triangleq\{1,\ldots,n\}$ is the node set, $E$ is the edge set, and $A\in\mathbb{R}_{+}^{n\times n}$ is 
the symmetric weight matrix with positive entry $A_{ij}$ denoting the weight of an edge $(i,j)\in E$.
If the node pair $(i,j)\notin E$, then $A_{ij}=0$. If the graph $G$ is unweighted, the weight matrix $A\in\{0,1\}^{n\times n}$
is called the adjacency matrix of $G$.
The degree of the node $i\in V$ in the graph $G$ is defined as 
$d_i=\sum_{j=1}^n A_{ij}$.
The Laplacian matrix of the graph $G$
is defined as $L\triangleq D-A$ where $D={\rm diag}(d_1,\ldots,d_n)$ is the degree matrix.
Let $\{\lambda_i\}_{i=1}^n$ be the sorted eigenvalues of $L$ such that $0=\lambda_1\leq\lambda_2\leq\cdots\leq\lambda_n$, which is called Laplacian spectrum.
We define $\vol(G)=\sum_{i=1}^n d_i$ as the volume of graph $G$, then $\vol(G)={\rm tr}(L)=\sum_{i=1}^n\lambda_i$ where ${\rm tr}(\cdot)$ is the trace operator.
For the convenience of delineation, we define a special function $f(x)\triangleq x\log_2 x$
on the support $[0,\infty)$ where $f(0)\triangleq\lim_{x\downarrow 0}f(x)=0$ by convention.
In the following, we present the formal definitions of the von Neumann graph entropy, the structural information, and the entropy gap.
Slightly different from the one-dimensional structural information proposed by Li et al. \cite{Li2016TIT},
our definition of structural information does not require the graph $G$ to be connected.

\begin{definition}[von Neumann graph entropy]\label{def:VNGE}
  The von Neumann graph entropy of an undirected graph $G=(V,E,A)$ is defined as 
  $\vnge(G)=-\sum_{i=1}^n f(\lambda_i/\vol(G))$,
  where $0=\lambda_1\leq\lambda_2\leq\cdots\leq\lambda_n$ are the eigenvalues of the Laplacian matrix $L=D-A$
  of the graph $G$, and $\vol(G)=\sum_{i=1}^n \lambda_i$ is the volume of $G$.
\end{definition}

\begin{definition}[Structural information]\label{def:structural-information}
    The structural information of an undirected graph $G=(V,E,A)$ is defined as 
    $\ose(G)=-\sum_{i=1}^n f(d_i/\vol(G))$,
    where $d_i$ is the degree of node $i$ in $G$ and $\vol(G)=\sum_{i=1}^n d_i$ is the volume of $G$.
\end{definition}

\begin{definition}[Entropy gap]\label{def:entropy-gap}
    The entropy gap of an undirected graph $G=(V,E,A)$ is defined as 
    $\Delta\calH(G)=\ose(G)-\vnge(G)$.
\end{definition}

The von Neumann graph entropy and the structural information are well-defined
for all the undirected graphs except for the graphs with empty edge set, in which
$\vol(G)=0$. When $E=\varnothing$, we take it for granted that $\ose(G)=\vnge(G)=0$.

    \section{Approximation Error Analysis}
\label{sec:bound-on-entropy-gap}

In this section we bound the entropy gap in the undirected graphs of order $n$. 
Since the nodes with degree $0$ have no contribution to both the structural information and the von Neumann graph entropy, without loss of generality we assume that 
$d_i>0$ for any node $i\in V$.

\subsection{Bounds on the Approximation Error}
We first provide bounds on the additive approximation error in \prettyref{thm:bounds}, \prettyref{cor:constant-bounds}, and \prettyref{cor:regular-graph}, then 
analyze the multiplicative approximation error in \prettyref{thm:multiplicative-error}.
\begin{theorem}[Bounds on the absolute approximation error]
    \label{thm:bounds}
    For any undirected graph $G=(V,E,A)$, the inequality
    \begin{equation}\label{eq:entropy-gap}
        0<\Delta\calH(G)\leq\frac{\log_2 e}{\delta}\cdot \frac{\mathrm{tr}(A^2)}{\vol(G)}
    \end{equation}
    holds, where $\delta=\min\{d_i|d_i>0\}$ is the minimum positive degree.
\end{theorem}

Before proving \prettyref{thm:bounds}, we introduce two techniques: the majorization and the Jensen's gap.
The former one is a preorder of the vector of reals, while the latter is an inverse version of the Jensen's inequality,
whose definitions are presented as follows.
\begin{definition}[Majorization \cite{Marshall2011}]
  For a vector $\mathbf{x}\in\mathbb{R}^d$, we denote 
by $\mathbf{x}^\downarrow\in\mathbb{R}^d$ the vector with the same components, but sorted in descending order.
Given $\mathbf{x},\mathbf{y}\in\mathbb{R}^d$, we say that $\mathbf{x}$ majorizes $\mathbf{y}$
(written as $\mathbf{x}\succ\mathbf{y}$) if and only if 
  $\sum_{i=1}^k x^\downarrow_i\geq\sum_{i=1}^k y^\downarrow_i$    
for any $k\in[d]$ and $\mathbf{x}^\intercal\mathbf{1}=\mathbf{y}^\intercal\mathbf{1}$.
\end{definition}
\begin{lemma}[Jensen's gap \cite{Liao2019}]\label{lem:Jensen-gap}
  Let $X$ be a one-dimensional random variable with the mean $\mu$ and the support $\Omega$.
  Let $\psi(x)$ be a twice differentiable function on $\Omega$ and define the function 
  $h(x)\triangleq\frac{\psi(x)-\psi(\mu)}{(x-\mu)^2}-\frac{\psi'(\mu)}{x-\mu}$, then 
  $\Expect[\psi(X)]-\psi(\Expect[X])\leq\sup_{x\in\Omega}\{h(x)\}\cdot\var(X)$.
  Additionally, if $\psi'(x)$ is convex, then $h(x)$ is monotonically increasing in $x$, and if $\psi'(x)$
  is concave, then $h(x)$ is monotonically decreasing in $x$.
\end{lemma}
\begin{lemma}\label{lem:convexity}
  The function $f(x)=x\log_2 x$ is convex, its first order derivative $f'(x)=\log_2 x+\log_2 e$
  is concave.
\end{lemma}
\begin{IEEEproof}
  The second order derivative $f''(x)=(\log_2 e)/x>0$, thus $f(x)=x\log_2 x$ is convex.
\end{IEEEproof}
We can see that the majorization characterizes the degree of concentration between the two vectors $\bx$ and $\by$.
Specifically,
$\bx\succ\by$ means that the entries of $\by$ are more concentrated on its mean $\by^\intercal\mathbf{1}/\mathbf{1}^\intercal\mathbf{1}$ than the entires of $\bx$.
An equivalent definition of the majorization \cite{Marshall2011} using linear algebra says that $\bx\succ\by$ if and only if 
there exists a doubly stochastic matrix $P$ such that $P\bx=\by$.
As a famous example of the majorization, the Schur-Horn theorem \cite{Marshall2011} says that
the diagonal elements of a positive semidefinite Hermitian matrix are majorized by its eigenvalues.
Since $\bx^T L\bx=\sum_{(i,j)\in E}A_{ij}(x_i-x_j)^2\geq0$ for any vector $\bx\in\mathbb{R}^n$, the Laplacian matrix $L$
is a positive semidefinite symmetric matrix whose diagonal elements form the degree sequence $\mathbf{d}$ and eigenvalues form the spectrum $\bm{\lambda}$.
Therefore, the majorization $\bm{\lambda}\succ\mathbf{d}$ implies that there exists some doubly stochastic matrix $P=(p_{ij})\in[0,1]^{n\times n}$ such that 
$P\bm{\lambda}=\mathbf{d}$.

Using the relation $P\bm{\lambda}=\mathbf{d}$ and the convexity of $f(x)$ in \prettyref{lem:convexity}, we can now proceed to prove \prettyref{thm:bounds}.

\begin{IEEEproof}[Proof of \prettyref{thm:bounds}]
  For each $i\in V$, we define a discrete random variable
  $X_i$ whose probability mass function is given by $\sum_{j=1}^n p_{ij}\delta_{\lambda_j}(x)$, where 
  $\delta_a(x)$ is the Kronecker delta function.
  Then the expectation $\Expect[X_i]=\sum_{j=1}^n p_{ij}\lambda_j= d_i$ and the variance 
  $\var(X_i)=\sum_{j=1}^n p_{ij}(\lambda_j-d_i)^2=\sum_{j=1}^n p_{ij}\lambda_j^2 - d_i^2$.

  First, we express the entropy gap in terms of the Lapalcian spectrum and the degree sequence.
  Since 
  \begin{equation}
    \begin{aligned}
      \ose(G)&=-\sum_{i=1}^n\left(\frac{d_i}{\vol(G)}\right)\log_2\left(\frac{d_i}{\vol(G)}\right) \\
      &=-\frac{1}{\vol(G)}\left(\sum_{i=1}^n f(d_i)-\sum_{i=1}^n d_i\log_2\left(\vol(G)\right)\right) \\
      &=\log_2\left(\vol(G)\right)-\frac{\sum_{i=1}^n f(d_i)}{\vol(G)},
    \end{aligned}
  \end{equation}
  and similarly 
  \begin{equation}\label{eq:vnge-simplification}
    \vnge(G)=\log_2(\vol(G))-\frac{\sum_{i=1}^n f(\lambda_i)}{\vol(G)}, 
  \end{equation}
  we have 
  \begin{equation}\label{eq:entropy-gap-expression}
    \Delta\calH(G)=\ose(G)-\vnge(G) 
    =\frac{\sum_{i=1}^n f(\lambda_i)-\sum_{i=1}^n f(d_i)}{\vol(G)}.
  \end{equation}
  
  Second, we use Jensen's inequality to prove $\Delta\calH(G)> 0$.
  Since $f(x)$ is convex, 
  $f(d_i)=f(\Expect[X_i])\leq\Expect[f(X_i)]$
  for any $i\in\{1,\ldots,n\}$. By summing over $i$, we have 
  \begin{equation*}
    \sum_{i=1}^n f(d_i)\leq\sum_{i=1}^n\Expect[f(X_i)]=\sum_{i=1}^n\sum_{j=1}^n p_{ij}f(\lambda_j)=\sum_{j=1}^n f(\lambda_j).
  \end{equation*}
  Therefore, $\Delta\calH(G)\geq 0$ for any undirected graphs.
  Actually, $\Delta\calH(G)$ cannot be $0$.
  To see this, suppose that the Laplacian matrix can be decomposed as $L=U\Lambda U^\intercal$ where $\Lambda={\rm diag}(\lambda_1,\ldots,\lambda_n)$ and 
  $U=(u_{\cdot,1}|\cdots|u_{\cdot,n})$ is a unitary matrix. Note that $\lambda_1=0$ and $u_{\cdot,1}=\mathbf{1}_n/\sqrt{n}$.
  The $i$-th diagonal element of the matrix $U\Lambda U^\intercal=\sum_{j=1}^n\lambda_j u_{\cdot,j}u_{\cdot,j}^\intercal$ is given by $\sum_{j=1}^n\lambda_j |u_{ij}|^2$.
  Since the $i$-th diagonal element of $L$ is $d_i$ and $L=U\Lambda U^\intercal$, we have $d_i=\sum_{j=1}^n\lambda_j |u_{ij}|^2$ for each $i\in[n]$. Recall that 
  $P\bm\lambda=\mathbf{d}$ where $P$ is a doubly-stochastic matrix, therefore $p_{ij}=|u_{ij}|^2$.
  Now suppose for contradiction that $\Delta\calH(G)=0$, which means that $f(\Expect[X_i])=\Expect[f(X_i)]$ for each $i\in [n]$.
  Due to the strict convexity of $f(\cdot)$, the discrete random variable $X_i$ has to be deterministic. Since $\Expect[X_i]=d_i>0$, we have $\mathbb{P}(X_i=d_i)=1$.
  However, it contradicts the fact that $\mathbb{P}(X_i=0)\geq\mathbb{P}(X_i=\lambda_1)=p_{i,1}=\frac{1}{n}$. Therefore the contradiction implies that $\Delta\calH(G)>0$ for any undirected graphs.
  
  Finally, we use the Jensen's gap to prove the upper bound on $\Delta\calH(G)$ in \prettyref{eq:entropy-gap}.
  Apply the Jensen's gap to $X_i$ and $f(x)$,
  \begin{equation}\label{eq:right-side-0}
      \Expect[f(X_i)]-f(\Expect[X_i])\leq\sup_{x\in[0, \vol(G)]}\{h_i(x)\}\cdot\var(X_i),
  \end{equation}
  where 
  \[
      h_i(x)=\frac{f(x)-f(\Expect[X_i])}{(x-\Expect[X_i])^2}-\frac{f'(\Expect[X_i])}{x-\Expect[X_i]}.    
  \]
  Since $f'(x)$ is concave, $h_i(x)$ is monotonically decreasing in $x$. Therefore,
  $\sup_{x\in[0,\vol(G)]}\{h_i(x)\}=h_i(0)$.
  Since 
  \begin{equation*}
      h_i(0)=\frac{f(0)-f(d_i)}{d_i^2}+\frac{f'(d_i)}{d_i}=\frac{\log_2 e}{d_i}\leq\frac{\log_2 e}{\delta},
  \end{equation*}
  the inequality in \prettyref{eq:right-side-0} can be simplified as 
  \begin{equation}\label{eq:right-side-4}
      \sum_{j=1}^n p_{ij}f(\lambda_j)-f(d_i)
      \leq\frac{\log_2 e}{\delta}\cdot\left(\sum_{j=1}^n p_{ij}\lambda_j^2 -d_i^2\right).
  \end{equation}

  By summing both sides of the inequality \prettyref{eq:right-side-4} over $i$, we get 
  an upper bound $\mathsf{UB}$ on $\sum_{j=1}^n f(\lambda_j)-\sum_{i=1}^n f(d_i)$ as 
  \begin{equation*}
    \begin{aligned}
      \mathsf{UB}&=\frac{\log_2 e}{\delta}\cdot\sum_{i=1}^n\left(\sum_{j=1}^n p_{ij}\lambda_j^2 - d_i^2\right) \\
      &=\frac{\log_2 e}{\delta}\cdot\left(\sum_{j=1}^n \lambda_j^2 - \sum_{i=1}^n d_i^2\right) \\
      &=\frac{\log_2 e}{\delta}\cdot\left(\text{tr}(L^2)-\text{tr}(D^2)\right)\\
      &=\frac{\log_2 e}{\delta}\cdot\left(\text{tr}(A^2)-\text{tr}(AD)-\text{tr}(DA)\right) \\
      &=\frac{\log_2 e}{\delta}\cdot\text{tr}(A^2).
    \end{aligned}
  \end{equation*}
  As a result, $\Delta\calH(G)=\frac{\sum_{i=1}^n f(\lambda_i)-\sum_{i=1}^n f(d_i)}{\vol(G)}\leq\frac{\log_2 e}{\delta}\frac{\mathrm{tr}(A^2)}{\vol(G)}$. 
\end{IEEEproof}

To illustrate the tightness of the bounds in \prettyref{thm:bounds}, we further derive bounds on the entropy gap for unweighted graphs, especially the regular graphs. Via multiplicative error analysis, we show that 
the structural information converges to the von Neumann graph entropy as the graph size grows.
\begin{corollary}[Constant bounds on the entropy gap]\label{cor:constant-bounds}
  For any unweighted, undirected graph $G$, $0<\Delta\calH(G)\leq\log_2 e$ holds.
\end{corollary}
\begin{IEEEproof}
  In unweighted graph $G$, 
  \begin{equation*}
    \begin{aligned}
    \mathrm{tr}(A^2)&=\sum_{i=1}^n\sum_{j=1}^n A_{ij}A_{ji} \\
    &=\sum_{i=1}^n\sum_{j=1}^n A_{ij} \\
    &=\sum_{i=1}^n d_i \\
    &=\vol(G),
    \end{aligned}
  \end{equation*}
   and $\delta\geq 1$, therefore
  $0<\Delta\calH(G)\leq\frac{\log_2 e}{\delta}\frac{\mathrm{tr}(A^2)}{\vol(G)}=\frac{\log_2 e}{\delta}\leq\log_2 e$.
\end{IEEEproof}
\begin{corollary}[Entropy gap of regular graphs]\label{cor:regular-graph}
  For any unweighted, undirected, regular graph $G_d$ of degree $d$, the inequality $0<\Delta\calH(G_d)\leq\frac{\log_2 e}{d}$ holds.
\end{corollary}
\begin{IEEEproof}[Proof sketch]
  In any unweighted, regular graph $G_d$, $\delta=d$.
\end{IEEEproof}

\begin{theorem}[Convergence of the multiplicative approximation error]\label{thm:multiplicative-error}
  For almost all unweighted graphs $G$ of order $n$, $\frac{\ose(G)}{\vnge(G)}-1> 0$ and 
  decays to $0$ at the rate of $o(1/\log_2(n))$.
\end{theorem}
\begin{IEEEproof}
  Dairyko et al. \cite{Dairyko2017} proved that for almost all unweighted graphs $G$ of order $n$, $\vnge(G)\geq\vnge(K_{1,n-1})$ where $K_{1,n-1}$ stands for the star graph.
  Since $\vnge(K_{1,n-1})=\log_2(2n-2)-\frac{n}{2n-2}\log_2 n=1+\frac{1}{2}\log_2 n+o(1)$, 
  $\frac{\ose(G)}{\vnge(G)}-1=\frac{\Delta\calH(G)}{\vnge(G)}\leq\frac{\log_2 e}{\vnge(K_{1,n-1})}=o(\frac{1}{\log_2 n})$.
\end{IEEEproof}

\subsection{Sharpened Bounds on the Entropy Gap}
Though the constant bounds on the entropy gap are tight enough for applications, we can still sharpen the bounds on the entropy gap in unweighted graphs
using more advanced majorizations.
\begin{theorem}[Sharpened lower bound on entropy gap]\label{thm:sharpened-lower-bound}
  For any unweighted, undirected graph $G$, $\Delta\calH(G)$ is lower bounded by 
  $(f(d_{\rm max}+1)-f(d_{\rm max})+f(\delta-1)-f(\delta))/\vol(G)$ where $d_{\rm max}$
  is the maximum degree and $\delta$ is the minimum positive degree.
\end{theorem}
\begin{IEEEproof}
  The proof is based on the advanced majorization \cite{Grone1995}
  $\bm{\lambda}\succ(d_1+1,d_2,\ldots,d_n-1)$ where $d_1\geq d_2\geq\cdots\geq d_n$ is the sorted degree sequence of the unweighted undirected graph $G$.
  Similar to the proof of \prettyref{thm:bounds}, we have
  $\sum_{i=1}^n f(\lambda_i)\geq f(d_1+1)+f(d_n-1)+\sum_{i=2}^{n-1}f(d_i)$. Then 
  the sharpened upper bound follows from the equation \prettyref{eq:entropy-gap-expression} since $d_1=d_{\mathrm{max}}$ and $d_n=\delta$.
\end{IEEEproof}

\begin{theorem}[Sharpened upper bound on entropy gap]
  For any unweighted, undirected graph $G=(V,E)$, $\Delta\calH(G)$ is upper bounded by $\min\{\log_2 e, b_1,b_2\}$
  where $b_1=\frac{\sum_{i=1}^n f(d_i^\ast)}{\vol(G)}-\frac{\sum_{i=1}^n f(d_i)}{\vol(G)}$ and 
  $b_2=\log_2(1+\sum_{i=1}^n d_i^2/\vol(G))-\frac{\sum_{i=1}^n f(d_i)}{\vol(G)}$. Here 
  $(d_1^\ast,\ldots,d_n^\ast)$ is the conjugate degree sequence of $G$ where $d_k^\ast=|\{i|d_i\geq k\}|$.
\end{theorem}
\begin{IEEEproof}
  We first prove $\Delta\calH(G)\leq b_1$ using the Grone-Merris majorization \cite{Bai2011TAMS}:
  $(d_1^\ast,\ldots,d_n^\ast)\succ\bm{\lambda}$. Similar to the proof of \prettyref{thm:bounds},
  we have $\sum_{i=1}^n f(d_i^\ast)\geq\sum_{i=1}^n f(\lambda_i)$, thus 
  $b_1\geq\frac{\sum_{i=1}^n f(\lambda_i)-\sum_{i=1}^n f(d_i)}{\vol(G)}=\Delta\calH(G)$.
  We then prove $\Delta\calH(G)\leq b_2$. Since
  \begin{equation*}
      \frac{\sum_{i=1}^n f(\lambda_i)}{\vol(G)}
      =\sum_{i=1}^n\left(\frac{\lambda_i}{\sum_{j=1}^n\lambda_j}\right)\log_2\lambda_i 
      \leq\log_2\left(\frac{\sum_{i=1}^n \lambda_i^2}{\sum_{j=1}^n \lambda_j}\right)
  \end{equation*}
  and 
  \[
    \frac{\sum_{i=1}^n\lambda_i^2}{\sum_{i=1}^n\lambda_i}=\frac{{\rm tr}(L^2)}{\vol(G)}=1+\frac{\sum_{i=1}^n d_i^2}{\vol(G)},
  \]
  we have $\Delta\calH(G)=\frac{\sum_{i=1}^n f(\lambda_i)-f(d_i)}{\vol(G)}\leq b_2$.
\end{IEEEproof}

\subsection{Entropy Gap of Various Types of Graphs}
As summarized in \prettyref{tab:various-graph-types}, we analyze the entropy gap of various types of graphs including 
the complete graph, the complete bipartite graph, the path graph, and the ring graph,
the proofs of which can be found in 
\prettyref{app:proof-specific-graphs}. 

\begin{table*}[t]
  \centering
  \caption{\label{tab:various-graph-types} Structural information, von Neumann graph entropy, and entropy gap of specific graphs.}
  \begin{tabular}{l||c|c|c}
    \toprule
    Graph Types & Structural information $\ose$ & von Neumann graph entropy $\vnge$ & Entropy gap $\Delta\calH$ \\
    \midrule
    Complete graph $K_n$ & $\log_2 n$ & $\log_2(n-1)$ & $\log_2(1+\frac{1}{n-1})$ \\
    Complete bipartite graph $K_{a,b}$ & $1+\frac{1}{2}\log_2(ab)$ & $1+\frac{1}{2}\log_2(ab)-\frac{\log_2(1+\frac{b}{a})}{2b}-\frac{\log_2(1+\frac{a}{b})}{2a}$ & $\frac{\log_2(1+\frac{b}{a})}{2b}+\frac{\log_2(1+\frac{a}{b})}{2a}$ \\
    Path $P_n$ & $\log_2(n-1)+\frac{1}{n-1}$ & $\log_2(n-1)+1-\log_2 e$ & $\log_2 e-1$ \\
    Ring $R_n$ & $\log_2 n$ & $\log_2 n+1-\log_2 e$ & $\log_2 e-1$ \\
    \bottomrule
  \end{tabular}
\end{table*}
    \section{Applications and Algorithms}
\label{sec:applications-and-algorithms}

As a measure of the structural complexity of a graph, the von Neumann entropy has been applied in a variety of applications.
For example, the von Neumann graph entropy is exploited to measure the edge centrality \cite{Joshua2016} and the vertex centrality \cite{Rossi2017} in complex networks.
As another example, the von Neumann graph entropy can also be used to measure the distance between graphs for graph classification and anomaly detection \cite{Chen2019ICML,Lu2011}.
In addition, the von Neumann graph entropy is used in the context of graph representation learning \cite{George2020} to learn low-dimensional feature representations of nodes.
We observe that, in these applications, the von Neumann graph entropy is used to address the following primitive tasks:

\begin{itemize}[leftmargin=*]
  \item {\bf Entropy-based network design}: 
  Design a new graph by minimally perturbing the existing graph to meet the entropic requirement.
  For example, Minello et al. \cite{Giorgia2018} use the von Neumann entropy to explore the potential network growth model via experiments.
  \item {\bf Graph similarity measure}: 
  Compute the similarity score between two graphs, which is represented by a real positive number.
  For example, Domenico et al. \cite{Domenico2015Nature} use the von Neumann graph entropy to compute the Jensen-Shannon distance between graphs for the purpose of compressing multilayer networks.
\end{itemize}

Both of the primitive tasks require exact computation of the von Neumann graph entropy.
To reduce the computational complexity and preserve the interpretability,
we can use the accurate proxy, structural information, to approximately solve 
these tasks.

\subsection{Entropy-based network design}
Network design aims to minimally perturb the network to fulfill some goals.
Consider such a goal to maximize the von Neumann entropy of a graph, it helps to understand how different structural patterns influence the entropy value.
The entropy-based network design problem is formulated as follows,

\begin{problem}[MaxEntropy]\label{prob:max-entropy}
  Given an unweighted, undirected graph $G=(V,E)$ of order $n$ and an integer budget $k$, find a set $F$ of non-existing edges of $G$
  whose addition to $G$ creates the largest increase of the von Neumann graph entropy and $|F|\leq k$.
\end{problem}

Due to the spectral nature of the von Neumann graph entropy, it is not easy to find an effective strategy to perturb the graph, especially
in the scenario where there are exponential number of combinations for the subset $F$.
If we use the structural information as a proxy of the von Neumann entropy, \prettyref{prob:max-entropy} can be reduced to maximizing
$\ose(G')$ where $G'=(V,E\cup F)$ such that $|F|\leq k$.
To further alleviate the computational pressure rooted in the exponential size of the search space for $F$,
we adopt the greedy method in which the new edges are added one by one until either the structural information attains its maximum value $\log_2 n$ or $k$ new edges have already been added.
We denote the graph with $l$ new edges as $G_l=(V,E_l)$, then $G_0=G$.
Now suppose that we have $G_l$ whose structural information is less than $\log_2 n$, then we want to find a new edge $e_{l+1}=(u,v)$ such that 
$\ose(G_{l+1})$ is maximized, where $G_{l+1}=(V,E_l\cup\{e_{l+1}\})$. Since $\ose(G_{l+1})$ can be rewritten as 
\[
  \log_2(2|E_l|+2)-\frac{f(d_u+1)+f(d_v+1)+\sum_{i\neq u,v}f(d_i)}{2|E_l|+2},
\]
the edge $e_{l+1}$ maximizing $\ose(G_{l+1})$ should also minimize the edge centrality 
$EC(u, v)=f(d_u+1)-f(d_u)+f(d_v+1)-f(d_v)$, 
where $d_i$ is the degree of node $i$ in $G_l$.

We present the pseudocode of our fast algorithm \textsf{EntropyAug} in \prettyref{alg:greedy}, which leverages
the pruning strategy to accelerate the process of finding a single new edge that creates a largest increase of the von Neumann entropy.
\textsf{EntropyAug} starts by initiating an empty set $F$ used to contain the node pairs to be found and an entropy value $\calH$ used to record the maximum structural information in the graph evolution process (line 1).
In each following iteration, it sorts the set of nodes $V$ in non-decreasing degree order (line 3). 
Note that the edge centrality $EC(u,v)$ has a nice monotonic property:
$EC(u_1,v_1)\leq EC(u_2,v_2)$ if $\min\{d_{u_1},d_{v_1}\}\leq\min\{d_{u_2},d_{v_2}\}$ and 
$\max\{d_{u_1},d_{v_1}\}\leq\max\{d_{u_2},d_{v_2}\}$.
With the sorted list of nodes $V_s$, the monotonicity of $EC(u,v)$ can be translated into 
  $EC(V_s[i_1],V_s[j_1])\leq EC(V_s[i_2],V_s[j_2])$ if the indices satisfy $i_1<j_1$, $i_2<j_2$, $i_1<i_2$, and $j_1<j_2$.
Thus, using the two pointers $\{{\rm head},{\rm tail}\}$ and a threshold $T$,
it can prune the search space and find the desired non-adjacent node pair as fast as possible (line 4-12).
It then adds the non-adjacent node pair minimizing $EC(u,v)$ into $F$ and update the graph $G$ (line 13).
The structural information of the updated graph is computed to determine whether $F$ is the optimal subset till current iteration (line 14-15).

\begin{algorithm}[t]
  \KwIn {The graph $G=(V,E)$ of order $n$, the budget $k$}
  \KwOut {A set of node pairs}
  $F\leftarrow\varnothing$, $\calH\leftarrow 0$\;
  \While{$|F|<k$}{
      $V_s$: list $\leftarrow$ sort $V$ in non-decreasing degree order\;
  ${\rm head}\leftarrow 0$, ${\rm tail}\leftarrow |V_s|-1$, $T\leftarrow +\infty$\;
  \While{${\rm head}<{\rm tail}$}{
      \For{$i={\rm head}+1,{\rm head}+2,\ldots,{\rm tail}$}{
          \If{$EC(V_s[{\rm head}],V_s[i])\geq T$}{
              ${\rm tail}\leftarrow i-1$; \Break\;
          }
          \If{$(V_s[{\rm head}], V_s[i])\notin E$}{
              $u\leftarrow V_s[{\rm head}]$, $v\leftarrow V_s[i]$, $T\leftarrow EC(u,v)$\;
              ${\rm tail}\leftarrow i-1$; \Break\;
          }
      }
      ${\rm head}\leftarrow{\rm head}+1$\;
  }
  $E\leftarrow E\cup\{(u,v)\}$, $F\leftarrow F\cup\{(u,v)\}$\;
  \lIf{$\ose(G)>\calH$}{$\calH\leftarrow\ose(G)$, $F^\ast\leftarrow F$}
  \lIf{$\calH=\log_2 n$}{\Break}
  }
  \Return $F^\ast$.
  \caption{\textsf{EntropyAug}}
  \label{alg:greedy}
\end{algorithm}

\subsection{Graph Similarity Measure}
Entropy based graph similarity measure aims to compare graphs using Jensen-Shannon divergence.
The Jensen-Shannon divergence, as a symmetrized and smoothed version of the Kullback-Leibler divergence,
is defined formally in the following \prettyref{def:Jensen-Shannon-divergence}.
\begin{definition}[Jensen-Shannon divergence]\label{def:Jensen-Shannon-divergence}
  Let $P$ and $Q$ be two probability distributions on the same support set $\Omega_N=\{1,\ldots,N\}$, 
  where $P=(p_1,\ldots,p_N)$ and $Q=(q_1,\ldots,q_N)$.
  The Jensen-Shannon divergence between $P$ and $Q$ is defined as 
  \[
    \calD_{\rm JS}(P,Q)=H((P+Q)/2)-H(P)/2-H(Q)/2,  
  \]
  where $H(P)=-\sum_{i=1}^N p_i\log p_i$ is the entropy of the distribution $P$.
\end{definition}

The square root of $\calD_{\rm JS}(P,Q)$, also known as Jensen-Shannon distance, has been proved \cite{Endres2003TIT} to be 
a bounded metric on the space of distributions over $\Omega_N$, with its maximum value $\sqrt{\log 2}$ being attained
when $\min\{p_i,q_i\}=0$ for each $i\in\Omega_N$.
However, the Jensen-Shannon distance cannot measure the similarity between high dimensional objects such as matrices and graphs.
Therefore, Majtey et al. \cite{Majtey2005PRA} introduce a quantum Jensen-Shannon divergence to measure the similarity between mixed quantum states,
the formal definition of which is presented in the following,
\begin{definition}[Quantum Jensen-Shannon divergence between quantum states]
  The quantum Jensen-Shannon divergence between two quantum states $\bm\rho$ and $\bm\sigma$ is defined as 
  \begin{equation*}
    \calD_{\rm QJS}(\bm\rho,\bm\sigma)=
    \tr\left(\frac{\bm\rho\log\bm\rho+\bm\sigma\log\bm\sigma}{2}-\frac{\bm\rho+\bm\sigma}{2}\log\frac{\bm\rho+\bm\sigma}{2}\right)
  \end{equation*}
  where $\bm\rho,\bm\sigma$ are symmetric and positive semi-definite density matrices 
  with $\tr(\bm\rho)=1$ and $\tr(\bm\sigma)=1$.
\end{definition}

The square root of $\calD_{\rm QJS}(\bm\rho,\bm\sigma)$ has also been proved to be a metric on the space of density matrices \cite{Lamberti2008PRA,VIROSZTEK2021107595}.
Since the Laplacian matrix $L$ of a weighted undirected graph $G$ is symmetric and positive semi-definite, 
we can view $\frac{L}{\tr(L)}$ as a density matrix.
Therefore, the quantum Jensen-Shannon divergence can be used to measure the similarity between two graphs $G_1$ and $G_2$, given that 
they share the same node set, \ie $V(G_1)=V(G_2)$.

\begin{definition}[Quantum Jensen-Shannon divergence between graphs]\label{def:Quantum-Jensen-Shannon-distance}
  The quantum Jensen-Shannon divergence between two weighted, undirected graphs $G_1=(V,E_1,A_1)$ and $G_2=(V,E_2,A_2)$
  on the same node set $V$ is defined as 
  \[
    \calD_{\rm QJS}(G_1,G_2)=\vnge(\overbar{G})-(\vnge(G_1)+\vnge(G_2))/2,  
  \]
  where $\overbar{G}=(V,E_1\cup E_2,\overbar{A})$ is an weighted graph 
  with $\overbar{A}=A_1/2\vol(G_1)+A_2/2\vol(G_2)$.
\end{definition}

Generally, the node sets of two graphs to be compared are not necessarily aligned.
For simplicity, we restrict our discussion to the aligned case and recommend \cite{LuBai2015} for a more detailed treatment of the unaligned case.

Based on the quantum Jensen-Shannon divergence between graphs, we consider the following problem that has found applications in anomaly detection and multiplex network compression.
\begin{problem}\label{prob:JSD}
  Compute the square root of the quantum Jensen-Shannon divergence between adjacent graphs in a stream of graphs $\{G_k=(V,E_k,t_k)\}_{k=1}^K$ where 
  $t_k$ is the timestamp of the graph $G_k$ and $t_k<t_{k+1}$.
\end{problem}

Since $\sqrt{\calD_{\rm QJS}(G_1,G_2)}$ is computationally expensive, 
we propose a new distance measure based on structural information
in \prettyref{def:structural-information-distance} and analyze its metric properties in \prettyref{thm:DSI}.

\begin{definition}[Structural information distance between two graphs]
  \label{def:structural-information-distance}
  The structural information distance between two weighted, undirected graphs $G_1=(V,E_1,A_1)$ and $G_2=(V,E_2,A_2)$ 
  on the same node set $V$
  is defined as 
  \[
    \calD_{\rm SI}(G_1,G_2)=\sqrt{\ose(\overbar{G})-\left(\ose(G_1)+\ose(G_2)\right)/2},
  \]
  where $\overbar{G}=(V,E_1\cup E_2,\overbar{A})$ is an weighted graph 
  with 
  $\overbar{A}=A_1/2\vol(G_1)+A_2/2\vol(G_2)$.
\end{definition}

\begin{theorem}[Properties of the distance measure $\calD_{\rm SI}$]\label{thm:DSI}
  The distance measure $\calD_{\rm SI}(G_1,G_2)$ 
  is a pseudometric on the space of undirected graphs:
  \begin{itemize}
    \item $\calD_{\rm SI}$ is symmetric, i.e., $\calD_{\rm SI}(G_1,G_2)=\calD_{\rm SI}(G_2,G_1)$;
    \item $\calD_{\rm SI}$ is non-negative, i.e., $\calD_{\rm SI}(G_1,G_2)\geq0$ where the equality holds if and only if $\frac{d_{i,1}}{\sum_{k=1}^n d_{k,1}}=\frac{d_{i,2}}{\sum_{k=1}^n d_{k,2}}$ for every node $i\in V$ where $d_{i,j}$ is the degree of node $i$ in $G_j$;
    \item $\calD_{\rm SI}$ obeys the triangle inequality, i.e., 
    \[
      \calD_{\rm SI}(G_1,G_2)+\calD_{\rm SI}(G_2,G_3)\geq\calD_{\rm SI}(G_1,G_3);
    \]
    \item $\calD_{\rm SI}$ is upper bounded by $1$, i.e., $\calD_{\rm SI}(G_1,G_2)\leq 1$ where the equality holds if and only if $\min\{d_{i,1},d_{i,2}\}=0$ for every node $i\in V$ where $d_{i,j}$ is the degree of node $i$ in $G_j$.
  \end{itemize}
\end{theorem}
Besides the metric properties, we further establish a connection between $\calD_{\rm SI}$ and $\sqrt{\calD_{\rm QJS}}$ by studying their extreme values, the results of which are summarized in \prettyref{thm:JS-and-DSI}.
\begin{theorem}[Connection between $\sqrt{\calD_{\rm QJS}}$ and $\calD_{\rm SI}$]\label{thm:JS-and-DSI}
  Both $\sqrt{\calD_{\rm QJS}(G_1,G_2)}$ and $\calD_{\rm SI}(G_1,G_2)$ attain the same maximum value of $1$ under the identical condition that 
  $\min\{d_{i,1},d_{i,2}\}=0$ for every node $i\in V$ where $d_{i,j}$ is the degree of node $i$ in $G_j$.
\end{theorem}

\begin{algorithm}[t]
  \KwIn{$G_1$ and $\{\Delta G_k\}_{k=1}^{K-1}$}
  \KwOut{$\{\calD_{\rm SI}(G_k,G_{k+1})\}_{k=1}^{K-1}$}
  $d\leftarrow$ the degree sequence of the graph $G_1$\;
  $m\leftarrow\sum_{i=1}^n d_i/2$\;
  $\ose(G_1)\leftarrow\log_2(2m)-\frac{1}{2m}\sum_{i=1}^n f(d_i)$\;
  \For{$k=1,\ldots,K-1$}{
    $\Delta d\leftarrow$ the degree sequence of the signed graph $\Delta G_k$\;
    $\Delta m\leftarrow\sum_{i\in V_k}\Delta d_i/2$\;
    Compute $a,b,y,z$ in \prettyref{lem:incremental} via iterating over $V_k$\;
    Compute $\ose(G_{k+1})$ and $\ose(\overbar{G}_k)$ based on \prettyref{lem:incremental}\;
    $\calD_{\rm SI}(G_k,G_{k+1})\leftarrow\sqrt{\ose(\overbar{G}_k)-(\ose(G_k)+\ose(G_{k+1}))/2}$\;
    $m\leftarrow m+\Delta m$\;
    \lForEach{$i\in V_k$}{$d_i\leftarrow d_i+\Delta d_i$}
  }
  \Return{$\{\calD_{\rm SI}(G_k,G_{k+1})\}_{k=1}^{K-1}$}
  \caption{\textsf{IncreSim}}
  \label{alg:incremental-DSI}
\end{algorithm}

In order to compute the structural information distance between adjacent graphs in the graph stream $\{G_k\}_{k=1}^K$ where $G_k=(V,E_k,t_k)$,
we first compute the structural information $\ose(G_k)$ for each $k\in[K]$, which takes $\Theta(Kn)$ time.
Then we compute the structural information of $\overbar{G}_k$ whose adjacent matrix $\overbar{A}_k=A_k/2\vol(G_k)+A_{k+1}/2\vol(G_{k+1})$ for each $k\in[K-1]$.
Since the degree of node $i$ in $\overbar{G}_k$ is $\overbar{d}_{i,k}=\frac{d_{i,k}}{2\vol(G_k)}+\frac{d_{i,k+1}}{2\vol(G_{k+1})}$ and $\sum_{i=1}^n\overbar{d}_{i,k}=1$, the structural information of $\overbar{G}_k$ is 
  $\ose(\overbar{G}_k)=-\sum_{i=1}^n f(\overbar{d}_{i,k})$ which takes $\Theta(n)$ time for each $k$.
Therefore, the total computational cost is $\Theta((2K-1)n)$.

In practice, the graph stream is fully dynamic such that it would be more efficient to represent the graph stream as a stream of operations over time, rather than a sequence of graphs.
Suppose that the graph stream is represented as an initial base graph $G_1=(V,E_1,t_1)$
and a sequence of operations $\{\Delta G_k=(V_k,E_{+,k},E_{-,k},t_k)\}_{k=1}^{K-1}$ on the graph
where $t_k$ is the timestamp of the set $E_{+,k}$ of edge insertions 
and the set $E_{-,k}$ of edge deletions, and $V_k$ is the subset of nodes covered by $E_{+,k}\cup E_{-,k}$.
We can view the operation $\Delta G_k$ as a signed network where the edge in $E_{+,k}$ has positive weight $+1$
and the edge in $E_{-,k}$ has negative weight $-1$.
The degree of node $i\in V_k$ in the operation $\Delta G_k$ refers to
$\sum_{j\in V_k}\mathbb{I}\{(i,j)\in E_{+,k}\}-\mathbb{I}\{(i,j)\in E_{-,k}\}$.
Using the information about previous graph $G_k$ and current operation $\Delta G_k$, we can compute 
the entropy statistics of the current graph $G_{k+1}$ incrementally and efficiently via the following lemma,
whose proof can be found in the appendix.

\begin{lemma}\label{lem:incremental}
  Using the degree sequence $d$ of the graph $G_k$, the structural information $\ose(G_k)$, and the degree sequence $\Delta d$ of the signed graph $\Delta G_k$,
  the structural information of the graph $G_{k+1}$ can be efficiently computed as 
  \[
    \ose(G_{k+1})=\frac{f(2(m+\Delta m))-a-f(2m)+2m\ose(G_k)}{2(m+\Delta m)},  
  \]
  where $m=\sum_{i=1}^n d_i/2$,
  $\Delta m=\sum_{i\in V_k}\Delta d_i/2$, and $a=\sum_{i\in V_k} f(d_i+\Delta d_i)-f(d_i)$.
  Moreover, the structural information of the averaged graph $\overbar{G}_k$ between $G_k$ and $G_{k+1}$ can be efficiently computed as 
  \[
    \ose(\overbar{G}_k)=-b-(2m-y)f(c)-c(f(2m)-2m\ose(G_k)-z),
  \]
  where 
  $y=\sum_{i\in V_k}d_i$, 
  $z=\sum_{i\in V_k}f(d_i)$,
  $c=\frac{2m+\Delta m}{4m(m+\Delta m)}$,
  and $b=\sum_{i\in V_k} f\left(\frac{d_i}{4m}+\frac{d_i+\Delta d_i}{4(m+\Delta m)}\right)$.
\end{lemma}

The pseudocode of our fast algorithm \textsf{IncreSim} for computing the structural information distance in a graph stream is shown in \prettyref{alg:incremental-DSI}.
It starts by computing the structural information of the base graph $G_1$ (line 1-3), which takes $\Theta(n)$ time.
In each following iteration, it first computes the value of $a,b,c,y,z$ (line 5-7), then calculates the 
structural information distance between two adjacent graphs (line 8-9), finally updates the edge count $m$ and the degree sequence $d$ (line 10-11).
The time cost of each iteration is $\Theta(|V_k|)$, consequently the total time complexity
is $\Theta(n+\sum_{k=1}^{K-1}|V_k|)$.

    \section{Connections with Community Structure}
\label{sec:comm}

In this section, we discuss the connections between the graph entropy and the community structure in graphs.

\subsection{Empirical Analysis of Stochastic Block Model}
\subsubsection{Preparations}
To study the connections between graph entropy and community structures in a specific ensemble of 
graphs, suppose that a graph $G$ is generated by the stochastic block model.
There are $q$ groups of nodes, and each node $v$ has a group label $g_v\in\{1,\ldots,q\}$.
Edges are generated independently according to a probability matrix $\mathbf{P}\in[0,1]^{q\times q}$,
with $\mathbb{P}(A_{uv}=1)=\mathbf{P}[g_u,g_v]$. In the sparse case, we have 
$\mathbf{P}[a,b]=\mathbf{C}[a,b]/n$, where the affinity matrix $\mathbf{C}$ stays constant in the limit $n\rightarrow\infty$.
For simplicity we make a common assumption that the affinity matrix 
$\mathbf{C}$ has two distinct entries $c_{\rm in}$ and $c_{\rm out}$ where 
$\mathbf{C}[a,b]=c_{\rm in}$ if $a=b$ and $c_{\rm out}$ if $a\neq b$.
For any graph generated from the stochastic block model with two(three) groups, we use 
the spectral algorithm in \prettyref{alg:2-spectral-clustering}(\prettyref{alg:3-spectral-clustering}) to detect the community structure.

\begin{algorithm}
    \KwIn{The graph $G=(V,E)$ of order $n$}
    \KwOut{A cluster membership vector $\mathbf{cl}\in\{0, 1\}^n$}
    $\mathbf{cl}\leftarrow \mathbf{0}$\;
    $L\leftarrow$ Laplacian matrix of the graph $G$\;
    $\mathbf{v}_2\leftarrow$ eigenvector corresponding to $\lambda_2$ of $L$\;
    \For{$i=1,\ldots,n$}{
        \lIf{$\mathbf{v}_2[i]<0$}{$\mathbf{cl}[i]=1$}
    }
    \Return{$\mathbf{cl}$}
    \caption{$2$-Spectral Clustering}
    \label{alg:2-spectral-clustering}
\end{algorithm}

\begin{algorithm}
    \KwIn{The graph $G=(V,E)$ of order $n$}
    \KwOut{A cluster membership vector $\mathbf{cl}\in\{0, 1, 2\}^n$}
    $L\leftarrow$ Laplacian matrix of the graph $G$\;
    $\mathbf{v}_2\leftarrow$ eigenvector corresponding to $\lambda_2$ of $L$\;
    $\mathbf{v}_3\leftarrow$ eigenvector corresponding to $\lambda_3$ of $L$\;
    $\mathbf{cl}\leftarrow$ $k$-means clustering of $[\mathbf{v}_2,\mathbf{v}_3]$ with $k=3$\;
    \Return{$\mathbf{cl}$}
    \caption{$3$-Spectral Clustering}
    \label{alg:3-spectral-clustering}
\end{algorithm}

\subsubsection{Evaluation Metrics}
For each synthetic graph, we compute the structural information, von Neumann graph entropy, Laplacian eigenvalues $\lambda_2,\lambda_3,\lambda_4,\ldots$ with small magnitude, spectral gaps $\lambda_{k+1}-\lambda_k$, and the detection error.

Specifically, let $\calP=\{P_1,\ldots,P_k\}$ and $\calQ=\{Q_1,\ldots,Q_k\}$ be two $k$-partitions of $V$.
We view $\calP$ as the ground-truth community structure and $\calQ$ as the detected community structure, then 
the detection error is 
\begin{equation*}
    \min_{\sigma}\sum_{i=1}^k|P_i\triangle Q_{\sigma(i)}|,
\end{equation*}
where $\sigma$ ranges over all bijections $\sigma:[k]\rightarrow[k]$
and $\triangle$ represents the symmetric difference.

\subsubsection{Empirical Results}
The results are shown in \prettyref{fig:bipartition} and \prettyref{fig:tripartition}, from which we have the following observations:

\begin{figure*}[!t]
    \begin{subfigure}{0.32\textwidth}
        \includegraphics[width=\textwidth]{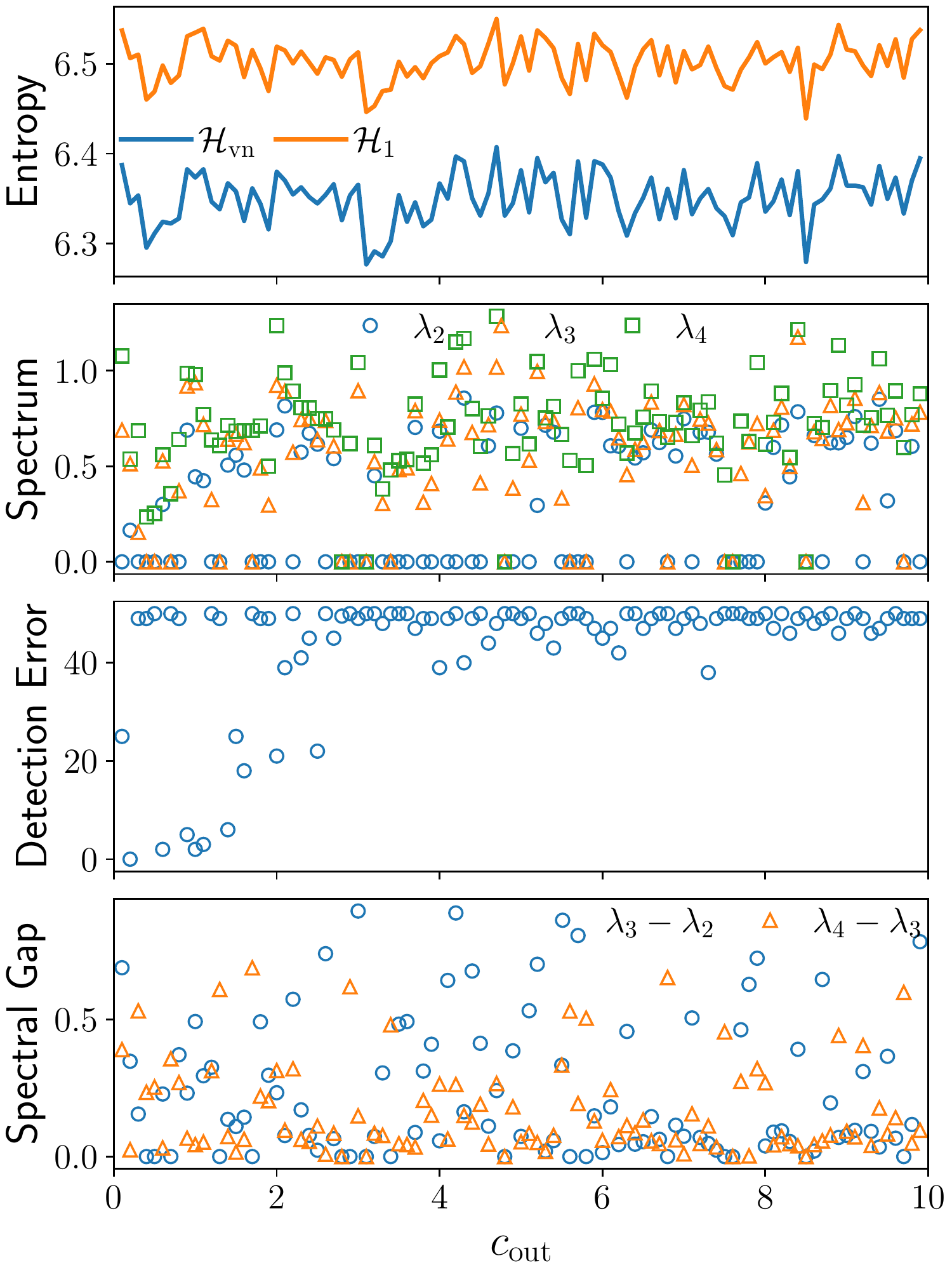}
        \caption{$\overbar{d}=5$, $c_{\rm in}+c_{\rm out}=10$}
    \end{subfigure}
    \hspace{2mm}
    \begin{subfigure}{0.32\textwidth}
        \includegraphics[width=\textwidth]{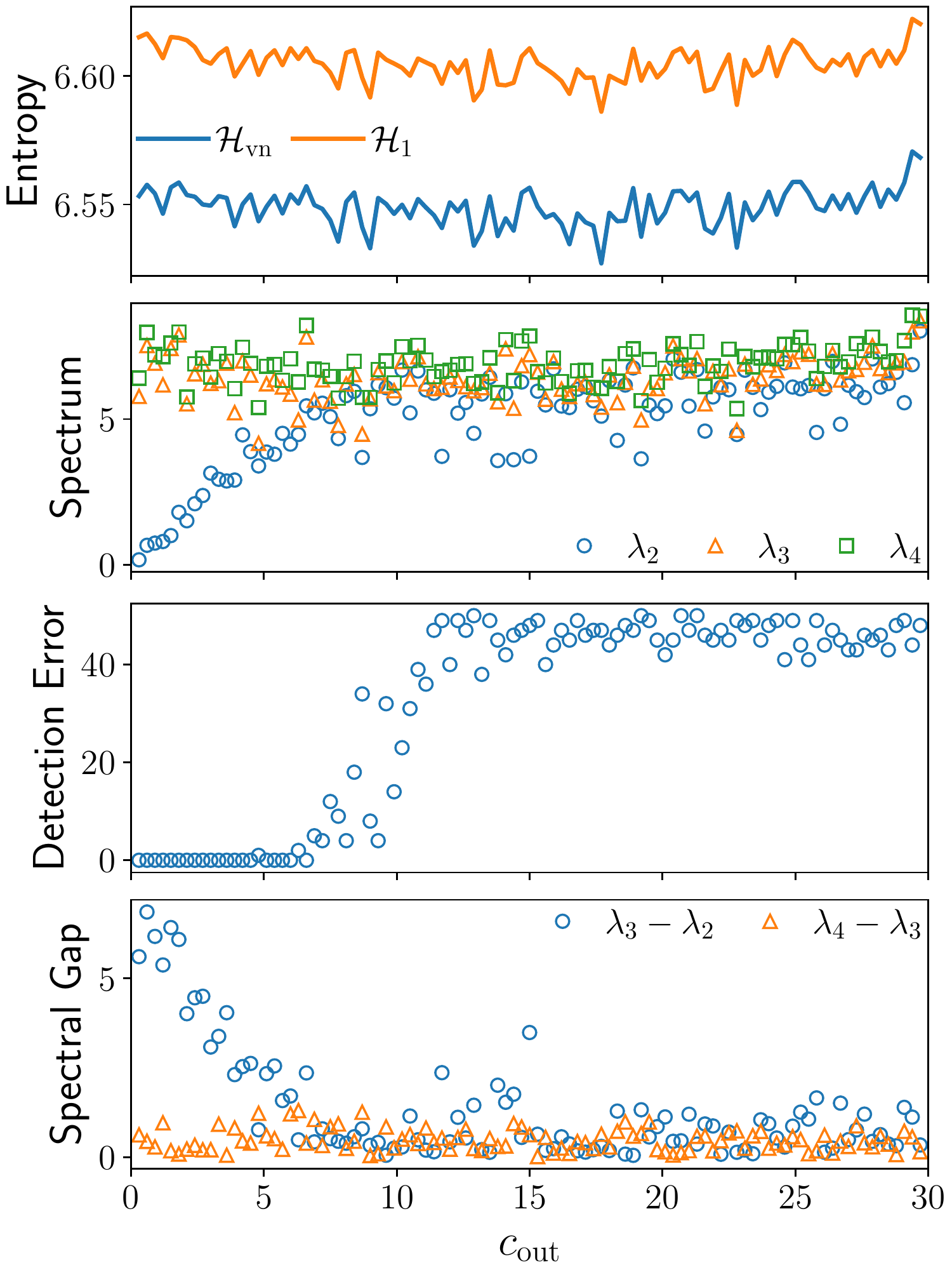}
        \caption{$\overbar{d}=15$, $c_{\rm in}+c_{\rm out}=30$}
        \label{fig:bipartition-b}
    \end{subfigure}
    \hspace{2mm}
    \begin{subfigure}{0.32\textwidth}
        \includegraphics[width=\textwidth]{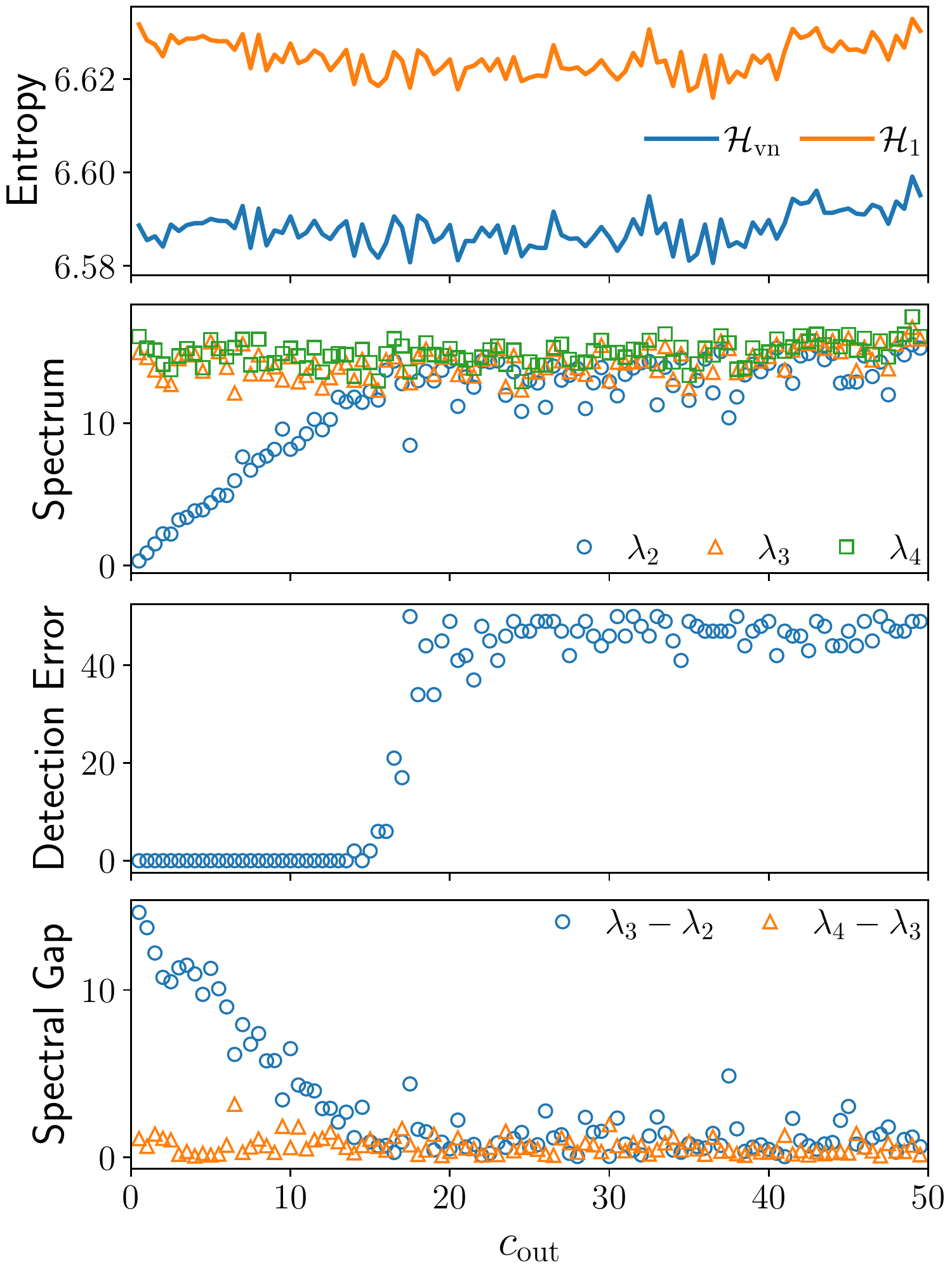}
        \caption{$\overbar{d}=25$, $c_{\rm in}+c_{\rm out}=50$}
    \end{subfigure}
    \caption{The structural information, von Neumann graph entropy, Laplacian spectrum, spectrum gap, and detection error of synthetic graphs from stochastic block model
    with $100$ nodes. There are two clusters of equal size $50$.}
    \label{fig:bipartition}
\end{figure*}

\begin{figure*}[!t]
    \begin{subfigure}{0.32\textwidth}
        \includegraphics[width=\textwidth]{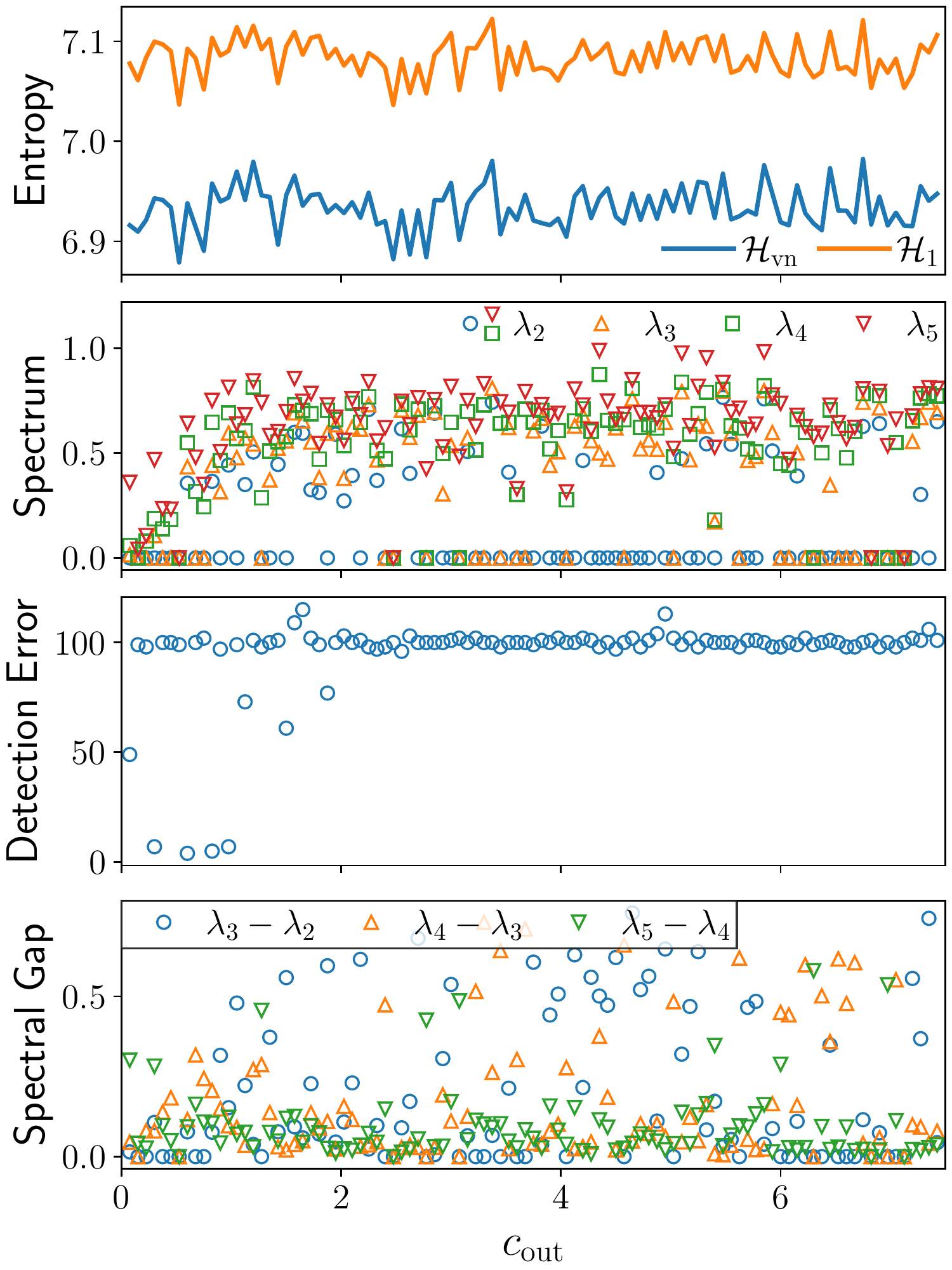}
        \caption{$\overbar{d}=5$, $c_{\rm in}+2c_{\rm out}=15$}
    \end{subfigure}
    \hspace{2mm}
    \begin{subfigure}{0.32\textwidth}
        \includegraphics[width=\textwidth]{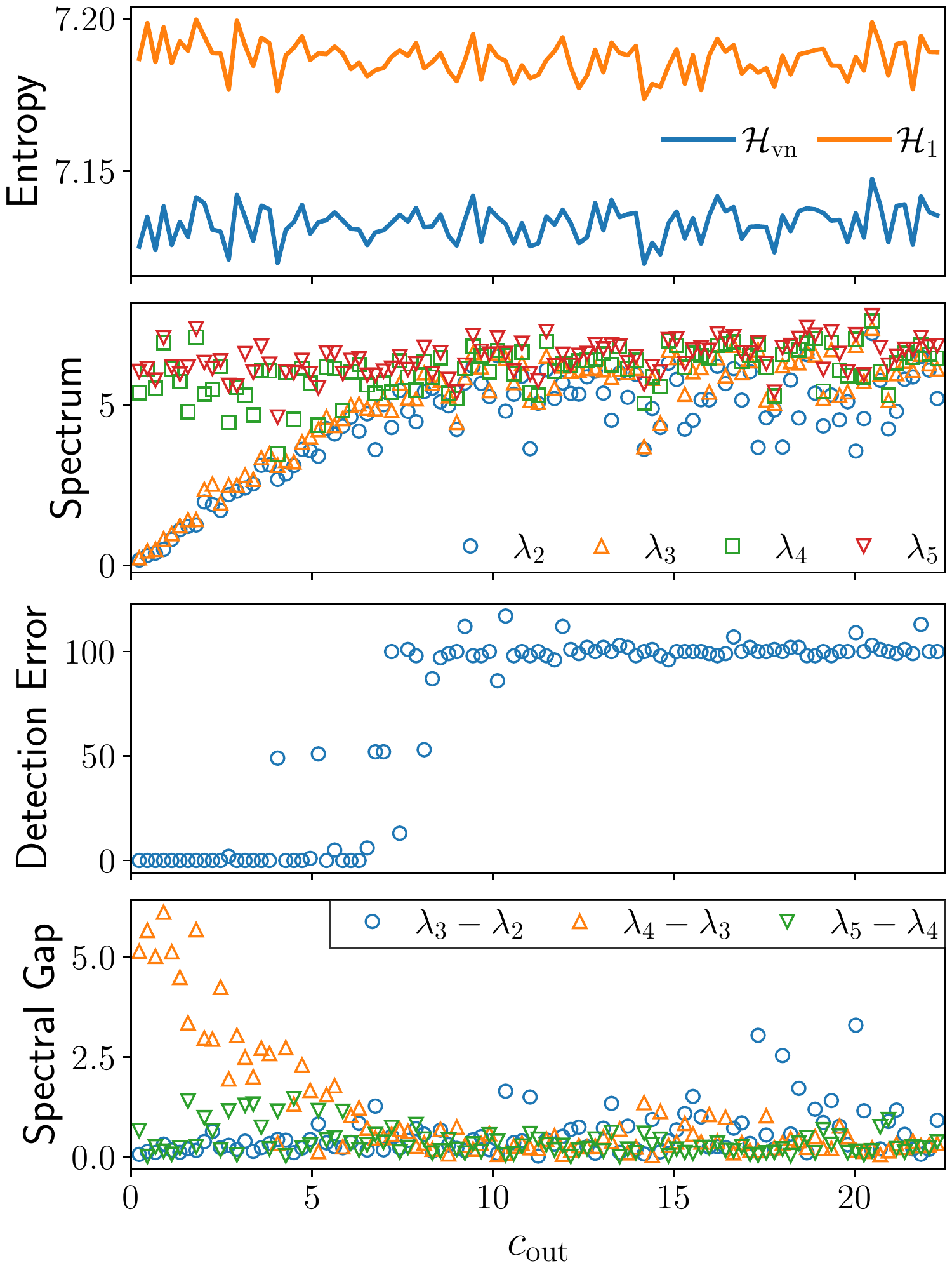}
        \caption{$\overbar{d}=15$, $c_{\rm in}+2c_{\rm out}=45$}
    \end{subfigure}
    \hspace{2mm}
    \begin{subfigure}{0.32\textwidth}
        \includegraphics[width=\textwidth]{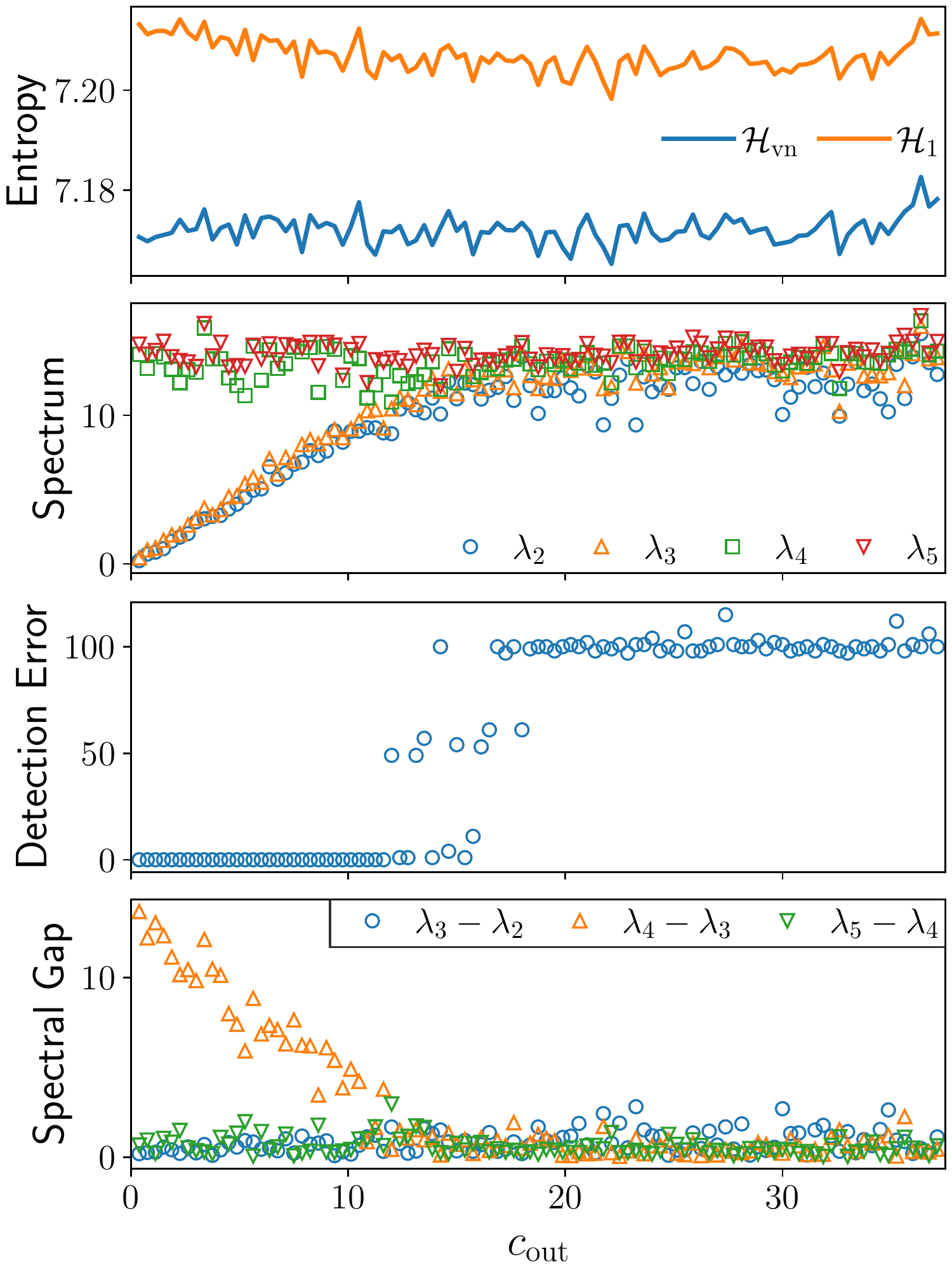}
        \caption{$\overbar{d}=25$, $c_{\rm in}+2c_{\rm out}=75$}
    \end{subfigure}
    \caption{The structural information, von Neumann graph entropy, Laplacian spectrum, spectrum gap, and detection error of synthetic graphs from stochastic block model
    with $150$ nodes. There are three clusters of equal size $50$.}
    \label{fig:tripartition}
\end{figure*}

\begin{itemize}
    \item {\em Observation 1 (Dynamics of graph entropy): } 
    Both the von Neumann graph entropy and structural information are stationary with small fluctuations and linearly correlated as $c_{\rm out}$ varies.
    \item {\em Observation 2 (Dynamics of eigenvalues): } In the stochastic block model with two clusters of equal size, the second smallest eigenvalue $\lambda_2$ linearly increases and finally reaches a steady state as $c_{\rm out}$ increases,
    while the eigenvalues $\lambda_3$ and $\lambda_4$ above it are stationary all the time.
    
    In the stochastic block model with three clusters of equal size, both the second smallest eigenvalue $\lambda_2$
    and the third smallest eigenvalue $\lambda_3$ linearly increase and finally reach a steady state as $c_{\rm out}$
    increases, while the eigenvalues $\lambda_4$ and $\lambda_5$ above them are stationary all the time.
    \item {\em Observation 3 (Phase transition): } In the stochastic block model with two clusters of equal size, both the detection error of the spectral algorithm and the spectral gap $\lambda_3-\lambda_2$ undergoes a same phase transition 
    as $c_{\rm out}$ varies. The spectral gap $\lambda_4-\lambda_3$ is stationary and close to $0$ all the time. For example, in \prettyref{fig:bipartition-b}  when $c_{\rm out}<7$ the spectral algorithm can discover the true clusters correctly and $\lambda_3-\lambda_2$ is significantly larger than $\lambda_4-\lambda_3$.
    When $c_{\rm out}>11$, the spectral algorithm works like a random guess and $\lambda_3-\lambda_2$ is mixed with $\lambda_4-\lambda_3$.
    
    In the stochastic block model with three clusters of equal size, both the detection error of the spectral algorithm and the spectral gap $\lambda_4-\lambda_3$
    undergo a same phase transition as $c_{\rm out}$ varies. The spectral gap $\lambda_5-\lambda_4$ is stationary and close to $0$ all the time.
\end{itemize}

Empirically, we conclude that
\begin{enumerate}
    \item {\em Graph entropy and community structure: } Both the von Neumann graph entropy and structural information reveal nothing about the community structure
    and the assortativity/disassortativity of graphs.
    \item {\em Spectral gaps and community structure: } If a graph has significant community structure with $k$ clusters, then the spectral gap $\lambda_{k+1}-\lambda_k$ should be significantly larger than 
    $\lambda_{k+2}-\lambda_{k+1}$ and $\lambda_k-\lambda_{k-1}$. Conversely, if there is a significant peak in the sequence of spectral gaps $\{\lambda_{i+1}-\lambda_i\}_{i=1}^{n-1}$
    of a graph, the graph should have significant community structure that could be easily detected by some algorithms.
\end{enumerate}

\subsection{Adversarial Attacks on Community Detection}

The empirical findings tell that the ground-truth 
community structure would not be easily detected if the spikes in the sequence of spectral gaps are suppressed. 
Therefore, we are interested in solving the following community obfuscation problem by exploiting the Laplacian spectrum.
\begin{problem}[Community Obfuscation]
    Minimally perturb the graph $G=(V,E)$ with community structure $\calP$ such that $\calP$ cannot be easily detected by algorithms.
\end{problem}
Unlike the graphs generated from stochastic block model, the real-world graphs have unknown number of clusters with varying sizes.
Therefore, it is hard to predict where the spike is in the sequence of spectral gaps. And it is computationally expensive to obtain the full spectral gaps.
Since the spikes represent the uneven distribution of spectrum, alternatively we can hide the community structure by maximizing some homogeneity measures on the Laplacian spectrum.
Besides the von Neumann graph entropy $\vnge(G)$, we propose another homogeneity measure called spectral polarization.
\begin{definition}[Spectral polarization]
    The spectral polarization $P(G)$ of a graph $G=(V,E)$ of order $n$ is defined as 
    \begin{equation*}
        P(G)=\sum_{i=1}^n\left(\frac{\lambda_i}{\vol(G)}-\frac{\overbar{\lambda}}{\vol(G)}\right)^2,
    \end{equation*}
    where $\lambda_1\leq\lambda_2\leq\cdots\leq\lambda_n$ are the eigenvalues of the Laplacian matrix of the graph $G$, $\overbar{\lambda}=\frac{1}{n}\sum_{i=1}^n\lambda_i$ is the average eigenvalue,
    and $\vol(G)=\sum_{i=1}^n\lambda_i$ is the volume of $G$.
\end{definition}

\begin{lemma}
    $P(G)=\frac{1}{\vol(G)}-\frac{1}{n}+\frac{\sum_{i=1}^n d_i^2}{\vol^2(G)}$.
\end{lemma}
\begin{IEEEproof}
    Simple calculus shows that
    \begin{equation*}
        \begin{aligned}
            P(G)&=\frac{1}{\vol^2(G)}\sum_{i=1}^n (\lambda_i-\overbar{\lambda})^2 \\
            &=\frac{1}{\vol^2(G)}\left(\sum_{i=1}^n \lambda_i^2-2\overbar{\lambda}\sum_{i=1}^n \lambda_i+n\overbar{\lambda}^2\right) \\
            &=\frac{1}{\vol^2(G)}\left(\sum_{i=1}^n d_i^2+\sum_{i=1}^n d_i-n\overbar{\lambda}^2\right) \\
            &=\frac{1}{\vol(G)}-\frac{1}{n}+\frac{\sum_{i=1}^n d_i^2}{\vol^2(G)}.
        \end{aligned}
    \end{equation*}
\end{IEEEproof}

Now suppose that we are allowed to add at most $k$ new edges to $G$ to hide 
the community structure, we can use \prettyref{alg:greedy} to approximately maximize spectral entropy $\vnge(G)$ or 
reset the edge centrality $EC(u,v)=d_u+d_v$ to minimize the spectral polarization $P(G)$.

\subsection{Effectiveness of von Neumann Graph Entropy and Spectral Polarization in Community Obfuscation}
We use differential analysis to show that both maximizing von Neumann graph entropy $\vnge(G)$ and minimizing spectral polarization $P(G)$
are effective in community obfuscation.

\begin{theorem}\label{thm:VNGE-comm}
    Minimally perturbing the graph $G=(V,E)$ by greedily maximizing the von Neumann graph entropy can effectively hide the community structure.
\end{theorem}
\begin{theorem}\label{thm:spectral-polarization}
    Minimally perturbing the graph $G=(V,E)$ by greedily minimizing the spectral polarization can effectively hide the community structure.
\end{theorem}
Since the proofs of \prettyref{thm:VNGE-comm} and \prettyref{thm:spectral-polarization} are similar, we only prove \prettyref{thm:VNGE-comm}
for reference.
\begin{IEEEproof}[Proof of \prettyref{thm:VNGE-comm}]
    Suppose that we minimally perturb the graph $G$ by adding a new edge $e$.
    The Laplacian spectrum of the original graph $G$ is denoted by $\bm{\lambda}(G)=(\lambda_1,\ldots,\lambda_n)$.
    The Laplacian spectrum of the perturbed graph $G'=(V,E\cup\{e\})$ is denoted by 
    $\bm{\lambda}(G')=(\lambda'_1,\ldots,\lambda'_n)$. According to the classic matrix perturbation theory, 
    $\lambda'_i=\lambda_i+\delta\lambda_i$ for any $i\in\{1,\ldots,n\}$ where 
    $\delta\lambda_i\geq 0$ is a very small increment. The sum of these increments is 
    \[
        \sum_{i=1}^n \delta\lambda_i=\sum_{i=1}^n \lambda'_i-\sum_{i=1}^n \lambda_i=2.        
    \]
    Since both $G$ and $G'$ are assumed to be connected, $\lambda'_1=\lambda_1=\delta\lambda_1=0$ and
    $\lambda'_2>\lambda_2>0$.

    According to \prettyref{eq:vnge-simplification}, maximizing $\vnge(G')$ is equivalent to minimize
    \begin{equation}
        \begin{aligned}
        \sum_{i=1}^n f(\lambda'_i)
        &=\sum_{i=1}^n f(\lambda_i+\delta\lambda_i) \\
        &=\sum_{i=2}^n f(\lambda_i)+f'(\lambda_i)\cdot\delta\lambda_i \\
        &=\sum_{i=2}^n f'(\lambda_i)\cdot\delta\lambda_i+\sum_{i=2}^n f(\lambda_i).
        \end{aligned}
    \end{equation}
    Therefore, the optimal edge $e$ can be found by minimizing $\sum_{i=2}^n f'(\lambda_i)\cdot\delta\lambda_i$
    subject to the constraints $\sum_{i=2}^n\delta\lambda_i=2$, $\lambda'_i\leq\lambda'_{i+1}$ for any $i\in\{1,\ldots,n-1\}$, and $\delta\lambda_i\geq 0$ for any $i\in\{2,\ldots,n\}$.
    Since $f'(\lambda_2)\leq f'(\lambda_3)\leq\ldots\leq f'(\lambda_n)$, the optimal edge $e$ maximizing $\vnge(G')$ assigns larger value to $\delta\lambda_2,\delta\lambda_3,\ldots$ than 
    $\delta\lambda_n,\delta\lambda_{n-1},\ldots$. Therefore, the spectral gaps indicating the community structure should disappear very quickly if 
    we greedily maximizing $\vnge(G)$ by adding edges one by one.
\end{IEEEproof}

\begin{corollary}\label{cor:SI-comm}
    Minimally perturbing the graph $G=(V,E)$ by maximizing the structural information $\ose(G)$ can effectively hide the community structure.
\end{corollary}

\begin{corollary}
    Detecting the community structure in $d$-regular graph $G_d$ is hard.
\end{corollary}
\begin{IEEEproof}
    According to \prettyref{cor:regular-graph}, $0<\Delta\calH(G_d)\leq\frac{\log_2 e}{d}$.
    Since $\ose(G_d)=\log_2 n$, we have 
    \[
        \vnge(G_d)=\ose(G_d)-\Delta\calH(G_d)\in\left[\log_2 n-\frac{\log_2 e}{d},\log_2 n\right).    
    \]
    Therefore, $\vnge(G_d)$ is close to its maximum value $\log_2 (n-1)$, implying that the spectral gaps $\lambda_{i+1}-\lambda_i\rightarrow 0$ for any $i$.
    According to the relation between spectral gaps and the significance of community structure, $G_d$ has no significant community structure.
\end{IEEEproof}

    \section{Experiments and Evaluations}
\label{sec:experiments}

We conduct extensive experiments over both synthetic and real-world datasets 
to answer the following questions:
\begin{enumerate}
  \item[Q1.] \uline{{\bf Universality} of the entropy gap over arbitrary simple graphs}: Is the entropy gap close to 0 for a wide range of graphs? Is the structural information a good proxy of the von Neumann graph entropy for a wide range of graphs?
  \item[Q2.] \uline{{\bf Sensitivity} of the entropy gap to graph properties}: How do graph properties affect the value of the entropy gap?
  \item[Q3.] \uline{{\bf Accuracy} of the approximation}: As a proxy of the von Neumann graph entropy, is the structural information more accurate than its prominent competitors?
  \item[Q4.] \uline{{\bf Speed} of the computation}: Is the computation of the structural information faster than its prominent competitors? 
  \item[Q5.] \uline{{\bf Extensibility} of the entropy gap to weighted graphs}: Is the entropy gap sensitive to the change of edge weights? Is the entropy gap still close to 0 for weighted graphs? 
  \item[Q6.] \uline{{\bf Performance analysis (\prettyref{app:further-experiments})}}: 
             What is the performance of \textsf{EntropyAug} (\prettyref{alg:greedy}) in maximizing the von Neumann graph entropy? 
             What is the performance of \textsf{IncreSim} (\prettyref{alg:incremental-DSI}) in analyzing graph streams?
             Can the structural information distance be further used to detect anomalies in a graph stream?  
             Are maximizing the von Neumann graph entropy and minimizing the spectral polarization effective in hiding community structure? 

\end{enumerate}

\subsection{Experimental Settings}
{\bf Datasets}: We consider both synthetic graphs and real-world graphs. The synthetic graphs are generated from three well-known random graph models:
Erd\"os-R\'enyi (ER) model, Barab\'asi-Albert (BA) model \cite{Barabasi1999Science}, and Watts-Strogatz (WS) model \cite{Watts1998Nature}.
The real-world graphs \cite{snap,konect,socfb-wosn-friends} used in our experiments are listed in \prettyref{tab:dataset}, which contain both static graphs with varying size and average degree, and temporal graphs with varying size and time span.
In every static graph, we ignore the direction and weight of all edges and remove both self-loops and multiple edges.
We treat every temporal graph as a stream of undirected weighted edges with timestamps.
For the convenience of analysis, we partition these edges into several groups where each group is within a certain time interval.

\begin{table}[t]
	\centering
	\small
  \caption{\label{tab:dataset} Real-world datasets used in our experiments.}
	\scalebox{0.85}{
	\begin{tabular}{l|c|c|c|c}
		\toprule
		{\bf Name} & {\bf $\#$Nodes} & {\bf $\#$Edges} & {\bf Category} & {\bf Statistics} \\
    \midrule
    \multicolumn{4}{l}{Static graphs without timestamps} & Avg. degree \\
    \midrule
		Zachary (ZA)  & 34 & 78 & Friendship &  4.59             \\
		Dolphins (DO) & 62 & 159  & Animal  &   5.13            \\
    Jazz (JA) & 198 & 2,742 & Contact  &     27.70           \\
    Skitter (SK) & 1,696,415 & 11,095,298 & Internet & 13.08 \\
    Brightkite (BK) & 58,228 & 214,078 & Friendship & 7.35 \\
    Caida (CA) & 26,475 & 53,381 & Internet & 4.03 \\
    YouTube (YT) & 1,134,890 & 2,987,624 & Friendship & 5.27 \\
    LiveJournal (LJ) & 3,997,962 & 34,681,189 & Friendship & 17.35 \\
    Pokec (PK) & 1,632,803 & 22,301,964 & Friendship & 27.32 \\
    \midrule
    \multicolumn{4}{l}{Dynamic graphs with timestamps} & $\#$Snapshots \\
    \midrule
    Wiki-IT (WK) & 1,204,009 & 34,826,283 & Hyperlink & 100 \\
    Facebook (FB) & 61,096 & 788,135 & Friendship & 29 \\
		\bottomrule
	\end{tabular}
	}
\end{table}	

\noindent{\bf Hardwares}: The experiments have been performed on a server with Intel(R) Xeon(R) CPU 2.40 GHz (32 virtual cores) and 256GB RAM, averaging 10 runs for random algorithms and random inputs unless stated otherwise.

\noindent{\bf Implementation}: All of the proposed algorithms and baselines are implemented in Python. 


\subsection{Q1. Universality (\prettyref{fig:universality})}
To evaluate the universality of the entropy gap, we measure
the structural information and the exact von Neumann entropy on a set of synthetic graphs with 2,000 nodes.
For the ER and BA models, we generate graphs with average degree in $\{2,4,\ldots,200\}$.
For the WS model, we generate graphs with edge rewiring probability in $\{0,1/20,\ldots,1\}$ for each 
average degree in $\{6,10,20,50\}$. We additionally measure the sharpened lower and upper bounds of the entropy gap.
The results are shown in \prettyref{fig:universality}.

The observations are three fold. First, {\bf the entropy gap is close to 0 for a wide range of graphs}.
The entropy gap of each synthetic graph is no more than $0.2$, whereas the exact von Neumann entropy is greater than $10$.
Second, {\bf the entropy gap is negatively correlated with the average degree}. Dense graph tends to have very small entropy gap.  
Third, {\bf the structural information is linearly correlated with the von Neumann graph entropy}, with only few exceptions.
There is no exception for the ER synthetic graphs. For the BA synthetic graphs, the exceptions are those graphs with extremely small average degree.
For the WS synthetic graphs, the exceptions are those graphs with extremely small edge rewiring probability.

\begin{figure*}
  \begin{subfigure}{0.32\textwidth}
    \includegraphics[width=\textwidth]{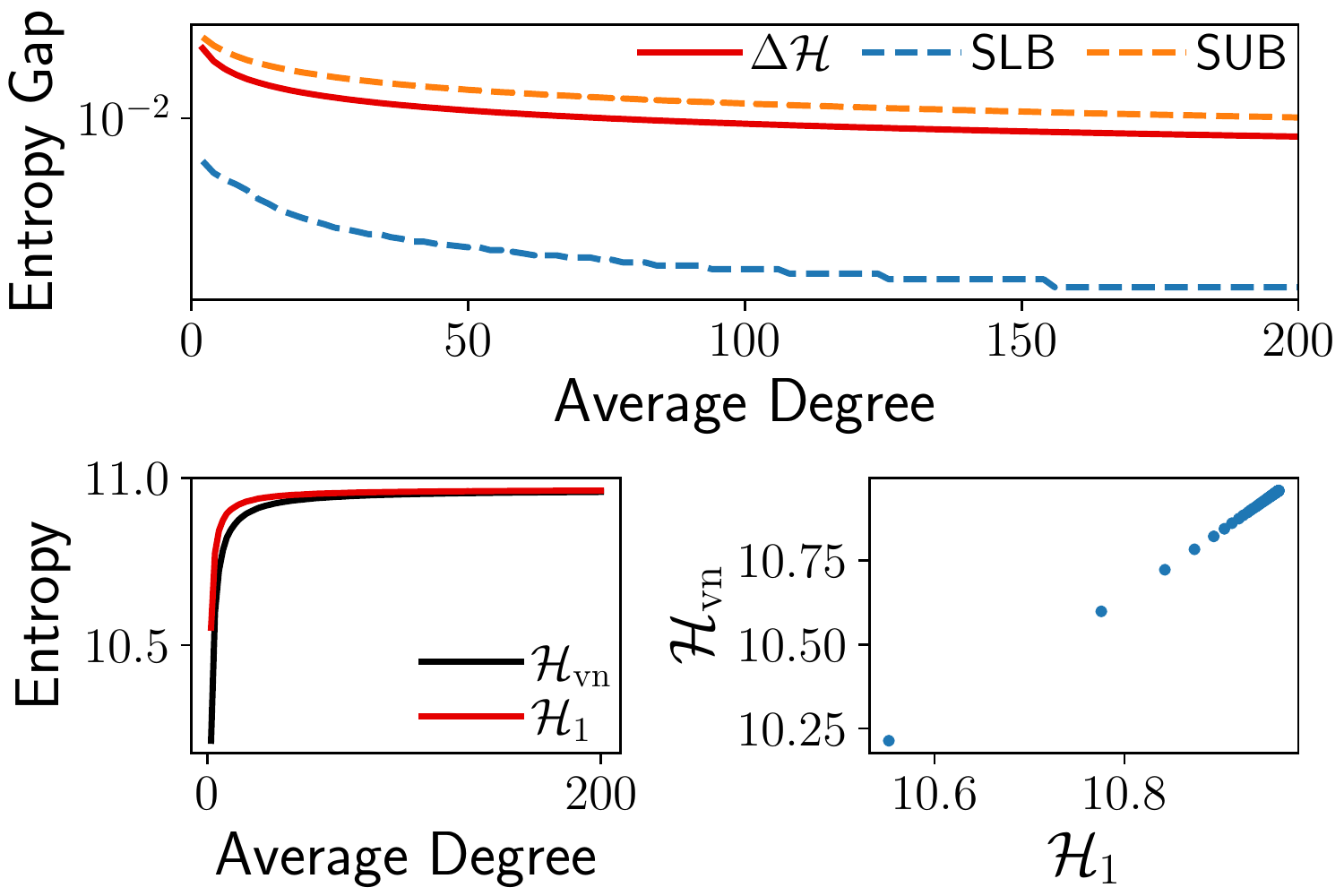}
    \caption{ER model}
    \label{fig:universality-ER}
  \end{subfigure}
  \hspace{2mm}
  \begin{subfigure}{0.32\textwidth}
    \includegraphics[width=\textwidth]{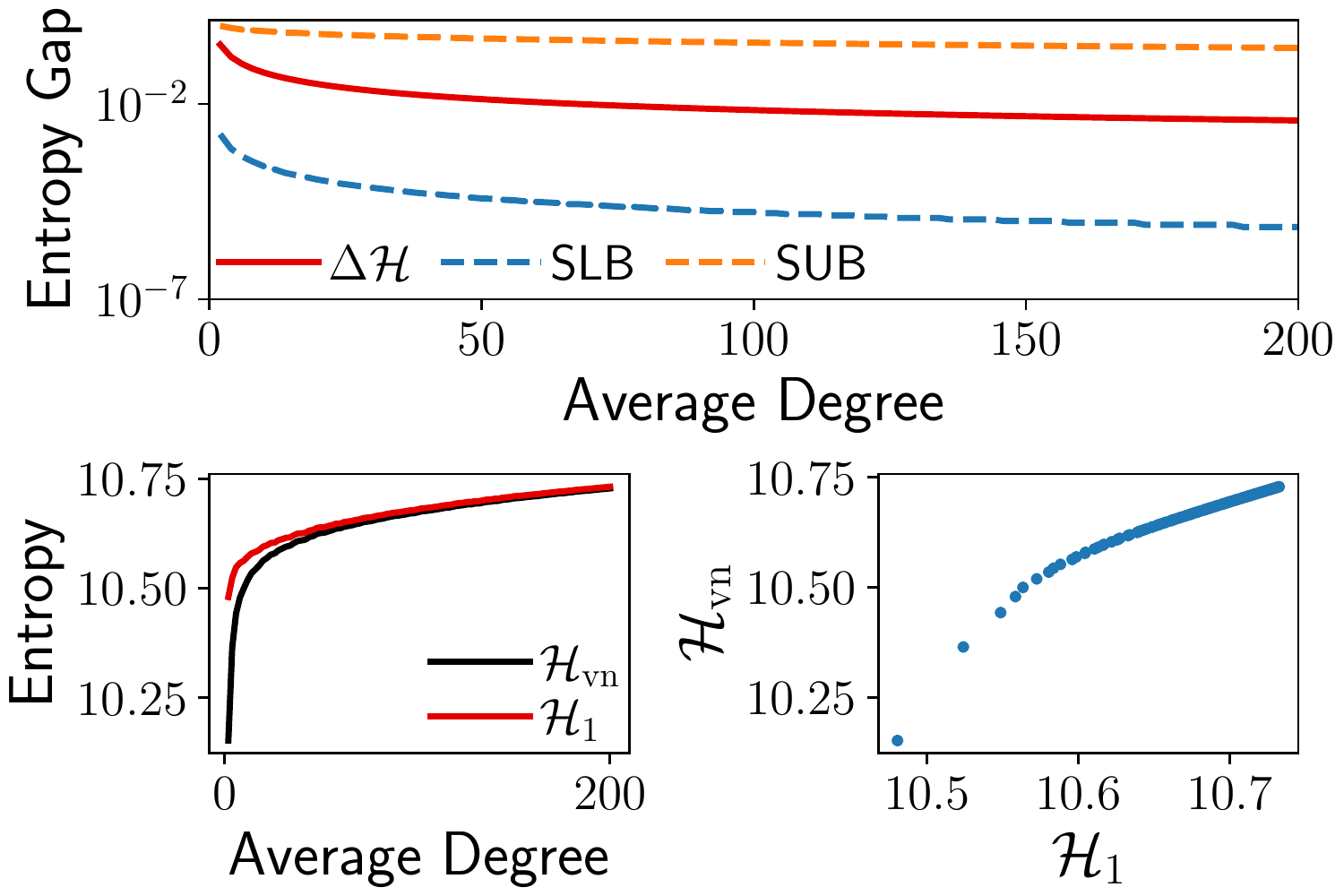}
    \caption{BA model}
    \label{fig:universality-BA}
  \end{subfigure}
  \hspace{2mm}
  \begin{subfigure}{0.32\textwidth}
    \includegraphics[width=\textwidth]{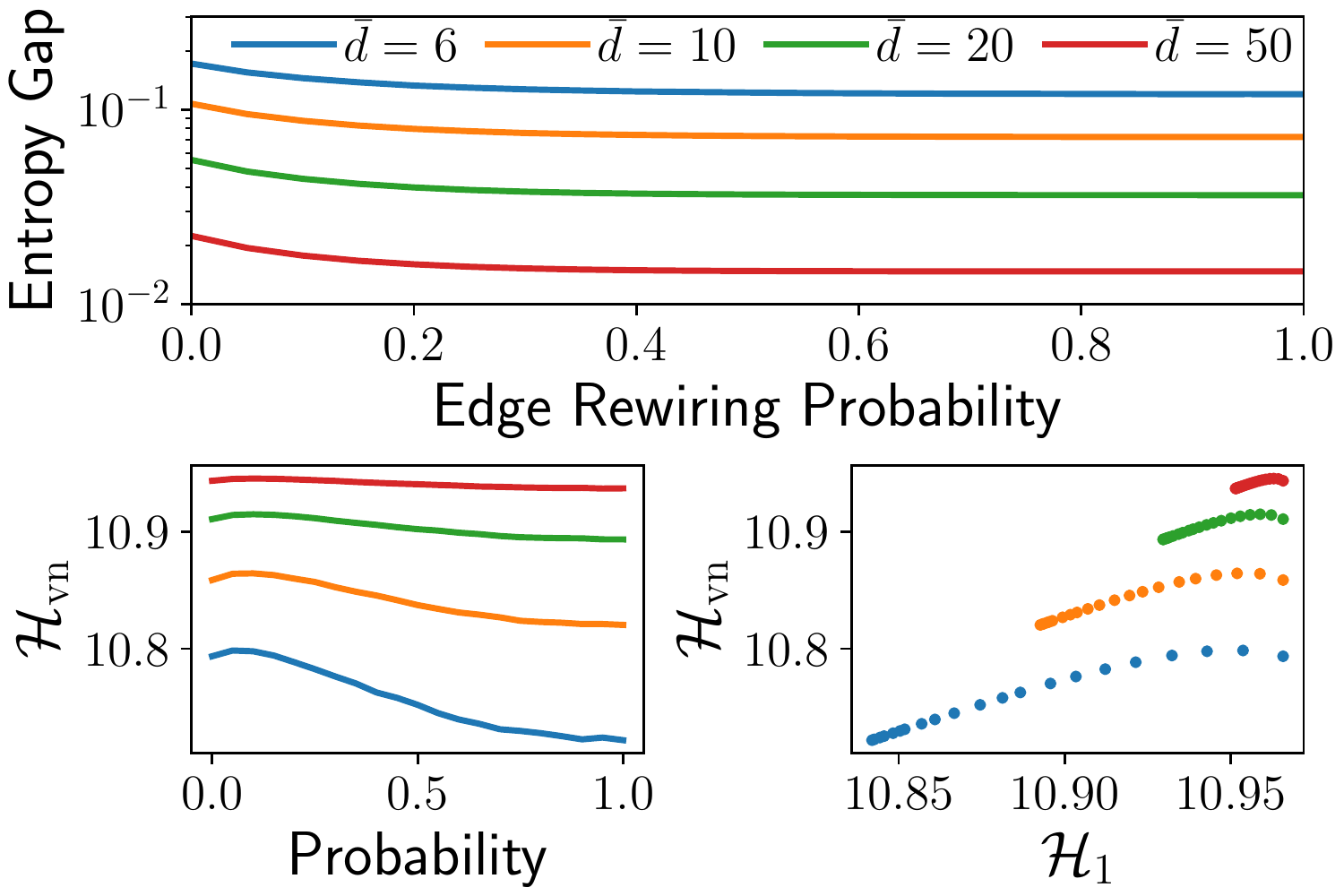}
    \caption{WS model}
    \label{fig:universality-WS}
  \end{subfigure}
  \caption{The structural information, von Neumann graph entropy, and entropy gap of synthetic graphs
  generated from three random graph models with $2,000$ nodes, varying average degree, and edge rewiring probability.}
  \label{fig:universality}
\end{figure*}

\subsection{Q2. Sensitivity (\prettyref{fig:universality}, \prettyref{fig:sensitivity})}
To evaluate the sensitivity of the entropy gap to graph properties such as average degree, graph size, and rewiring probability,
we further measure the entropy gap of 10 synthetic graphs with graph size in $\{500,1000,\ldots,5000\}$
for each random model. The average degree is chosen from $\{2,5,10,20,50,100\}$ for ER and BA models, and the edge rewiring probability
is chosen from $\{0,0.1,0.2,0.4,0.8,1\}$ for the WS model. 

The observations from \prettyref{fig:universality} and \prettyref{fig:sensitivity} are three fold.
First, the entropy gap decreases as the average degree increases for all the three random graph models.
Second, the entropy gap decreases as the edge rewiring probability increases for the WS model.
Third, {\bf the entropy gap is nearly insensitive to the change of graph size}.

\begin{figure*}
  \begin{subfigure}{0.32\textwidth}
    \includegraphics[width=\textwidth]{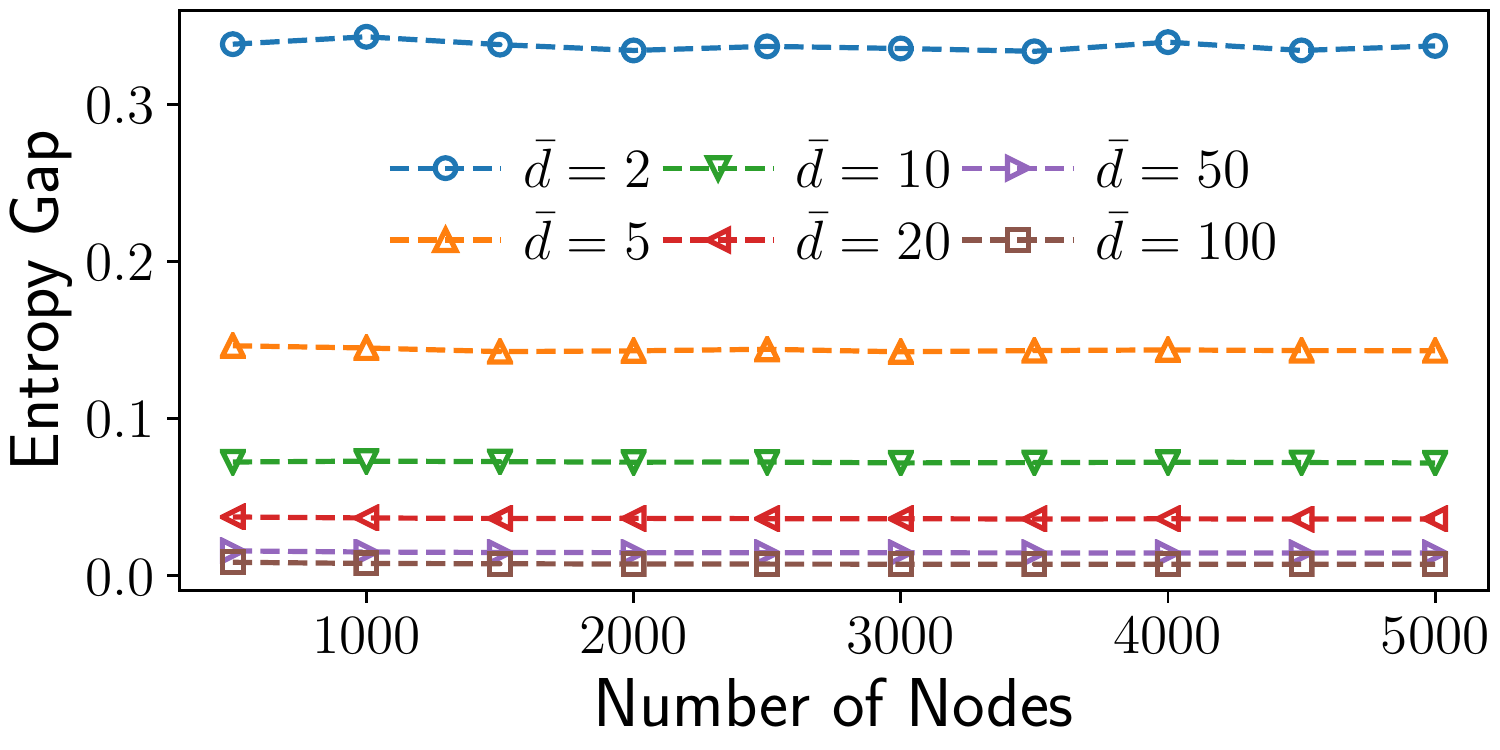}
    \caption{ER model}
  \end{subfigure}
  \hspace{2mm}
  \begin{subfigure}{0.32\textwidth}
    \includegraphics[width=\textwidth]{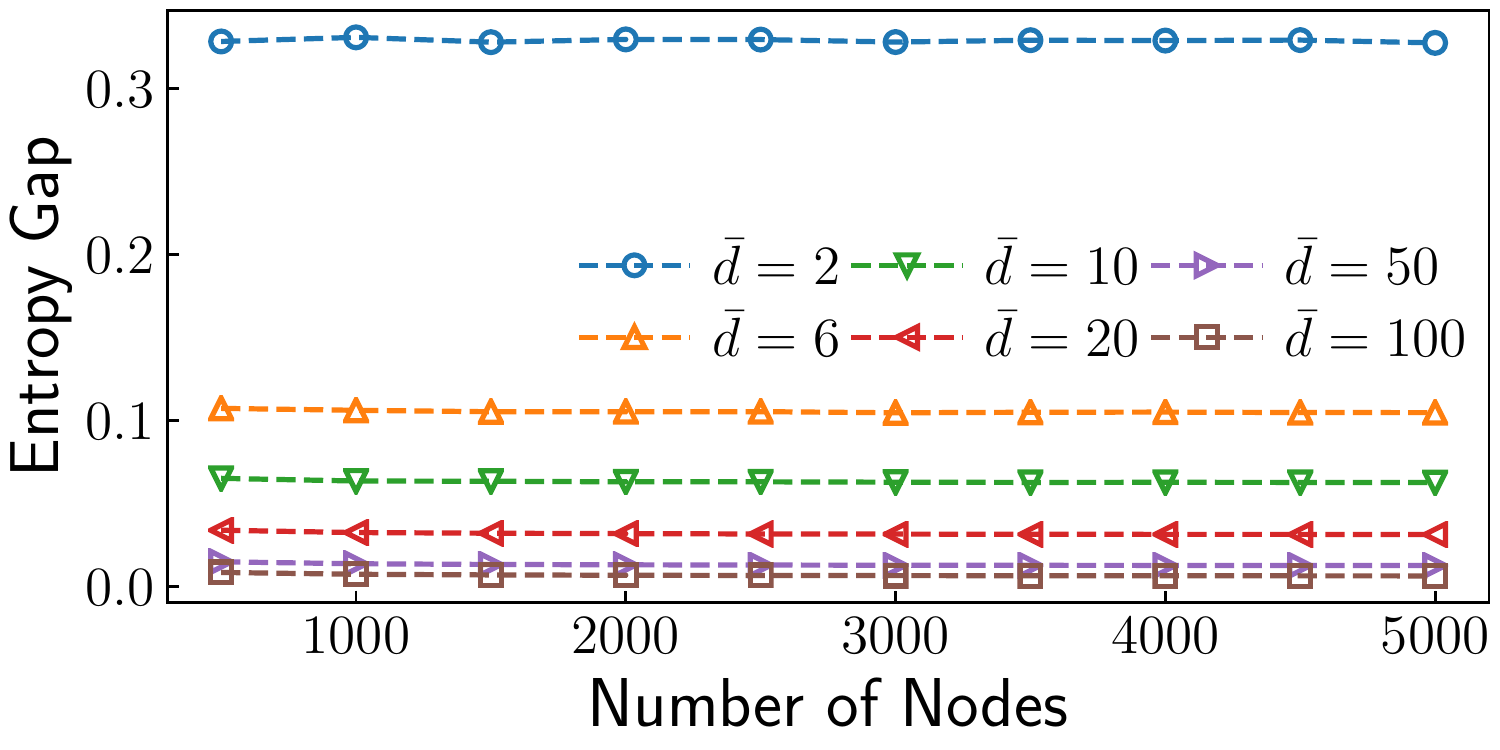}
    \caption{BA model}
  \end{subfigure}
  \hspace{2mm}
  \begin{subfigure}{0.32\textwidth}
    \includegraphics[width=\textwidth]{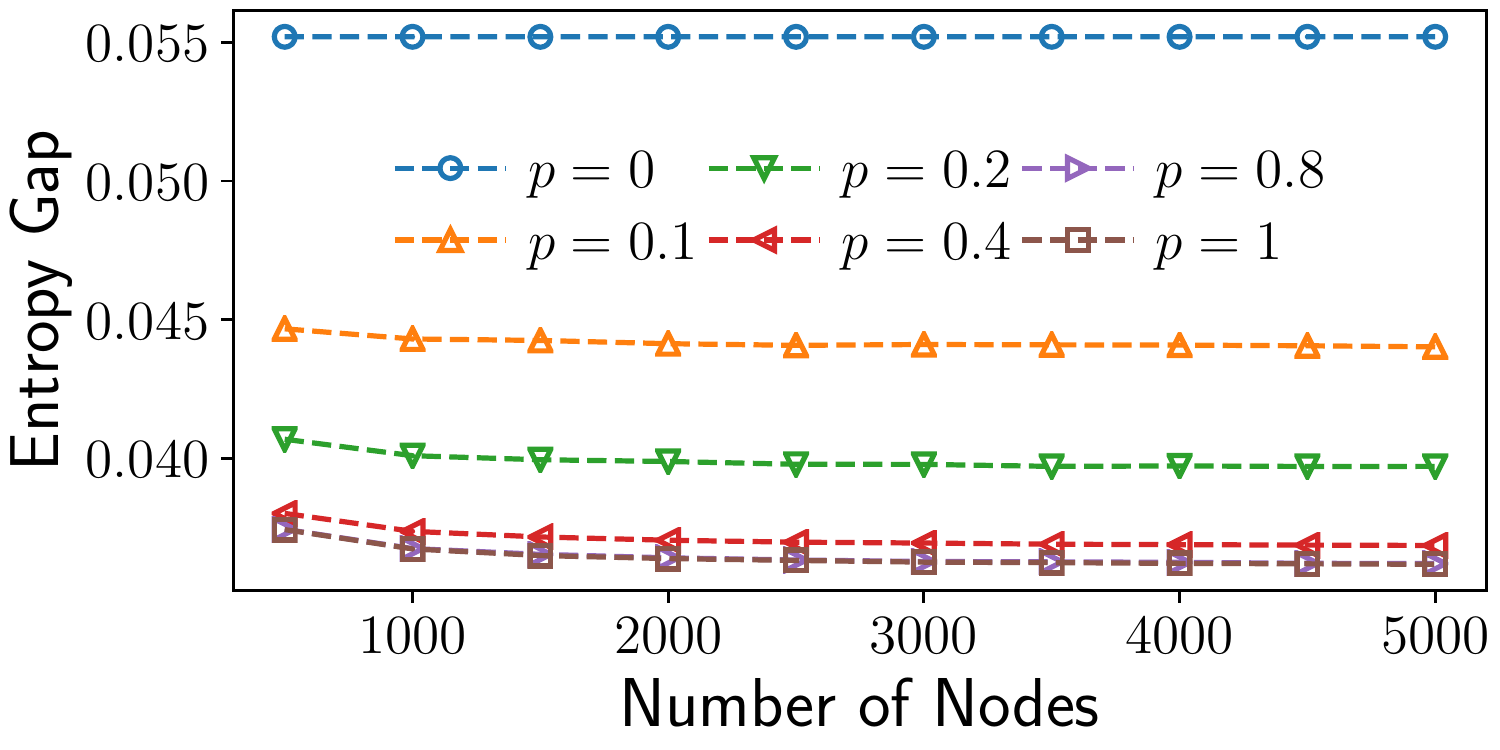}
    \caption{WS model ($\bar{d}=20$)}
  \end{subfigure}
  \caption{Effects of input graph properties on the entropy gap for three random graph models.}
  \label{fig:sensitivity}
\end{figure*}

\subsection{Q3. Accuracy (\prettyref{fig:accuracy})}
To evaluate the accuracy of the structural information as an approximation of the von Neumann graph entropy, 
we measure the structural information, exact von Neumann entropy (when the graph size is small), and three prominent approximations (as competitors) in 9 real-world static graphs.
The competitors are 
1) FINGER-$\widehat{H}$ \cite{Chen2019ICML} defined as $\widehat{\calH}_{\rm F}(G)=-Q\log_2(\lambda_{\rm max}/{\rm tr}(L))$ where 
$Q=1-{\rm tr}(L^2)/{\rm tr}^2(L)$,
2) FINGER-$\widetilde{H}$ \cite{Chen2019ICML} defined as $\widetilde{\calH}_{\rm F}(G)=-Q\log_2(2d_{\rm max}/{\rm tr}(L))$, 
and 3) SLaQ \cite{Anton2020WWW}.
The results in \prettyref{fig:accuracy} show that 
{\bf the structural information is an accurate approximation of the von Neumann graph entropy}.
The approximation error of structural information is obviously
much smaller than $\widehat{\calH}_{\rm F}$ and $\widetilde{\calH}_{\rm F}$.
And it is comparable to the approximation error of SLaQ with only few exceptions such as YT and SK 
where the structural information is slightly better.

\begin{figure}
  \includegraphics[width=0.9\linewidth]{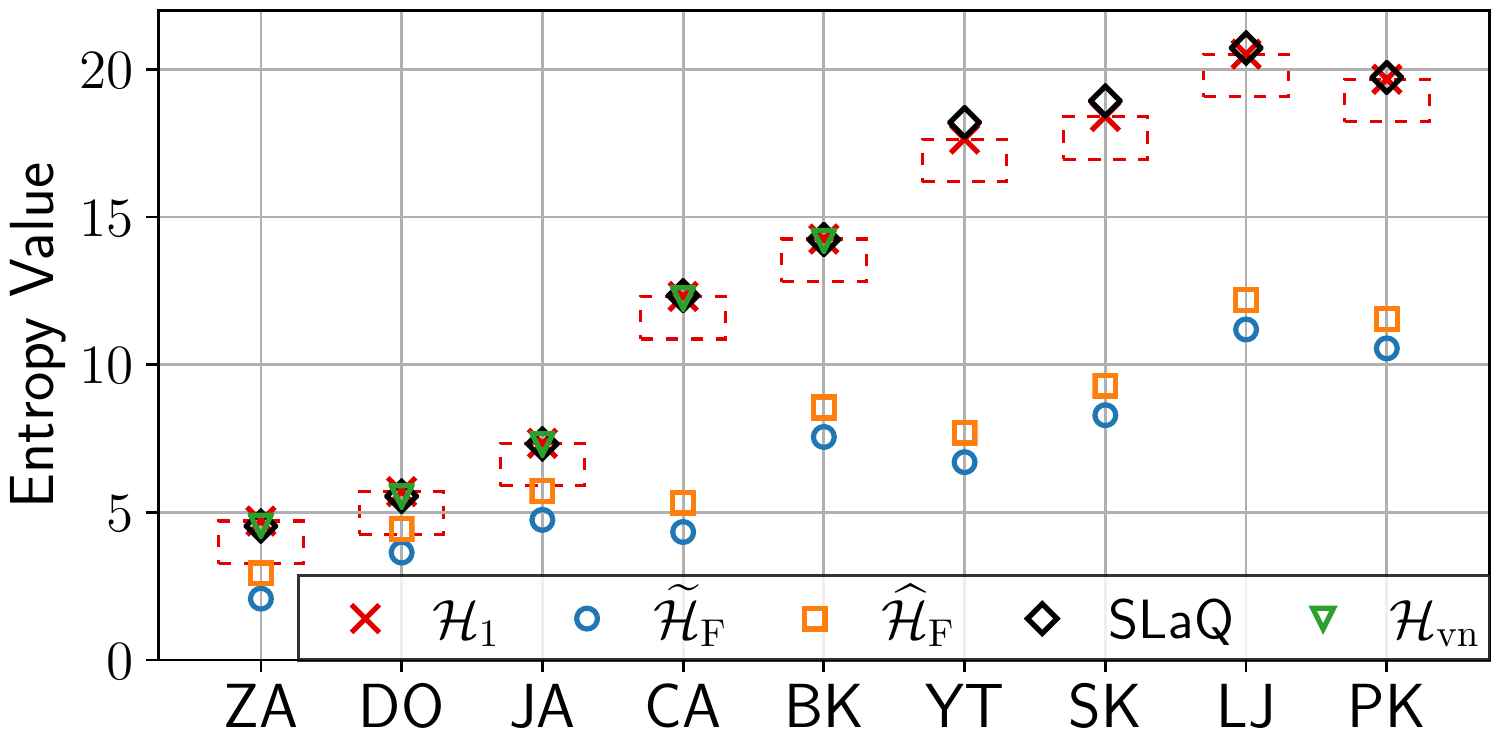}
  \caption{\uline{Structural information is an accurate proxy of von Neumann graph entropy.} The exact von Neumann graph entropy lies in the red dotted box whose height is $\log_2 e$.}
  \label{fig:accuracy}
\end{figure}

\subsection{Q4. Speed (\prettyref{fig:speed})}
To evaluate the computational speed of the structural information, we measure the running time of structural information and its three competitors in 9 real-world static graphs.
The results in \prettyref{fig:speed} show that 
{\bf the computation of structural information is fast}.
It is about 2 orders of magnitude faster than $\widehat{\calH}_{\rm F}$, at least 2 orders of magnitude faster than SLaQ,
and comparable to $\widetilde{\calH}_{\rm F}$.
Combining \prettyref{fig:accuracy} and \prettyref{fig:speed}, we conclude that 
{\bf the structural information is the only one that achieves both high efficiency and high accuracy among the prominent methods}.

\begin{figure}
  \includegraphics[width=0.9\linewidth]{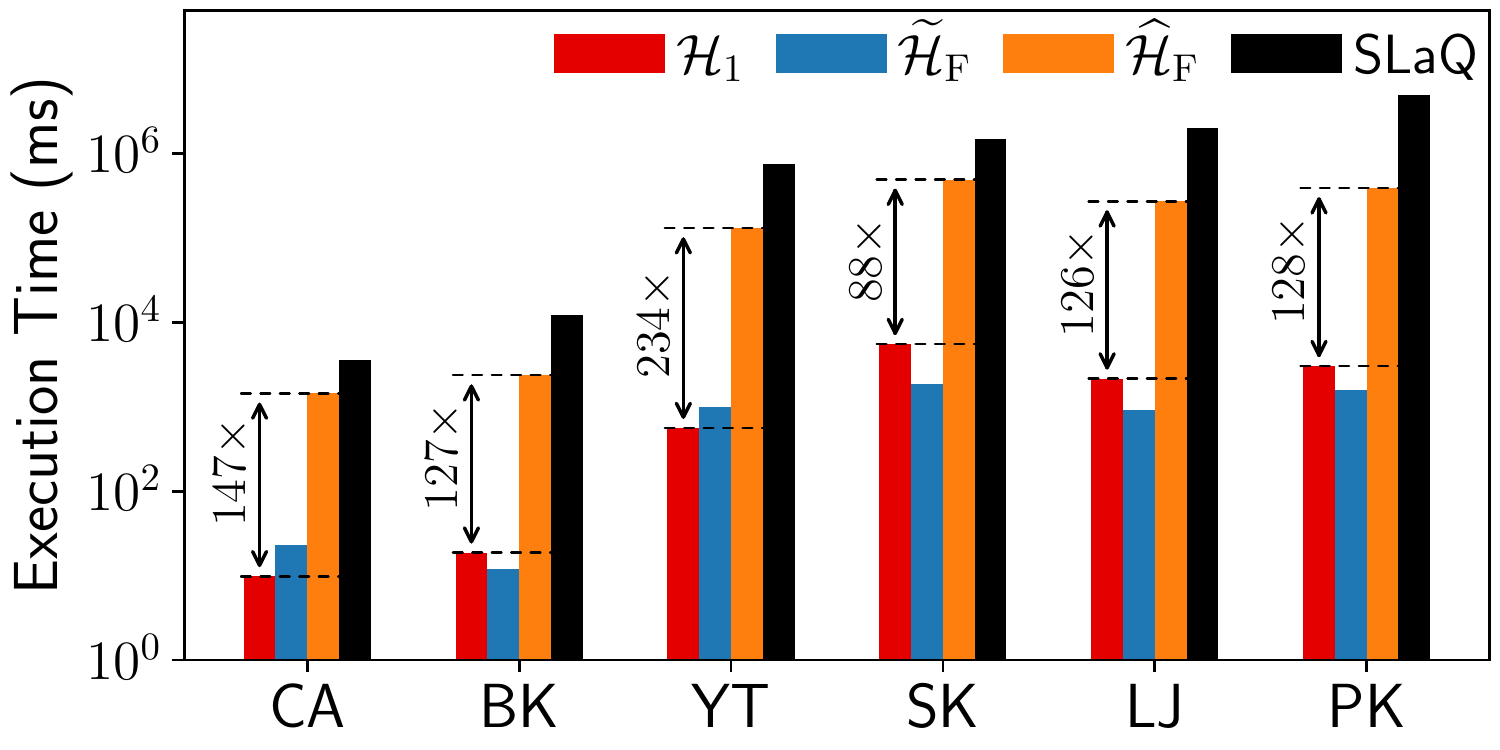}
  \caption{\uline{The computation of structural information is fast.}}
  \label{fig:speed}
\end{figure}

\subsection{Q5. Extensibility (\prettyref{fig:extensibility})}
To evaluate the extensibility of the entropy gap to weighted graphs, we measure the 
entropy gap of synthetic weighted graphs. Specifically, we choose 3 real-world graphs (ZA, DO, JA) with small size, a complete graph $K_{1000}$ and ring graph $R_{1000}$ 
each with 1000 nodes. 
The weight of each edge is set uniformly at random in the range $[1,w]$.
We repeat the experiments for each $w\in\{1,2,\ldots,20\}$.
The results in \prettyref{fig:extensibility} show that 
{\bf the entropy gap is insensitive to the change of edge weights in these graphs}.
Therefore, it is of high probability that the entropy gap is still very small for a wide range of weighted graphs.

\begin{figure}
  \includegraphics[width=0.9\linewidth]{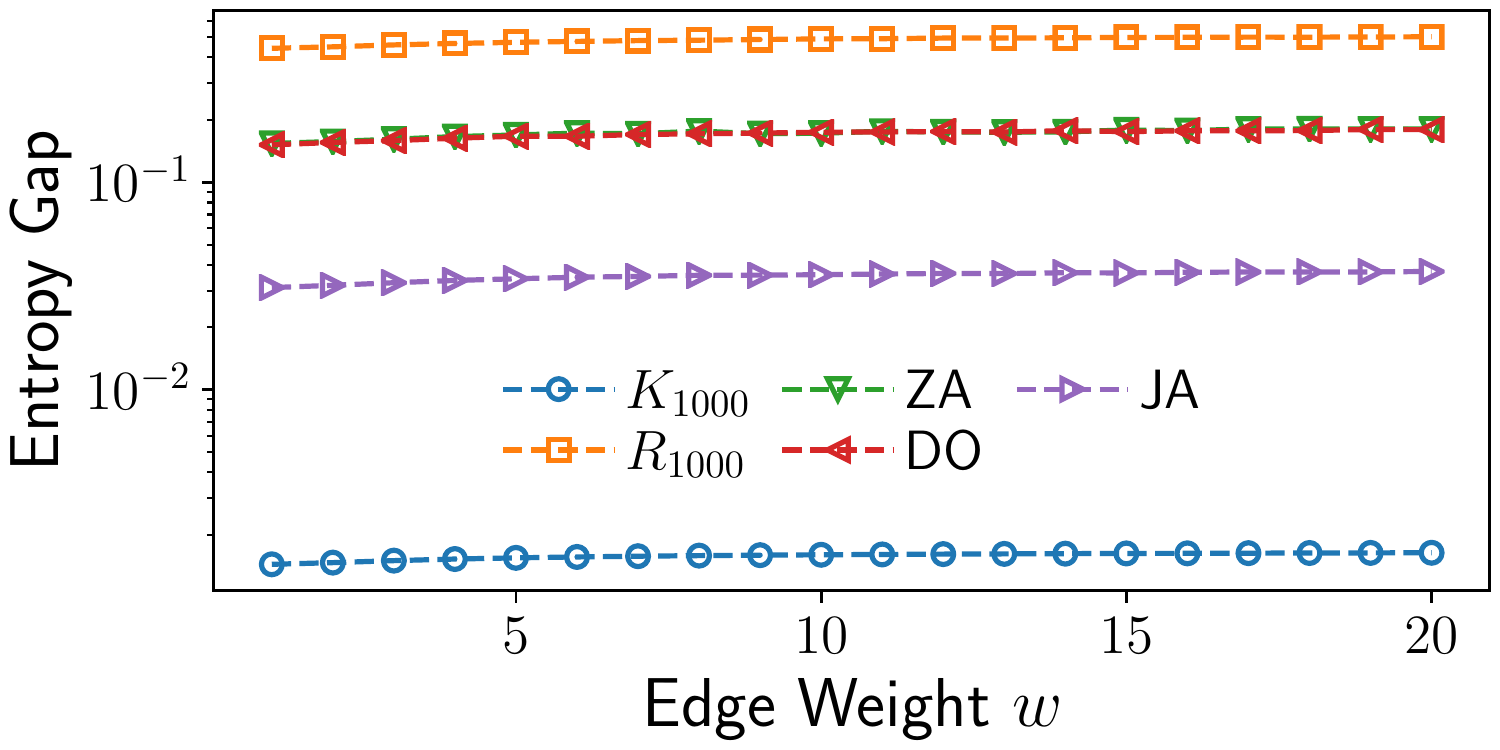}
  \caption{\uline{The entropy gap is insensitive to the edge weights.}}
  \label{fig:extensibility}
\end{figure}
    \section{Conclusions and Future Work}
\label{sec:conclusion-and-future-work}

In this work, we suggest to use the structural information as a proxy of the von Neumann
graph entropy such that provable accuracy, scalability, and interpretability are achieved 
at the same time. 
Since the experimental results show that the entropy gap is insensitive to the graph size, we can estimate the entropy gap of a very large graph
using small graphs generated from the same generative random graph model.
We believe that our idea also provides new insights into approximations of graph spectral descriptors:
besides function approximation, we can try to approximate the graph spectrum
using simple and easily available graph statistics, such as the degree sequence.

There are multiple tangible research fronts we can pursue.
First, in some access limited scenarios such as the World Wide Web, the complete degree sequence
is often not available, therefore we need to develop sampling-based methods to estimate the structural information.
Second, both the von Neumann graph entropy and the structural information can be viewed as a function on the edge set.
Their properties such as the submodularity and monotonicity are under exploration.
Last, the approximation of the von Neumann graph entropy defined on the eigenvalues of \emph{normalized} Laplacian matrix
is still in its infancy.

    \appendices

    \section{Additional Experiments}
\label{app:further-experiments}

\subsection{Performance of EntropyAug (\prettyref{fig:entropy-augmentation})}
To evaluate the performance of \textsf{EntropyAug} (\prettyref{alg:greedy}) in maximizing the von Neumann graph entropy, 
we measure the running time and dynamics of the von Neumann graph entropy for \textsf{EntropyAug} and two competitors in three small real-world graphs
ZA, DO, and JA. The two baselines are 1) ``\textsf{random}'' referring to the random addition of $k$ non-existing edges, 
and 2) ``\textsf{algebraic}'' \cite{Ghosh2006CDC} referring to the greedy addition of $k$ non-existing edges that leads to the largest 
increase of the algebraic connectivity $\lambda_{n-1}$, which is the second smallest eigenvalue of the Laplacian matrix.
We believe the ``\textsf{algebraic}'' algorithm is a competent competitor since 
maximizing $\lambda_{n-1}$ would make the Laplacian spectrum concentrated on its mean, thereby maximizing the von Neumann entropy.
The results in \prettyref{fig:entropy-augmentation} show that 
{\bf EntropyAug is the only one that achieves both high efficiency and large increments of von Neumann graph entropy}.

\begin{figure*}
  \begin{subfigure}{0.32\textwidth}
    \includegraphics[width=\textwidth]{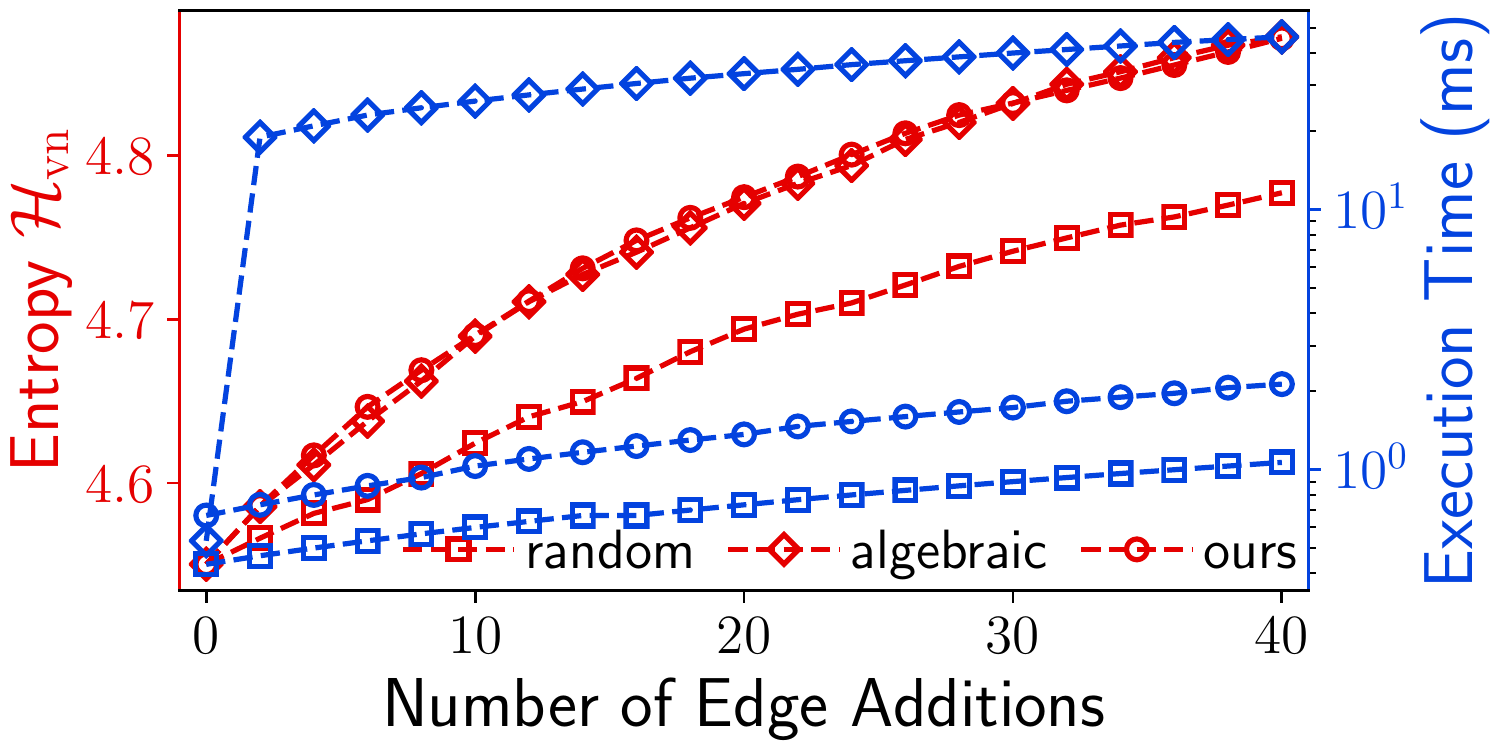}
    \caption{Zachary (ZA)}
    \label{fig:greedy-zachary}
  \end{subfigure}
  \hspace{2mm}
    \begin{subfigure}{0.32\textwidth}
        \includegraphics[width=\textwidth]{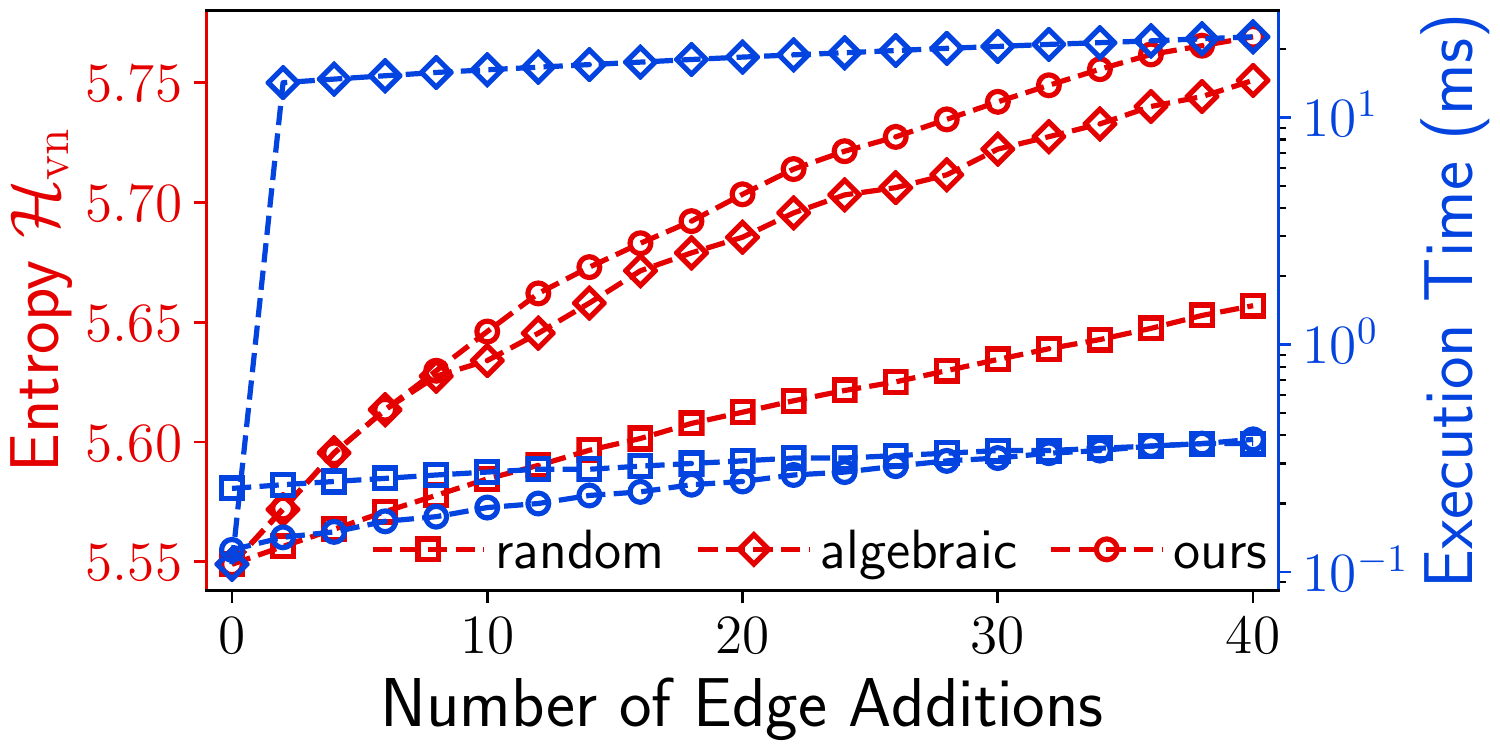}
        \caption{Dolphins (DO)}
        \label{fig:greedy-dolphins}
    \end{subfigure}
    \hspace{2mm}
    \begin{subfigure}{0.32\textwidth}
        \includegraphics[width=\textwidth]{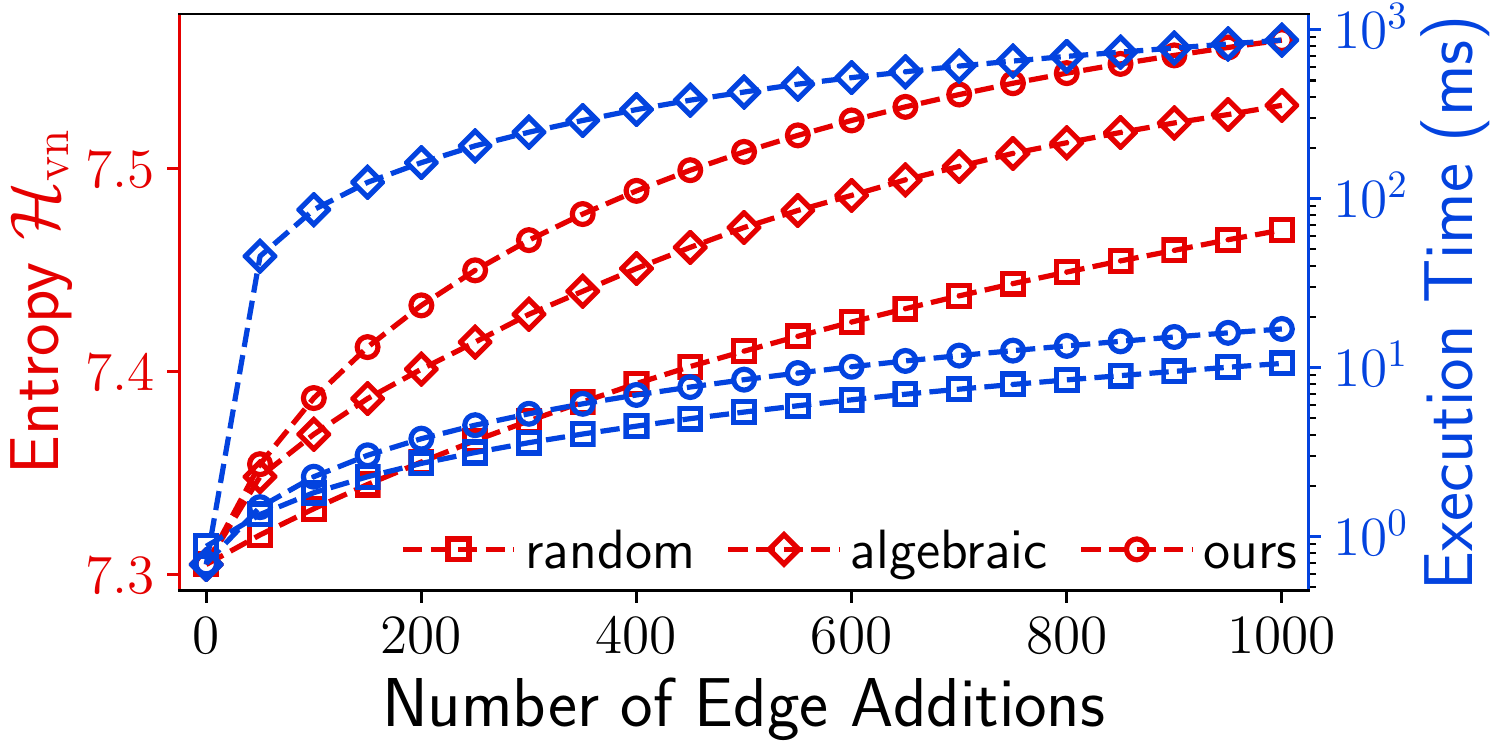}
        \caption{Jazz (JA)}
        \label{fig:greedy-jazz}
    \end{subfigure}
  \caption{Compared with the other two methods, our structural information based method is the only one that achieves both high efficiency and large increments of von Neumann graph entropy.}
  \label{fig:entropy-augmentation}
\end{figure*}

\subsection{Performance of IncreSim (\prettyref{fig:similarity})}
To evaluate the performance of \textsf{IncreSim} (\prettyref{alg:incremental-DSI}) and its relation with the VEO score, we measure the distance
between two adjacent graphs in two real-world temporal graphs. We choose three methods (\textsf{IncreSim}, VEO score, and deltaCon) along with two simple measures (the number of added edges and the number of deleted edges).
The VEO score \cite{Papadimitriou2010} between two adjacent graphs $G_t$ and $G_{t+1}$
is defined as $1-\frac{2(|V_t\cap V_{t+1}|+|E_t\cap E_{t+1}|)}{|V_t|+|V_{t+1}|+|E_t|+|E_{t+1}|}$, which measures the change rate of edge set and node set.
The deltaCon \cite{Koutra2016TKDD} is a prominent method to measure graph similarity
based on fast belief propagation. The results are shown in \prettyref{fig:similarity}. 

The observations are two fold.
First, {\bf the structural information distance is linearly correlated with the VEO score}, 
indicating that the structural information distance is not dominated by only local information, but rather a 
global measure on the graphs. For the FB temporal graph, the Pearson correlation coefficient and Spearman rank-order correlation coefficient
of $\calD_{\rm SI}$ with the VEO score are $(0.95, 0.97)$ respectively, which is much higher than $(0.70, 0.77)$ with deltaCon.
For the WK temporal graph, the two correlation coefficients of $\calD_{\rm SI}$
with the VEO score are $(0.96, 0.96)$ respectively, which is also much higher than $(-0.14,0.00)$ with deltaCon.
Second, {\bf all of the three methods effectively capture the dynamics of graph streams}.
For the FB temporal graph, the trends of the three distance measures are similar.
For the WK temporal graph, we can see that the distance measure changes dramatically in the beginning, then gradually 
turns to be flat, which implies that the structure of WK temporal graph gradually becomes stable.

\begin{figure}
  \includegraphics[width=\linewidth]{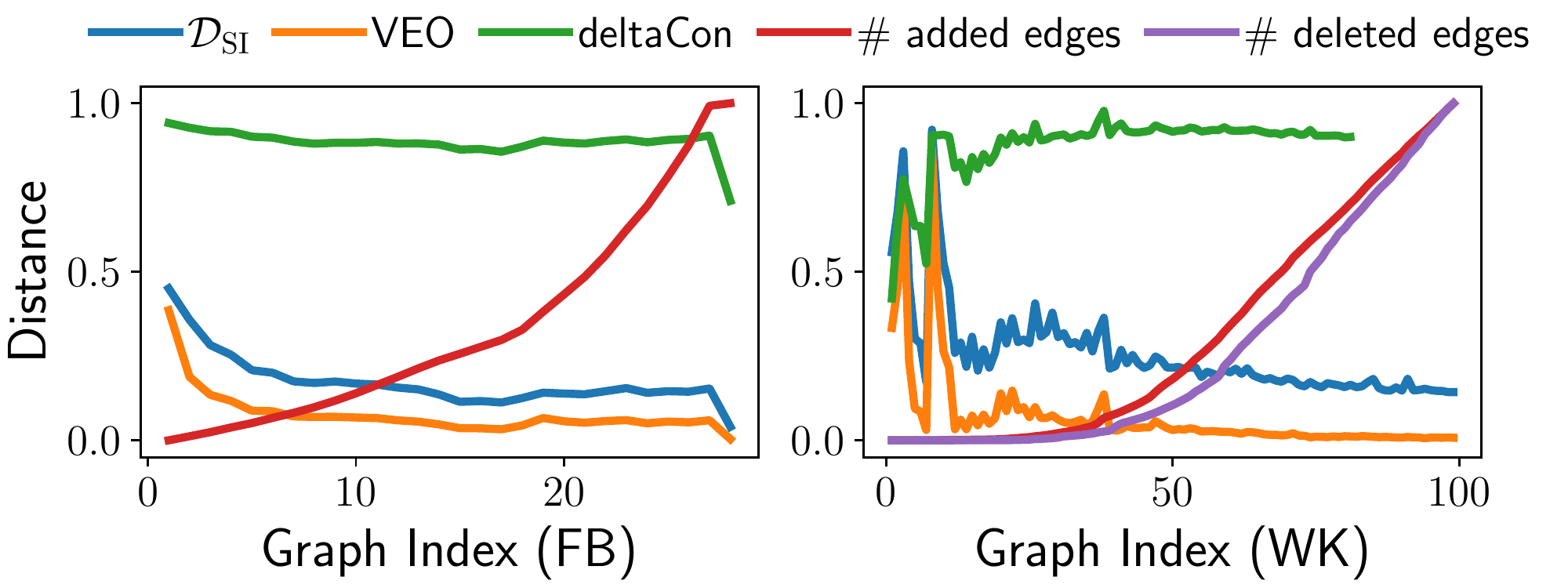}
  \caption{Distance between adjacent graphs in graph streams.
  The number of added/deleted edges is divided by the total number of added/deleted edges.}
  \label{fig:similarity}
\end{figure}

\subsection{Performance in Anomaly Detection (\prettyref{fig:anomaly})}
We further evaluate the effectiveness of the structural information distance
in detecting the distributed denial-of-service (DDoS) attacks in a graph stream.
We first generate $10$ synthetic graphs $\calG=\{G_t\}_{t=1}^{10}$ from the BA model, each of which has 100 nodes and 
average degree $\overbar{d}=4$. We believe that the synthetic graph stream $\calG$ is a good representative of the real-world
scale-free graph streams.
Then we model the DDoS attack with strength $k$ as follows:
  (1) Randomly select a graph $G_{t^\ast}$ from $\calG$.
  (2) Transform $G_{t^\ast}$ into an anomalous graph $G'_{t^\ast}$. Specifically, we first randomly select a target node $v$, then randomly select $k$ source nodes $\calS=\{s_i\}_{i=1}^k$.
        Finally, we connect the target node $v$ with the source node $s_i$ for each $i\in\{1,\ldots,k\}$.
  (3) Generate the anomalous graph stream $\calG'$ via replacing the graph $G_{t^\ast}$ from $\calG$ with $G'_{t^\ast}$.

We use a graph distance measure to rank the anomalous graph in a graph stream. Suppose that 
the distance between $G_t$ and $G_{t+1}$ is $\theta_{t,t+1}$, then the anomalous score for $G_t$
is $\frac{\theta_{t-1,t}+\theta_{t,t+1}}{2}$. We rank the graphs according to their anomalous scores in descending order.
Then we use the rank of the true synthetic anomalous graph to measure the effectiveness of 
the graph distance measure in detecting DDoS attacks.
We choose four candidates for the graph similarity measure:
$\calD_{\rm SI}$, $\calD_{\rm QJS}$, VEO score, and deltaCon.
And we repeat the random DDoS attacks for 100 times for each attack strength
$k\in\{5,10,20,30,40\}$. 
The results are shown in \prettyref{fig:anomaly}. 

The observations are two fold. 
First, $\calD_{\rm SI}$ and $\calD_{\rm QJS}$ have similar behaviors in analyzing graph streams.
Their trends in analyzing the synthetic graph stream $\calG$ are nearly identical.
Second, the structural information distance $\calD_{\rm SI}$ is very suitable for detecting DDoS attacks in a graph stream.
The structural information distance $\calD_{\rm SI}$ behaves better than the other competitors for the attack strength $k\in\{20,30,40,50\}$.
When $k\in\{5,10\}$, the performance of all the distance measures are mainly affected by the properties of the original normal graph stream.

\begin{figure*}[!t]
  \begin{subfigure}{0.28\textwidth}
    \includegraphics[width=\textwidth]{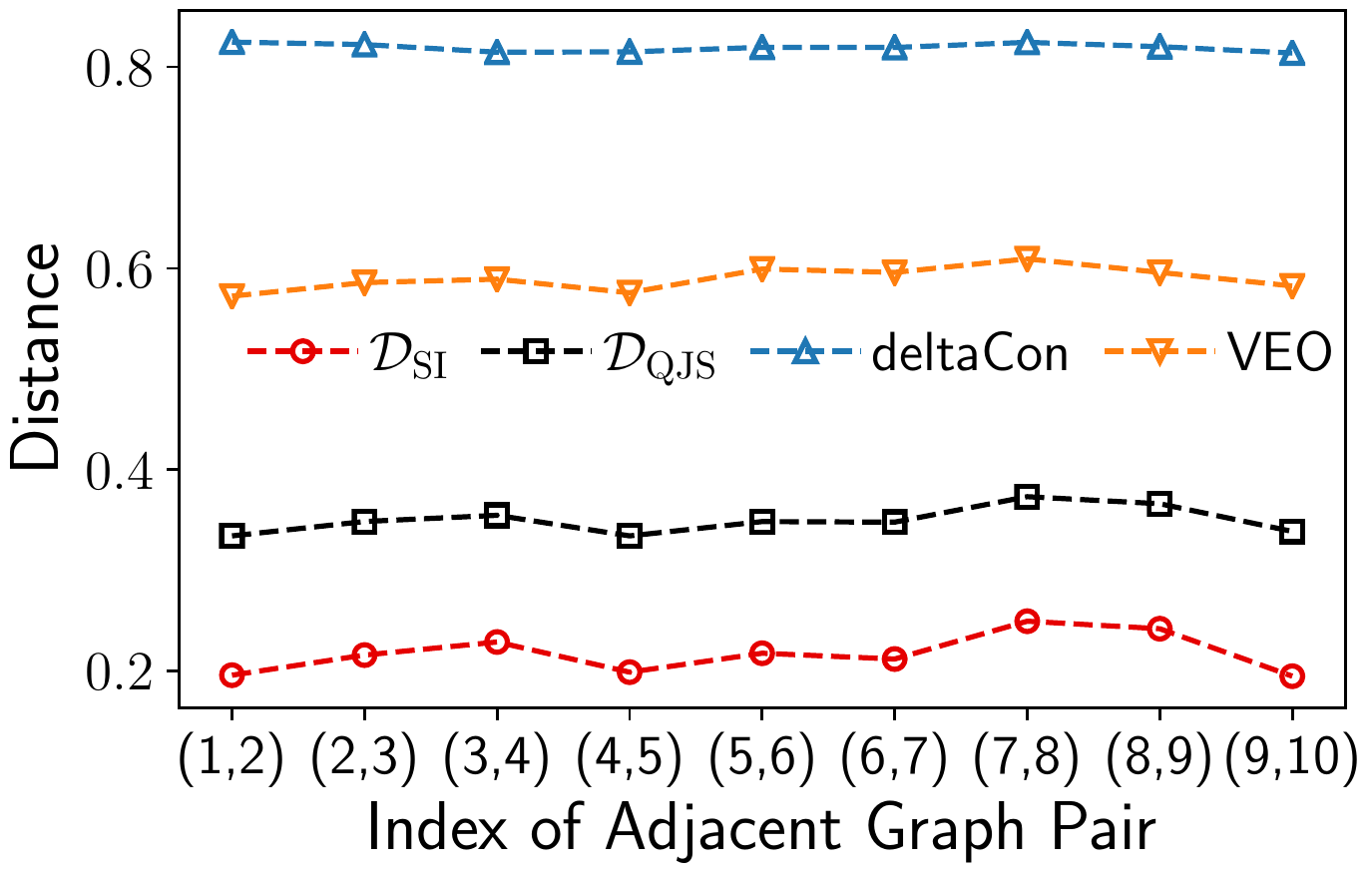}
    \caption{Distances between adjacent graphs in a synthetic graph stream $\calG$ from BA model with 100 nodes and $\overbar{d}=4$.}
  \end{subfigure}
  \hspace{2mm}
  \begin{subfigure}{0.68\textwidth}
    \includegraphics[width=\textwidth]{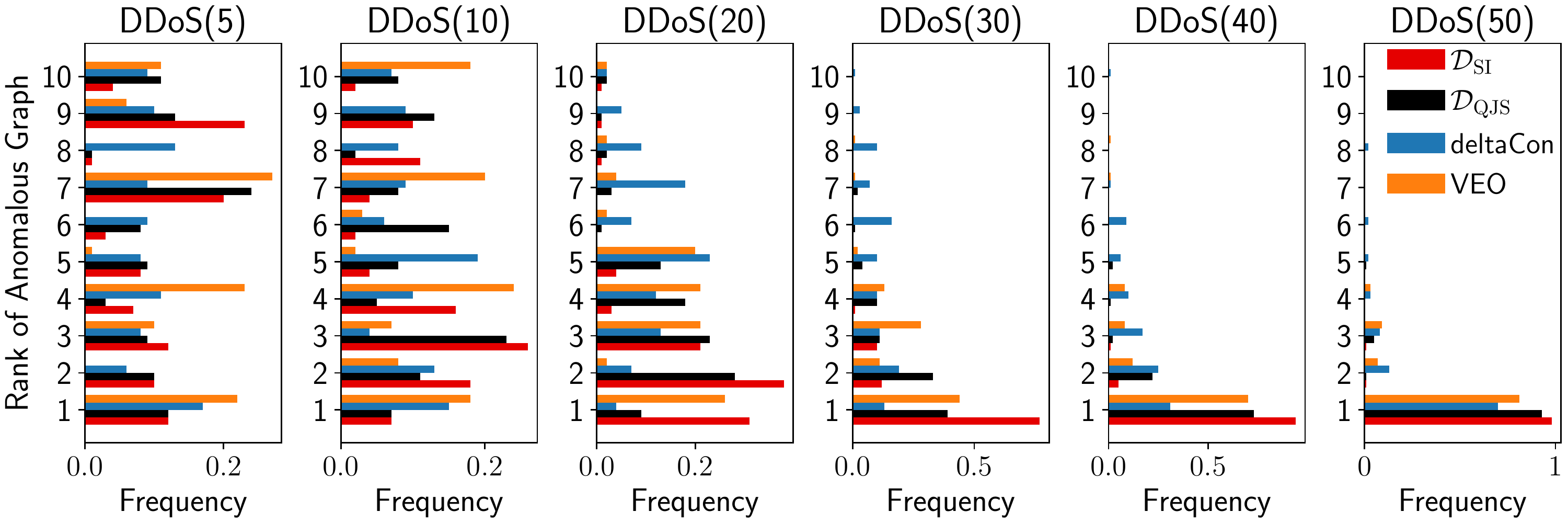}
    \caption{Rank of the anomalous graph under DDoS attack in the synthetic graph stream.}
  \end{subfigure}
  \caption{Structural information distance is well suited for detecting DDoS attacks in a graph stream.}
  \label{fig:anomaly}
\end{figure*}

\subsection{Performance in Community Obfuscation}
\label{app:appendix-comm}
To evaluate the performance of maximizing the von Neumann graph entropy 
(denoted as $\calA_0$) and minimizing the spectral polarization (denoted as $\calA_1$) in community obfuscation, we measure the dynamics of 
spectral gaps, detection error, graph entropy, and spectral polarization in the greedy edge addition process.
We evenly allocate the budget of edge additions among all the community pairs.
The results are shown in \prettyref{fig:deception-two} and \prettyref{fig:deception-three}.

The observations are four folds. 
First, the graph entropy is monotonically increasing w.r.t. the number of added edges.
Second, the spectral polarization is monotonically decreasing w.r.t. the number of added edges.
Third, the detection error is monotonically increasing w.r.t. the number of added edges. 
Therefore, both $\calA_0$ and $\calA_1$ are effective in community obfuscation.
Fourth, the spectral gap indicating the existence of community structure slightly increases in the beginning and then goes down quickly as more edges are added.

\begin{figure}[!t]
  \begin{subfigure}{0.49\linewidth}
      \includegraphics[width=\textwidth]{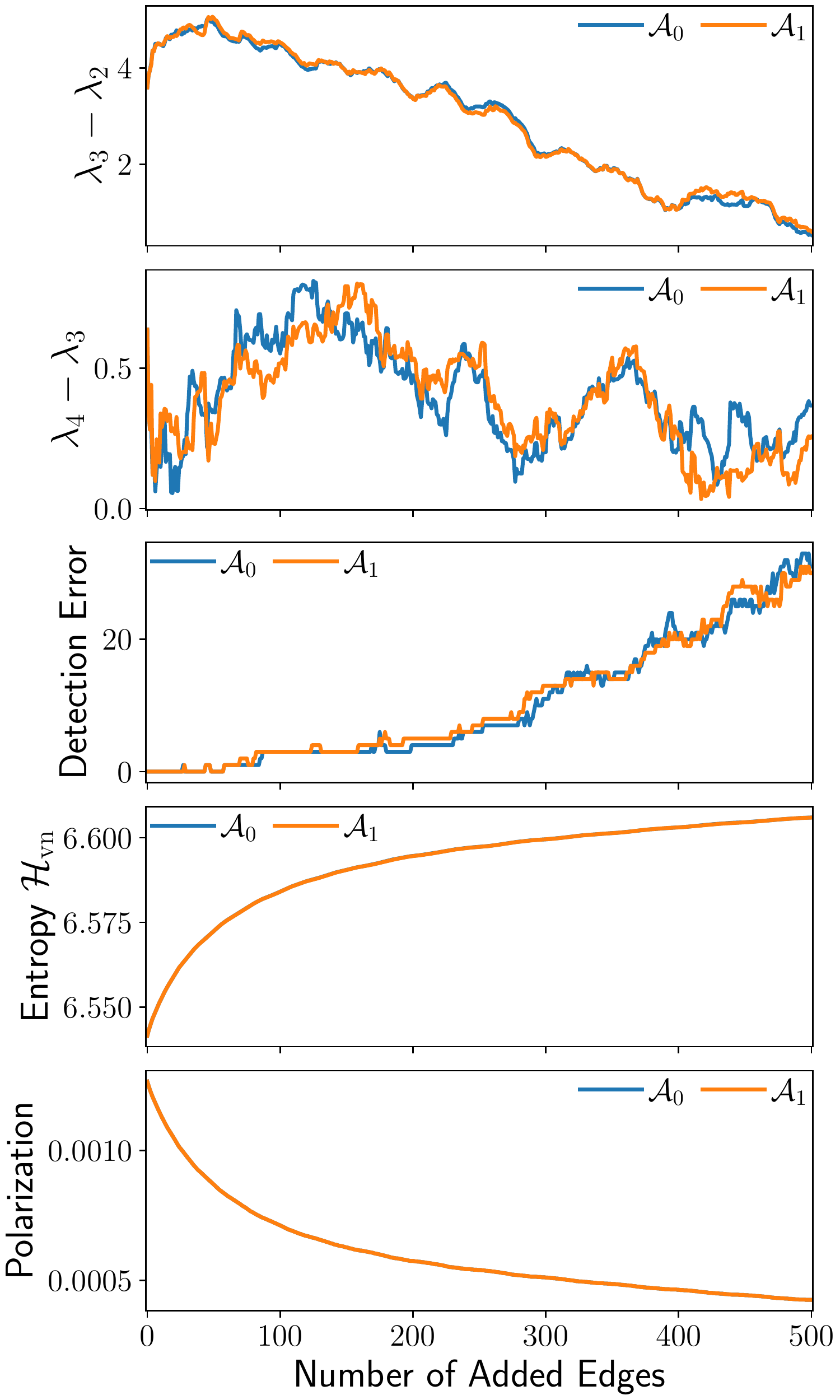}
      \caption{$c_{\rm in}=28, c_{\rm out}=2$}
  \end{subfigure}    
  \begin{subfigure}{0.49\linewidth}
      \includegraphics[width=\textwidth]{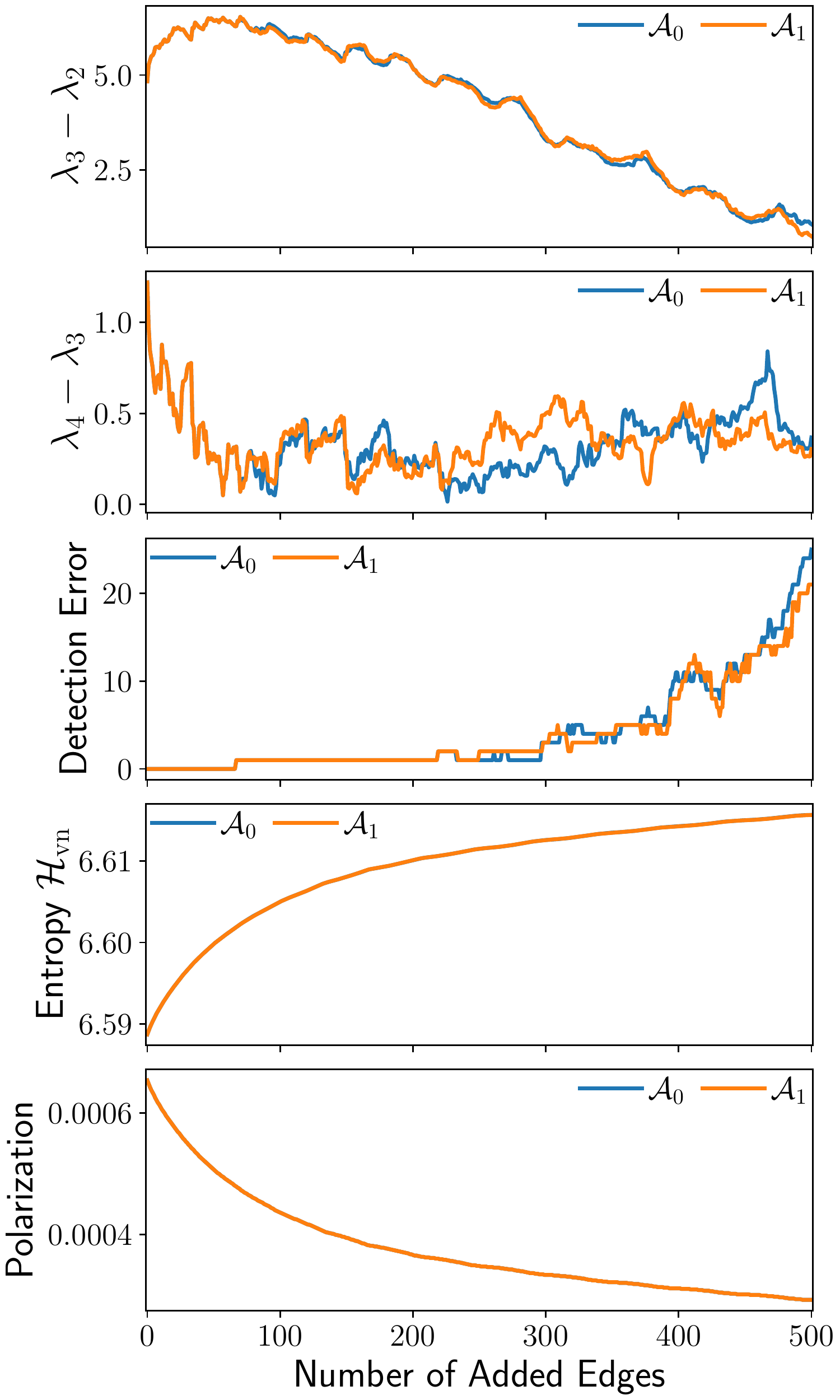}
      \caption{$c_{\rm in}=10, c_{\rm out}=40$}
  \end{subfigure}
  \caption{Community obfuscation on two graphs generated from stochastic block model with $100$ nodes. There are two clusters of equal size $50$.}
  \label{fig:deception-two}
\end{figure}

\begin{figure}[!t]
  \begin{subfigure}{0.49\linewidth}
      \includegraphics[width=\textwidth]{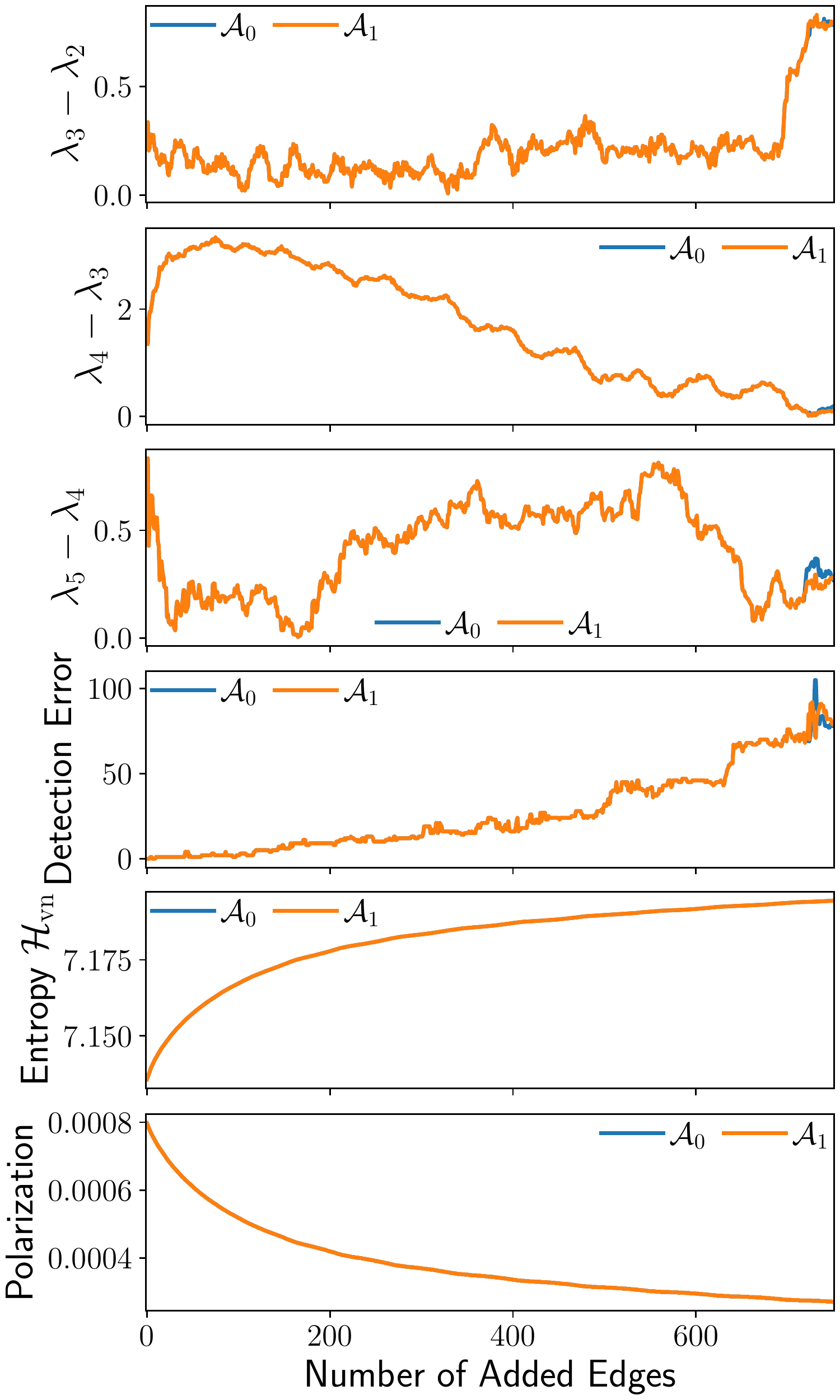}
      \caption{$c_{\rm in}=35, c_{\rm out}=5$}
  \end{subfigure}
  \begin{subfigure}{0.49\linewidth}
      \includegraphics[width=\textwidth]{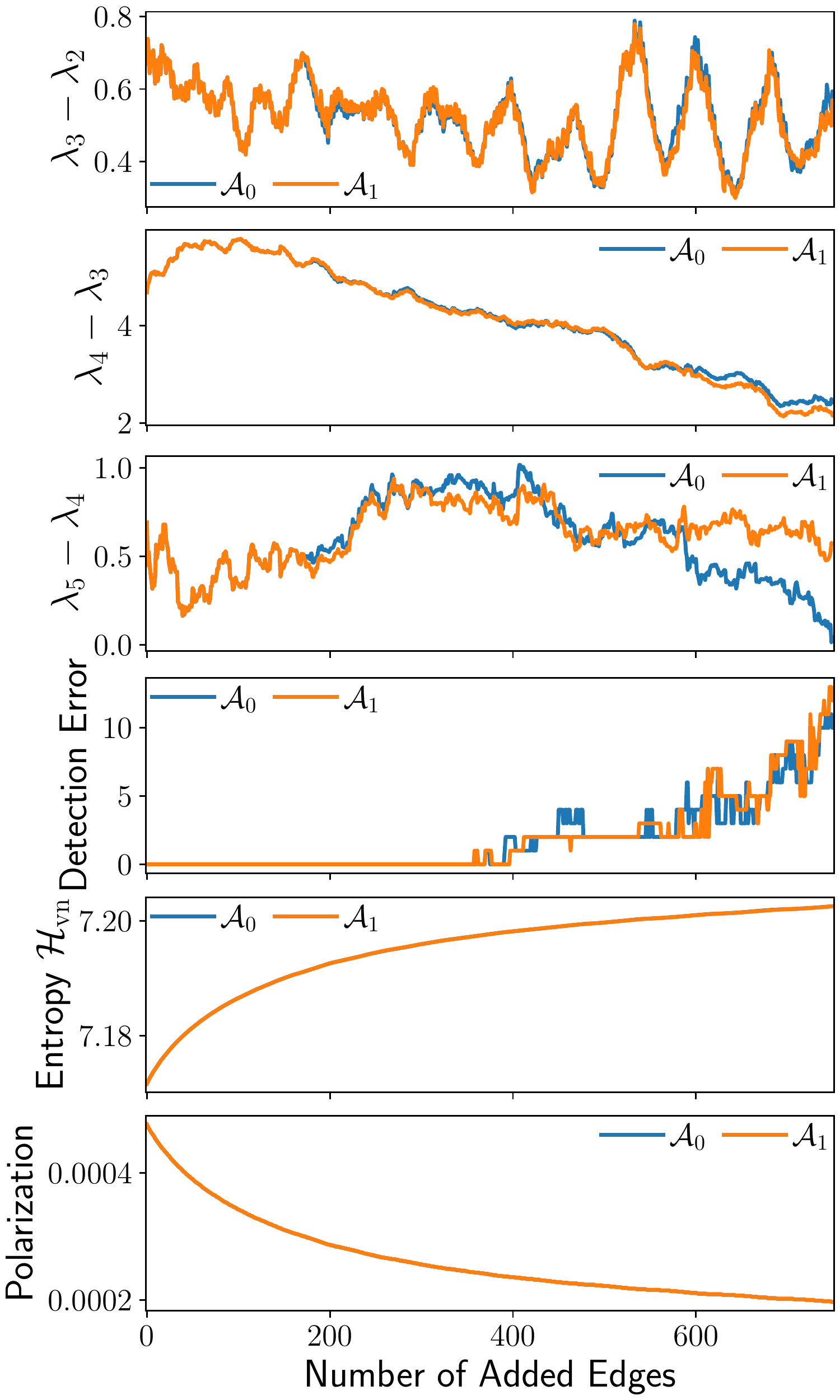}
      \caption{$c_{\rm in}=55, c_{\rm out}=10$}
  \end{subfigure}
  \caption{Community obfuscation on three graphs generated from stochastic block model with $150$ nodes. There are three clusters of equal size $50$.}
  \label{fig:deception-three}
\end{figure}

    \section{Proof of \prettyref{tab:various-graph-types}}
\label{app:proof-specific-graphs}

\subsection{Preliminaries: Several Integrations}
\begin{lemma}
  The integration 
  \begin{equation*}
    \calI_1\triangleq\int_0^\pi \log_2(1-\cos(x))\diff x=-\pi.
  \end{equation*}
\end{lemma}
\begin{IEEEproof}
  Let $x=t+\pi/2$, then $\diff x=\diff t$ and 
  \begin{equation*}
    \begin{aligned}
    \int_{\pi/2}^\pi \log_2(1-\cos(x))\diff x&=\int_0^{\pi/2}\log_2(1-\cos(t+\pi/2))\diff t \\
    &=\int_0^{\pi/2} \log_2(1+\cos(\pi/2-t))\diff t.
    \end{aligned}
  \end{equation*}
  Let $z=\pi/2-t$, then $\diff z=-\diff t$ and 
  \begin{equation*}
    \int_0^{\pi/2}\log_2(1+\cos(\pi/2-t))\diff t=-\int_{\pi/2}^0\log_2(1+\cos(z))\diff z.
  \end{equation*}
  Therefore, 
  \begin{equation*}
    \int_{\pi/2}^\pi\log_2(1-\cos(x))\diff x=\int_0^{\pi/2}\log_2(1+\cos(x))\diff x.
  \end{equation*}
  Then 
  \begin{equation*}
    \begin{aligned}
    \calI_1&=\int_0^{\pi/2}\log_2(1-\cos(x))\diff x + \int_{\pi/2}^\pi\log_2(1-\cos(x))\diff x \\
    &=\int_0^{\pi/2}\log_2(1-\cos(x))\diff x + \int_0^{\pi/2}\log_2(1+\cos(x))\diff x \\
    &=\int_0^{\pi/2}\log_2(\sin^2(x))\diff x \\
    &=2\int_0^{\pi/2}\log_2(\sin(x))\diff x.
    \end{aligned}
  \end{equation*}

  Let $x=\pi/2-t$, then $\diff x=-\diff t$ and 
  \begin{equation*}
    \begin{aligned}
      \int_0^{\pi/2}\log_2(\sin(x))\diff x&=-\int_{\pi/2}^0\log_2(\sin(\pi/2-t))\diff t \\
      &=\int_0^{\pi/2}\log_2(\sin(\pi/2-t))\diff t \\
      &=\int_0^{\pi/2}\log_2(\sin(\pi/2+t))\diff t \\
      &=\int_{\pi/2}^\pi\log_2(\sin(x))\diff x.
    \end{aligned}
  \end{equation*}
  Therefore,
  \begin{equation*}
    \begin{aligned}
      \calI_1&=\int_0^\pi\log_2(\sin(x)) \diff x \\
      &=\int_0^{\pi/2}\log_2(\sin(2t))\diff(2t) \\
      &=2\int_0^{\pi/2}\log_2(2\sin(t)\cos(t))\diff t \\
      &=2\left(\frac{\pi}{2}+\int_0^{\pi/2}\log_2(\sin(t))\diff t+\int_0^{\pi/2}\log_2(\cos(t))\diff t\right) \\
      &=\pi+\calI_1+2\int_0^{\pi/2}\log_2(\cos(t))\diff t.
    \end{aligned}
  \end{equation*}
  As a result,
  \begin{equation*}
    \begin{aligned}
      0&=\pi+2\int_0^{\pi/2}\log_2(\cos(x))\diff x \\
      &=\pi+2\int_0^{\pi/2}\log_2(\sin(\pi/2-x))\diff x \\
      &=\pi-2\int_{\pi/2}^0\log_2(\sin(z))\diff z \\
      &=\pi+2\int_0^{\pi/2}\log_2(\sin(z))\diff z \\
      &=\pi+\calI_1.
    \end{aligned}
  \end{equation*}
  Therefore, $\calI_1=-\pi$.
\end{IEEEproof}
\begin{lemma}
  The integration 
  \begin{equation*}
    \calI_2\triangleq\int_0^\pi \cos(x)\log_2(1-\cos(x))\diff x=-\pi\log_2 e.
  \end{equation*}
\end{lemma}
\begin{IEEEproof}
  Let $t=\sin(x)$, then $\diff t=\cos(x)\diff x$, $\cos(x)=\sqrt{1-t^2}$ for $x\in(0,\pi/2)$, and 
  $\cos(x)=-\sqrt{1-t^2}$ for $x\in(\pi/2,\pi)$. Therefore
  \begin{equation*}
    \begin{aligned}
    \calI_2&=\int_0^{\pi/2}\cos(x)\log_2(1-\cos(x))\diff x \\
    &\quad+\int_{\pi/2}^\pi \cos(x)\log_2(1-\cos(x))\diff x \\
    &=\int_0^1 \log_2(1-\sqrt{1-t^2})\diff t + \int_1^0 \log_2(1+\sqrt{1-t^2})\diff t \\
    &=\int_0^1 \log_2\left(\frac{1-\sqrt{1-t^2}}{1+\sqrt{1-t^2}}\right) \diff t \\
    &=\int_0^1 \log_2\left(\frac{(1-\sqrt{1-t^2})^2}{t^2}\right) \diff t \\
    &=2\int_0^1 \log_2(1-\sqrt{1-t^2})\diff t - 2\int_0^1\log_2 t\diff t.
    \end{aligned}
  \end{equation*}
  
  Define $\calI_3\triangleq\int_0^1 \log_2(1-\sqrt{1-t^2})\diff t$, $\calI_4\triangleq\int_0^1 \log_2 t\diff t$, and 
  a new function $G(t)\triangleq t\ln(1-\sqrt{1-t^2})-t-\sin^{-1}(t)$. Then 
  \begin{equation*}
    \begin{aligned}
    \frac{\diff G(t)}{\diff t}&=\ln(1-\sqrt{1-t^2})+\frac{t^2}{\sqrt{1-t^2}(1-\sqrt{1-t^2})} \\
    &\quad -1-\frac{1}{\sqrt{1-t^2}} \\
    &=\ln(1-\sqrt{1-t^2}),
    \end{aligned}
  \end{equation*}
  therefore,
  \begin{equation*}
    \begin{aligned}
    \calI_3&=\frac{1}{\ln 2}\int_0^1 \ln(1-\sqrt{1-t^2})\diff t \\
    &=\log_2 e(G(1)-G(0)) \\
    &=-\left(1+\frac{\pi}{2}\right)\log_2 e.
    \end{aligned}
  \end{equation*} 
  \begin{equation*}
    \begin{aligned}
      \calI_4&= \frac{1}{\ln 2}\int_0^1 \ln t\diff t \\
      &=\log_2 e\left(t\ln t|_0^1-\int_0^1 t\diff \ln t\right) \\
      &=-\log_2 e\int_0^1 t\cdot \frac{1}{t}\diff t \\
      &=-\log_2 e.
    \end{aligned}
  \end{equation*}
  Finally, 
  \begin{equation*}
    \calI_2=2\calI_3-2\calI_4=-\pi\log_2 e.
  \end{equation*}
\end{IEEEproof}
Define a new function $g(x)\triangleq f(2-2\cos(x))$, then we have the following corollary.
\begin{corollary}
  The integration
  \begin{equation*}
    \calI_5\triangleq\int_0^\pi g(x)\diff x=2\pi\log_2 e.
  \end{equation*}
\end{corollary}
\begin{IEEEproof}
  \begin{equation*}
    \begin{aligned}
      \calI_5&=\int_0^\pi(2-2\cos(x))\log_2(2-2\cos(x))\diff x \\
      &=2\int_0^\pi \log_2(2-2\cos(x))\diff x \\
      &\quad -2\int_0^\pi \cos(x)\log_2(2-2\cos(x))\diff x \\
      &=2(\pi+\calI_1)-2\left(\int_0^\pi\cos(x)\diff x+\calI_2\right) \\
      &=2(\pi+\calI_1-\calI_2)\\
      &=2\pi\log_2 e.
    \end{aligned}
  \end{equation*}
\end{IEEEproof}

\begin{corollary}
  \begin{equation*}
    \int_0^{2\pi}g(x)\diff x=2\int_0^\pi g(x)\diff x=2\int_\pi^{2\pi} g(x)\diff x.
  \end{equation*}
\end{corollary}
\begin{IEEEproof}
  Let $x=\pi-t$, then $\diff x=-\diff t$ and 
  \begin{equation*}
    \begin{aligned}
      \int_0^\pi g(x)\diff x=-\int_{\pi}^0 g(\pi-t)\diff t
      =\int_0^\pi g(\pi-t)\diff t.
    \end{aligned}
  \end{equation*}

  Let $x=\pi+t$, then $\diff x=\diff t$ and 
  \begin{equation*}
    \int_\pi^{2\pi} g(x)\diff x=\int_0^\pi g(\pi+t)\diff t.
  \end{equation*}
  Since $\cos(\pi-t)=\cos(\pi+t)$, $g(\pi-t)=g(\pi+t)$. Thus 
  \begin{equation*}
    \int_0^\pi g(\pi-t)\diff t=\int_0^\pi g(\pi+t)\diff t.
  \end{equation*}
  Therefore
  \begin{equation*}
    \int_0^\pi g(x)\diff x=\int_\pi^{2\pi}g(x)\diff x=\frac{1}{2}\int_0^{2\pi} g(x)\diff x.
  \end{equation*}
\end{IEEEproof}

\subsection{Complete Graph}
  The Laplacian spectrum of complete graph $K_n$ is $\bm{\lambda}=(n,n,\ldots,n,0)$ and the degree sequence of $K_n$ is 
  $\mathbf{d}=(n-1,n-1,\ldots,n-1)$, thus $\ose(K_n)=\log_2 n$ and $\vnge(K_n)=\log_2 (n-1)$ yielding 
  $\Delta\calH(K_n)=\ose(K_n)-\vnge(K_n)=\log_2(1+\frac{1}{n-1})$.

\subsection{Complete Bipartite Graph}
  The Laplacian spectrum of complete bipartite graph $K_{a,b}$ is 
  \[
  \bm{\lambda}=(a+b,\underbrace{a,\ldots,a}_{b-1},\underbrace{b,\ldots,b}_{a-1},0),
  \]
  and the degree sequence of $K_{a,b}$ is 
  \[
  \mathbf{d}=(\underbrace{a,\ldots,a}_{b},\underbrace{b,\ldots,b}_{a}),
  \]
  therefore
  \begin{equation*}
    \begin{aligned}
      \ose(K_{a,b})&=\log_2(2ab)-\frac{ba\log_2 a+ab\log_2 b}{2ab} \\
      &=1+\frac{1}{2}\log_2(ab),
    \end{aligned}
  \end{equation*}
  and 
  \begin{equation*}
    \begin{aligned}
      \vnge(K_{a,b})&=\log_2(2ab)-\frac{ba\log_2 a+ab\log_2 b}{2ab} \\
      &\quad -\frac{(a+b)\log_2(a+b)-a\log_2 a-b\log_2 b}{2ab} \\
      &=1+\frac{1}{2}\log_2(ab)-\frac{\log_2(1+\frac{b}{a})}{2b}-\frac{\log_2(1+\frac{a}{b})}{2a}.
    \end{aligned}
  \end{equation*}
  The entropy gap 
  \begin{equation*}
    \begin{aligned}
      \Delta\calH(K_{a,b})&=\ose(K_{a,b})-\vnge(K_{a,b}) \\
      &=\frac{\log_2(1+\frac{b}{a})}{2b}+\frac{\log_2(1+\frac{a}{b})}{2a}.
    \end{aligned}
  \end{equation*}

\subsection{Path Graph}
  The Laplacian spectrum of path graph $P_n$ is 
  \[
    \bm{\lambda}=\left(2-2\cos\left(\frac{\pi k}{n}\right),k=0,\ldots,n-1\right),  
  \]
  and the degree sequence of $P_n$ is 
  \[
    \mathbf{d}=(1,\underbrace{2,\ldots,2}_{n-2},1),  
  \]
  therefore 
  \begin{equation*}
    \begin{aligned}
    \ose(P_n)&=\log_2(2n-2)-\frac{(n-2)\cdot 2\log_2 2}{2n-2} \\
    &=\log_2(n-1)+\frac{1}{n-1},
    \end{aligned}
  \end{equation*}
  and 
  \begin{equation*}
    \begin{aligned}
      \vnge(P_n)=\log_2(2n-2)-\frac{\sum_{k=0}^{n-1} f(2-2\cos(\frac{\pi k}{n}))}{2n-2}.
    \end{aligned}
  \end{equation*}
  Then $\vnge(P_n)-\log_2(2n-2)$ can be expressed as 
  \begin{equation*}
    \begin{aligned}
      -\frac{\sum_{k=0}^{n-1}g(\frac{\pi k}{n})}{2n-2}
      &=\left[\frac{\pi}{n}\sum_{k=0}^{n-1}g\left(\frac{\pi k}{n}\right)\right]\cdot\frac{n}{\pi}\cdot\frac{-1}{2n-2} \\
      &\xrightarrow{n\rightarrow\infty}-\frac{1}{2\pi}\int_0^\pi g(x)\diff x = -\log_2 e.
    \end{aligned}
  \end{equation*}
  Therefore, $\vnge(P_n)-\log_2(n-1)\xrightarrow{n\rightarrow\infty}1-\log_2 e$.

\subsection{Ring}
  The Laplacian spectrum of ring graph $R_n$ is 
  \[
    \bm{\lambda}=\left(2-2\cos\left(\frac{2\pi k}{n}\right),k=0,\ldots,n-1\right),  
  \]
  and the degree sequence of $R_n$ is $\mathbf{d}=(2,2,\ldots,2)$, therefore
  $\ose(R_n)=\log_2 n$ and 
  \begin{equation*}
    \begin{aligned}
      \vnge(R_n)=\log_2(2n)-\frac{\sum_{k=0}^{n-1}f(2-2\cos(\frac{2\pi k}{n}))}{2n}.
    \end{aligned}
  \end{equation*}
  Then $\vnge(R_n)-\log_2(2n)$ can be expressed as  
  \begin{equation*}
    \begin{aligned}
      -\frac{\sum_{k=0}^{n-1}g(\frac{2\pi k}{n})}{2n}
      &=\left[\frac{2\pi}{n}\sum_{k=0}^{n-1}g\left(\frac{2\pi k}{n}\right)\right]\cdot\frac{n}{2\pi}\cdot\frac{-1}{2n} \\
      &\xrightarrow{n\rightarrow\infty}-\frac{1}{4\pi}\int_0^{2\pi}g(x)\diff x = -\log_2 e.
    \end{aligned}
  \end{equation*}
  Therefore, $\vnge(R_n)-\log_2 n\xrightarrow{n\rightarrow\infty}1-\log_2 e$.

\section{Proof of \prettyref{thm:DSI}}
We are going to establish a close relation between $\calD_{\rm SI}$
and the Jensen-Shannon divergence $\calD_{\rm JS}$, then the pseudometric properties of $\calD_{\rm SI}$
simply follow from the metric properties of $\sqrt{\calD_{\rm JS}}$.

The structural information  
\[  
  \ose(G_j)=-\sum_{i=1}^n f\left(\frac{d_{i,j}}{\vol(G_j)}\right)=H(P_j)
\] 
where $P_j=\left(\frac{d_{1,j}}{\vol(G_j)},\ldots,\frac{d_{n,j}}{\vol(G_j)}\right)$ is a distribution on the set $V$.

In the graph $\overbar{G}=(V,E_1\cup E_2,\overbar{A})$, the degree $\overbar{d}_i$ of node $i$ is 
\begin{equation*}
  \begin{aligned}
  \overbar{d}_i&=\sum_{j=1}^n\overbar{A}_{ij} \\
  &=\sum_{j=1}^n \frac{A_{ij,1}}{2\vol(G_1)}+\frac{A_{ij,2}}{2\vol(G_2)} \\
  &=\frac{d_{i,1}}{2\vol(G_1)}+\frac{d_{i,2}}{2\vol(G_2)}.  
  \end{aligned}
\end{equation*}
Then the volume of $\overbar{G}$ is $\vol(\overbar{G})=\sum_{i=1}^n \overbar{d}_i=1$. Therefore the structural information of $\overbar{G}$ is 
\begin{equation*}
  \begin{aligned}
    \ose(\overbar{G})&=-\sum_{i=1}^n f\left(\frac{\overbar{d}_i}{\vol(\overbar{G})}\right) \\
  &=-\sum_{i=1}^n f\left(\frac{d_{i,1}}{2\vol(G_1)}+\frac{d_{i,2}}{2\vol(G_2)}\right),
  \end{aligned}
\end{equation*}  
which is equivalent to the entropy of the distribution $(P_1+P_2)/2$.
As a result, $\calD_{\rm SI}(G_1,G_2)=\sqrt{\calD_{\rm JS}(P_1,P_2)}$.


\section{Proof of \prettyref{thm:JS-and-DSI}}
As shown in \prettyref{thm:DSI}, the claim is true for $\calD_{\rm SI}$.
It remains to prove that $\calD_{\rm QJS}(G_1,G_2)\leq 1$, and if $\min\{d_{i,1},d_{i,2}\}=0$
for every node $i\in V$ then $\calD_{\rm QJS}(G_1,G_2)=1$. 

We prove 
$\calD_{\rm QJS}(G_1,G_2)\leq 1$ using the inequality \cite{nielsen_chuang_2010,Majtey2005PRA} for the von Neumann entropy:
if $\rho=\sum_{i}p_i\rho_i$ is a mixture of density matrix $\rho_i$ with $p_i$ a set of positive real numbers such that $\sum_{i}p_i=1$, then 
  $\vnge(\sum_i p_i\rho_i)\leq\sum_i p_i\vnge(\rho_i)+H(\{p_i\})$. 
Note that the scaled Laplacian matrix $L_i/\tr(L_i)$ of the graph $G_i$ can be viewed as a density matrix.
Then 
\begin{equation*}
  \begin{aligned}
  \calD_{\rm QJS}(G_1,G_2)&=\vnge(\overbar{G})-(\vnge(G_1)+\vnge(G_2))/2 \\
  &=\vnge(\tilde{L}_1+\tilde{L}_2)-(\vnge(\tilde{L}_1)+\vnge(\tilde{L}_2))/2 \\
  &\leq\vnge(\tilde{L}_1+\tilde{L}_2)-\vnge((\tilde{L}_1+\tilde{L}_2)/2)+1 \\
  &=1.
  \end{aligned}
\end{equation*}

We denote by $S_j$ the set of singletons in the graph $G_j$ for $j\in\{1,2\}$.
Since $\min\{d_{i,1},d_{i,2}\}=0$ for every node $i\in V$, we have $S_1\cup S_2=V$
which implies that $(V\backslash S_1)\cap (V\backslash S_2)=\varnothing$ by the De Morgan's laws.
Therefore, the node set $V$ can be partitioned into three disjoint subsets $V\backslash S_1$, $V\backslash S_2$, and $S_1\cap S_2$.
Notice that one singleton contributes one eigenvalue of $0$ to the Laplacian spectrum, and the Laplacian
spectrum of a graph is composed of the Laplacian spectrum of its each connected components.
We denote by $\lambda_{j,1},\ldots,\lambda_{j,n-s_j},0,\ldots,0$ the Laplacian spectrum of $G_j$,
where $s_j=|S_j|$ for $j\in\{1,2\}$. It follows that $\sum_{i=1}^{n-s_j}\lambda_{j,i}=\vol(G_j)$.
Since $\overbar{A}=A_1/2\vol(G_1)+A_2/2\vol(G_2)$, $\overbar{L}=\overbar{L}_1/2\vol(G_1)+\overbar{L}_2/2\vol(G_2)$.
Then the Laplacian spectrum of $\overbar{G}$ is 
composed of Laplacian spectrum of $G_j$ divided by $2\vol(G_j)$ for $j\in\{1,2\}$ and zeros.
As a result, 
\begin{equation*}
  \begin{aligned}
    &\calD_{\rm QJS}(G_1,G_2) \\
    &=-\sum_{j=1}^{2}\sum_{i=1}^{n-s_j}f\left(\frac{\lambda_{j,i}}{2\vol(G_j)}\right)
    +\frac{1}{2}\sum_{j=1}^{2}\sum_{i=1}^{n-s_j}f\left(\frac{\lambda_{j,i}}{\vol(G_j)}\right) \\
    &=\sum_{j=1}^{2}\sum_{i=1}^{n-s_j}\frac{\lambda_{j,i}}{2\vol(G_j)}\log_2 2
    =\sum_{j=1}^2\frac{\vol(G_j)}{2\vol(G_j)}=1.
  \end{aligned}
\end{equation*}

\section{Proof of \prettyref{lem:incremental}}
Denote by $\tilde{d}$ the degree sequence of $G_{k+1}$, then 
\begin{equation*}
  \begin{aligned}
    &\quad\ose(G_{k+1})=-\sum_{i=1}^n f\left(\frac{\tilde{d}_i}{2(m+\Delta m)}\right) \\
    &=\frac{f(2(m+\Delta m))-\sum_{i=1}^n f(\tilde{d}_i)}{2(m+\Delta m)} \\
    &=\frac{f(2(m+\Delta m))-\sum_{i\in V_k}f(d_i+\Delta d_i)-\sum_{i\in\overbar{V}_k}f(d_i)}{2(m+\Delta m)} \\
    &=\frac{f(2(m+\Delta m))-a-\sum_{i=1}^n f(d_i)}{2(m+\Delta m)} \\
    &=\frac{f(2(m+\Delta m))-a-f(2m)+2m\ose(G_k)}{2(m+\Delta m)}.
  \end{aligned}
\end{equation*}

The structural information $\ose(\overbar{G}_k)$ is equal to 
\begin{equation*}
  \begin{aligned}
    &\quad-\sum_{i=1}^n f\left(\frac{d_i}{4m}+\frac{\tilde{d}_i}{4(m+\Delta m)}\right) \\
    &=-b-\sum_{i\in\overbar{V}_k}f\left(\frac{2m+\Delta m}{4m(m+\Delta m)}d_i\right) \\
    &=-b-\sum_{i\in\overbar{V}_k}cd_i(\log_2 c+\log_2 d_i) \\
    &=-b-f(c)\sum_{i\in \overbar{V}_k}d_i-c\sum_{i\in \overbar{V}_k}f(d_i) \\
    &=-b-f(c)(2m-y)-c(f(2m)-2m\ose(G_k)-z)
  \end{aligned}
\end{equation*}

    \section*{Acknowledgement}
    The authors would like to thank the reviewers and the Associated Editor, Prof. Ronen Talmon, for their detailed reading and 
    valuable comments that greatly improved the quality of the paper.

    \bibliographystyle{IEEEtran}
    \bibliography{citation}

    \begin{IEEEbiographynophoto}{Xuecheng Liu}
        received the bachelor's degree in information engineering from Shanghai Jiao Tong University in 2017.
        He is currently pursuing the Ph.D. degree in information and communication engineering from Shanghai Jiao Tong University.
        His research interests include network science and graph algorithms, with applications on big data analytics.
    \end{IEEEbiographynophoto}

    \begin{IEEEbiographynophoto}{Luoyi Fu}
        received her B.E. degree in Electronic Engineering in 2009 and Ph.D. degree in Computer Science and Engineering from Shanghai
        Jiao Tong University, China, in 2015. She is currently an assistant professor in Department of Computer Science and Engineering in Shanghai
        Jiao Tong University. Her research of interests are in the area of social networking and big data,
        scaling laws analysis in wireless networks, connectivity analysis and random graphs. She has
        been a member of the Technical Program Committees of several conferences including ACM MobiHoc 2018--2020,
        IEEE INFOCOM 2018--2020.
    \end{IEEEbiographynophoto}

    \begin{IEEEbiographynophoto}{Xinbing Wang}
        received the B.S. degree (Hons.) from the Department of Automation, Shanghai Jiao Tong University, Shanghai, China, in 1998,
        the M.S. degree from the Department of Computer Science and Technology, Tsinghua University, Beijing, China, in 2001, and the Ph.D.
        degree, majoring in the electrical and computer engineering and minoring in mathematics,
        from North Carolina State University, Raleigh, NC, USA, in 2006. He is currently a professor with the Department of Electronic Engineering,
        Shanghai Jiao Tong University. He has been an Associate Editor for the IEEE/ACM Transactions on Networking and the IEEE Transactions on
        Mobile Computing, and a member of the Technical Program Committees of several conferences including the ACM MobiCom 2012, the ACM
        MobiHoc 2012--2014, and the IEEE INFOCOM 2009--2017.
    \end{IEEEbiographynophoto}

    \begin{IEEEbiographynophoto}{Chenghu Zhou}
        received the B.S. degree in geography from Nanjing University, Nanjing, China,
        in 1984, and the M.S. and Ph.D. degrees in
        geographic information system from the Chinese
        Academy of Sciences (CAS), Beijing, China, in
        1987 and 1992, respectively. He is currently an
        academician with CAS, where he is also a research professor with the Institute of Geographical Sciences and Natural Resources Research,
        and a professor with the School of Geography
        and Ocean Science, Nanjing University. His research interests include spatial and temporal data mining, geographic
        modeling, hydrology and water resources, and geographic information
        systems and remote sensing applications.
    \end{IEEEbiographynophoto}
\end{NoHyper}
\end{document}